\documentclass[namedreferences]{solarphysics}
\usepackage[optionalrh,solaromanenum]{spr-sola-addons} 
\usepackage{graphicx}                    
\usepackage{amssymb}                    
\usepackage{color}                       
\usepackage{url}                         

\begin{document}

\begin{article}

\begin{opening}

\title{The Signature of Flare Activity   in Multifractal Measurements  of Active Regions Observed by SDO/HMI}
%
\author{F.~\surname{Giorgi}$^{1}$\sep
        I.~\surname{Ermolli}$^{1}$\sep
        P.~\surname{Romano}$^{2}$\sep     
                  M.~\surname{Stangalini}$^{1}$\sep
 F.~\surname{Zuccarello}$^{3}$\sep   
        S.~\surname{Criscuoli}$^{4}$}
%
\runningauthor{Giorgi et al.}
\runningtitle{Multifractal Parameters of  Flaring Regions on SDO/HMI Magnetograms}

%
  \institute{$^{1}$ INAF Osservatorio Astronomico di Roma, via Frascati 33, 00040 Monte Porzio Catone, Italy\\
                     email: \url{giorgi@oaroma.inaf.it} \\ 
            $^{2}$ INAF - Osservatorio Astrofisico di Catania,
              Via S. Sofia 78, 95125 Catania, Italy \\
            $^{3}$ Dipartimento di Fisica e Astronomia - Sezione Astrofisica, Universit\`{a} di Catania,
			 Via S. Sofia 78, 95125 Catania, Italy \\
             $^{4}$ National Solar Observatory
			Sacramento Peak P.O. Box 62 Sunspot, NM 88349-0062, United States 
		}

\begin{abstract}
Recent studies indicate that measurements of  fractal and  multifractal parameters of active regions (ARs)  are not efficient tools  to discriminate ARs on the basis of the flare activity, as well as to predict flare events.  Attempting validation  of this result on  a large data set of higher spatial and temporal resolution, as well as higher flux sensitivity, observations than employed in previous studies,  we analyzed  high-cadence time series of   line-of-sight magnetograms of  43 ARs characterized by different flare activity, which  were observed with SDO/HMI from May 2010 to December  2013.  On these data, we estimated four  parameters, the generalized fractal dimensions $D_0$ and  $D_8$, and the multifractal parameters $C_{\mathrm{div}}$ and $D_{\mathrm{div}}$.
{ 
We found distinct average values of  the parameters measured on ARs that have hosted flares of different class. However,   the dispersion of  values measured on  ARs that have produced same class events  is such that the parameters deduced from distinct classes  of flaring regions  can also largely overlap.
Based on the results of our measurements, C- and M-class flaring ARs are practically indistinguishable, as well as M- and X-class flaring ARs. 
We found 
consistent changes on the time series of the  measured parameters only on $\approx$ 50\% of the studied ARs and $\approx$ 50\% of the M- and X-class events considered. We show that these results hold for fractal and multifractal parameter estimates based on both total unsigned and signed flux data of the analyzed ARs.
}
\end{abstract}

%

\end{opening}

%
\section{Introduction}
Solar flares are among the most energetic events in the solar system.  
They 
occur in active regions (ARs), by involving  processes  at various heights in the solar atmosphere that are powered by the magnetic field (Shibata and Magara, 2011).    Observations show that  the  photospheric topology of  magnetic field is one of the key factors in determining the evolution of ARs. There is a general trend for large regions to produce large flares, and for more complex regions to generate more numerous and larger flares than other regions of comparable size (\opencite{Sammis_etal2000}).  Emergence of a  bipolar region  that  interacts with preexisting magnetic field in the corona, and activation of a filament with destabilization in the subsisting magnetic region, are known as precursors events preceding the onset of flares. However, the precise conditions required to create the enormously energetic (up to 10$^{32}$ erg on the timescale of hours) flare events are as yet unknown (\inlinecite{Leka_Barnes_2013}, and references therein).  
Because interplanetary disturbances associated with solar flares can impact  space- and ground-based infrastructures,  
identification of precursor conditions leading to solar flares is getting a growing interest  for   practical applications of space weather forecast (\opencite{Hapgood_2012}).

Any measurement of magnetic energy and complexity in an AR  deserves interest for testing its sensitivity to single out the flare activity of the region for space weather forecasts. However, among the various measurements,  
those concerning multifractal properties of the  region in the photosphere have the potential of providing information about the interaction between magnetic field and plasma motions that 
may lead to flare events.  { Indeed,}  these  parameters are thought to describe AR properties that can be consequence of a self-organized evolution (see discussion by {\it e.g.} \opencite{Georgoulis_2012}).

Several studies  in the literature indicate  that measurements  of fractal and multifractal properties of ARs may help assessing the eruptive potential  of the regions and also predicting their  flare activity (for a list of studies carried out during the last decade see {\it e.g.} \inlinecite{Ermolli_etal2013}, hereafter Paper I). { Nevertheless, the  literature also shows  conflicting conclusions on the efficiency of these measurements to discriminate flaring ARs. For example, \inlinecite{Mcateer_etal2005} reported no discernible
differences between the fractal dimension distributions of 
different Mount Wilson classes of ARs that were studied on  {\it Solar and Heliospheric Observatory} (SOHO)/{\it Michelson-Doppler Imager} (MDI;  \opencite{Scherrer_etal1995}) line-of-sight (LOS) magnetograms.
 However, they also showed that there is a good correlation between the increased flare productivity of ARs, in terms of both
flare size and frequency, and increased
fractal dimension measured in the  ARs. Besides, they identified a
lower threshold fractal dimension value  which an
AR must possess in order to have the potential to
produce an M-class (X-class) flare.
More recently,}  \inlinecite{Georgoulis_2012} reported that  none of the  widely-used multifractal parameters  measured  in his study  {  on a large data set of SOHO/MDI LOS magnetograms}
allows to distinguish AR with major flares from flare-quiet ones, though both classes of solar regions show significant fractality and multifractality.  Then, from  analysis of higher resolution observations than employed in the previous study, \inlinecite{Georgoulis_2013} also found that the multiscale parameters measured on {\it Solar Dynamics Observatory} (SDO)/{\it Heliospheric and Magnetic Imager} (HMI; \opencite{Scherrer_etal2012}; \opencite{Schou_etal2012}; \opencite{Wachter_etal2012})  LOS magnetograms of  an intensely eruptive  AR and of  a flare-quiet AR  have fairly similar values and temporal evolutions. 

Recent results from an  analysis of  near-si\-mul\-ta\-neous 
observations of a flaring AR by SOHO/MDI and SDO/HMI
(Paper I)  
suggested us to further investigate the response of  the fractal and multifractal parameters  to the flare activity of ARs. 
In particular,  Paper I  
shows that the measurements   derived from   SOHO/MDI observations  are affected by larger and spurious variations  over time than those deduced from analysis of the higher-quality SDO/HMI data. Therefore, analysis of higher quality  observations of a larger sample of ARs than analyzed by \inlinecite{Georgoulis_2013} may produce measurements of the multifractal parameters characterized by a significantly reduced dispersion than derived by  \inlinecite{Georgoulis_2012}. 
 The study presented herein investigates the possibility that a reduced dispersion of multifractal parameters measured on SDO/HMI observations of ARs may help discriminating  flaring from flare-quiet regions.
To this purpose, we analyzed time series of magnetograms of { 43}  ARs with different flare activity that were observed with SDO/HMI from May 2010 to { December}  2013. 
We also noticed 
that almost all the studies  of multifractal properties in flaring ARs were carried out on the total unsigned flux measurement of the regions, which is considered a  standard-flare forecasting tool (\opencite{Barnes_Leka_2008}). 
To the knowledge of these authors, only \inlinecite{Georgoulis_2005} reported that discriminating between the two polarities in the studied ARs  do not alter the results derived from  analysis of total unsigned flux measurements of  ARs. 
 New observational evidences on the physical mechanisms driving the  evolution of magnetic field in more flaring ARs, concerning {\it e.g.} the photospheric magnetic flux change and cancellation     (\opencite{Burtseva_etal2013}; \opencite{Cliver_etal2012}) and the fragmentation of magnetic helicity patches (\opencite{Romano_etal2011}), 
suggested us to study  multifractal parameters inside ARs, by taking into account the evolution of magnetic regions with different polarities, thus using signed magnetic flux measurements  for multifractal parameters estimates rather than total unsigned flux data.  
It has been also shown that the regions with different polarity in an  AR have distinct morphology (see {\it e.g.} \opencite{Yamamoto_2012}) and dynamical behavior (\opencite{Giannattasio_etal2013}).
Therefore, we also tested  the response of  measurements of multifractal parameters  based on signed flux measurements to the flare activity of the ARs. 

The paper is organized as follows. The data sets and analysis methods employed in our study are described in Section 2. Our main findings are presented in Section 3, whereas the  conclusions of this study are summarized  in Section 4.

\section{Data and Methods}
\label{s2}

\begin{table}
\caption{ ARs analyzed in  this study accordingly to the NOAA catalog, with the number of B-, C-, M-, and X-class events hosted by the region during its  transit over the solar disc,  the  flare index (FI) computed during the same time interval and the maximum value of the flare index (FI max) for the most intense event produced by the region.  The FI values are given using $\times 10^{-3}$~erg~cm$^{-2}$s$^{-1}$  units. Details are given in the text. Date (year, month, day) and time (hour, minutes) indicate  the AR transit at the central meridian, while Hem is the solar hemisphere hosting the AR. M- and X-class flaring ARs are marked bold.
}
\label{tbl:0}
\begin{tabular}{llllllllll}     
\hline
AR & 	B  &	C	&M	&X &    FI  &     FI  max & date & time (UT) &Hem\\
\hline
11072  &  6   &    0   &   0   &   0   &     1.3  &     0.7  &  2010 05 23 & 00:24 & S \\
11117  & 55   &    5   &   0   &   0   &    28.1  &     5.7  &  2010 10 25 & 19:24 & N \\
11143  &  2   &    0   &    0   &    0   &     0.6   &     0.4  &  2011 01 08 &22:24  & S \\
 11146  &  2   &    0   &    0   &    0   &     0.4   &     0.2  &  2011 01 14 &14:48  & N \\     
{\bf 11158}  &  6   &   54   &   7   &   1   &   615.6  &   220.0  &  2011 02 13 & 23:59  & S\\
{\bf 11166}  &  7   &   25   &    4   &    1   &   303.6   &   150.0  &  2011 03 08 &20:24  & N \\
{\bf 11283}  & 10   &   16   &    5   &    2   &   602.7   &   210.0  &  2011 09 05 &20:24  & N \\
 11318  &  0   &    1   &    0   &    0   &     1.4   &     1.4  &  2011 10 12 & 17:24  & N \\
{\bf 11429}  &  3   &   47   &  15   &   3   &  1342.6  &   540.0  &  2012 03 08 & 22:48  & N\\
{\bf 11504}  &  8   &   42   &   5   &   0   &   178.9  &    19.0  &  2012 06 15 & 00:24  & S\\
{\bf 11513}  &  8   &   25   &   7   &   0   &   205.7  &    28.0  &  2012 07 02 & 04:24  & N\\
{\bf 11515}  &  5   &   77   &  24   &   0   &  1064.1  &    69.0  &  2012 07 02 & 23:24  & S\\
{\bf 11520}  &  1   &   32   &   8   &   1   &   444.3  &   140.0  &  2012 07 12 & 11:36  & S\\
{\bf 11536}  &  2   &   20   &   4   &   0   &   166.3  &    61.0  &  2012 08 01& 11:24  & S\\
{\bf 11613}  &  1   &   21   &   5   &   0   &   184.9  &    60.0  &  2012 11 16 &01:36  & S\\
{\bf 11618}  &  5   &   23   &   5   &   0   &   152.8  &    35.0  &  2012 11 21 &17:00  & N\\
11640  &  7   &    6   &   0   &   0   &    15.9  &     4.0  &  2013 01 01 &05:24  & N\\
{\bf 11652}  &  3   &   23   &   3   &   0   &    96.9  &    17.0  &  2013 01 11 &23:24  & N\\
11660  & 13   &    3   &   0   &   0   &    16.4  &     5.8  &  2013 01 20 &17:24  & N\\
11663  &  5   &    1   &   0   &   0   &     2.8  &     1.1  &  2013 01 30 &04:36  & S\\
11667  & 19   &    6   &   0   &   0   &    34.8  &     8.7  &  2013 02 07 &08:24  & N\\
11670  & 13   &    3   &   0   &   0   &    10.2  &     1.5  &  2013 02 09 &21:00  & N\\
{\bf 11675}  &  9   &    5   &   1   &   0   &    31.6  &    19.0  &  2013 02 19 &06:24  & N\\
11689  &  5   &    3   &   0   &   0   &     7.2  &     2.2  &  2013 03 10 &08:48  & S\\
11691  &  8   &    1   &   0   &   0   &     6.8  &     2.1  &  2013 03 12 &08:12  & N\\
{\bf 11692}  & 15   &   14   &   2   &   0   &    64.1  &    16.0  &  2013 03 15 &22:00  & N\\
11696  &  6   &    7   &   0   &   0   &    12.2  &     1.4  &  2013 03 14 &12:36  & N\\
11708  & 10   &    2   &   0   &   0   &     8.4  &     2.7  &  2013 04 03 &00:24  & N\\
{\bf 11719}  & 10   &   16   &   2   &   0   &   125.3  &    65.0  &  2013 04 12 &08:36  & N\\
{\bf 11726}  & 16   &   72   &   1   &   0   &   194.5  &    10.0  &  2013 04 20 &14:24  & N\\
{\bf 11731}  &  1   &   35   &   3   &   0   &   115.0  &    14.0  &  2013 04 30 &21:36  & N\\
11734  &  2   &   27   &   0   &   0   &    51.0  &     3.4  &  2013 05 05 &01:36  & S\\
{\bf 11739}  &  2   &   28   &   2   &   0   &   149.8  &    57.0  &  2013 05 09 &04:24  & N\\
11756  &  5   &   15   &   0   &   0   &    34.6  &     3.9  &  2013 05 25 &21:24  & S\\
{\bf 11778}  &  3   &   11   &   1   &   0   &    65.1  &    29.0  &  2013 06 28 &22:36  & S\\
 11863  &  1   &    0   &    0   &    0   &     0.0   &     0.0  &  2013 10 10 & 14:48  & S\\
11868  &  0   &    2   &    0   &    0   &     4.9   &     2.9  &  2013 10 18 & 02:36  & N\\
{\bf 11875}  &  1   &   66   &  11   &   2   &   829.8  &   230.0  &  2013 10 23 &09:24  & N\\
{\bf 11882}  &  0   &   10   &  13   &   2   &   695.6  &   210.0  &  2013 10 30 &10:24  & S\\
{\bf 11890}  &  2   &   51   &   5   &   3   &   840.2  &   330.0  &  2013 11 08 &18:00  & S\\
 11896  &  0   &    1   &    0   &    0   &     1.9   &     1.9  &  2013 11 16 &20:24  & N\\
 11918  &  0   &    4   &    0   &    0   &     7.3   &     2.2  &  2013 12 14 &18:24  & S\\
11923  &  1   &    0   &    0   &    0   &     0.0   &     0.0  &  2013 12 12 &22:24  & N\\
\hline
\end{tabular}
\end{table}

We analyzed { 43}   ARs  characterized by different flare activity that were  observed by SDO/HMI 
from May 2010  to { December}  2013. Table \ref{tbl:0} lists the ARs in our sample, by providing information about the flare events hosted by the region and its flaring level. The latter is described by  the flare index  (FI; \opencite{LI_etal2004}), which was computed assuming the flare history of the region depicted by the NOAA's GOES X-ray archive during the  disc transit of the region, as well as by considering the most intense event hosted by the region {  (hereafter referred to as maximum FI, FI max). FI max  is employed in the following to discriminate the analyzed ARs depending on their flaring activity.  In particular, the  X-, M-, and B or C-class flaring ARs discussed below correspond to  the regions whose FI max$>$100, 10$<$FI max$<$100, and   FI max$<$10, respectively.}

{ The AR sample analyzed in our study contains 22 regions that have  hosted $\ge$ M-class flares and 21 regions that have produced from a single B-class  to a few B- or C-class events. 
The analyzed  ARs
include all the regions  observed from February 2011 to December 2013  that  have produced either M- or X-class flares,  with 
the disc position specified in the following,  and 17  ARs  that were either flare-quiet (FI max$<$1) or  less flaring (10$<$FI max$<$100)  than  the flaring regions, observed over the time interval and disc position of the flaring ARs. The  sample also contains 4 flare-quiet ARs observed from May 2010 to January 2011, which were considered for comparison with results of earlier
studies.}

According to the data and methods applied in  Paper I, the selected ARs were  analyzed using the time series of SDO/HMI Level 1.5 LOS full-disc magnetograms when  the AR longitudinal distance was within  $\pm 30^\circ$ of the central meridian.  The whole data set consists of   { 20571}  magnetograms, each of 4096 $\times$4096 pixels,  with a pixel size of 0.505  arcsec and cadence  of 720 s.  The spatial resolution is set by the instrument aperture to 1 arcsec.
The data set was processed as described in Paper I. The main steps are summarized below, following the description given in the original paper.

From each magnetogram, we extracted a sub-array centered on the AR baricenter, whose size slightly exceeds 512$\times$512  pixels.  On these data, according to results presented in Paper I and other studies in the literature ({\it e.g.} \opencite{Georgoulis_2012}), geometrical projection effects were not compensated for. { Indeed, Paper I shows  that 
additional de-projection correction only slightly alter the fractal
and multifractal values obtained for the AR, with differences of measurement results that are well within uncertainties of
values.
However, it is worth noting that the geometrical de-projection of the data does not allow one to recover the loss of spatial resolution due to the increasing solar  area subtended by the resolution element of the recording device when one moves from the disc center to the limb. This resolution loss  
cannot be  compensated
for when estimating the fractal and multifractal parameters, but it  can be minimized with restriction of the AR positions analyzed. These 
considerations  motivate the restriction of our analysis to un-projected  ARs with longitudinal distance  
within  $\pm 30^\circ$ of the central meridian. 
}
  

On each sub-array, we computed the four parameters describing the fractal and multifractal properties of the photospheric magnetic field: 
the generalized fractal dimension $D_0$ and $D_8$, and the multifractal Contribution Diversity $C_{\mathrm{div}}$ and Dimensional Diversity $D_{\mathrm{div}}$.  
These parameters are  extensively described by {\it e.g.} \inlinecite{Mcateer_etal2005}, \inlinecite{Criscuoli_etal2009},  \inlinecite{Georgoulis_2012}, and in Paper I.
We investigated the variation of these parameters with respect to  the flare activity of the analyzed ARs, by considering results derived from both the unsigned and signed flux measurements of the photospheric magnetic field in the observations. 
The parameters were computed on the region of interest derived from  application of a threshold in the  line-of-sight field component in each sub-array. According to results in Paper I, the threshold values employed  are $\pm$ 3 times  the standard deviation of quiet sun measurements  in the data, specifically $\pm$   20 G (gauss). 
All the sub-arrays employed in our study  are unaffected by  flux saturation.

The accuracy of the methods and algorithms  employed in our study, as well the effects on the measured values of using different thresholds for the analysis, are discussed in Paper I. { The uncertainty associated with the measured values is equal to the 2-sigma uncertainty for the parameters derived as in Paper I. This uncertainty  is shown in {\it e.g.} Figure \ref{f8}, for the results of the parameter measurements from unsigned flux data of the analyzed ARs.}


\section{Results}
     \begin{figure*}
      \centerline{\includegraphics[trim={0.5cm 2cm  0.5cm 0.2cm},clip,width=5.2cm]{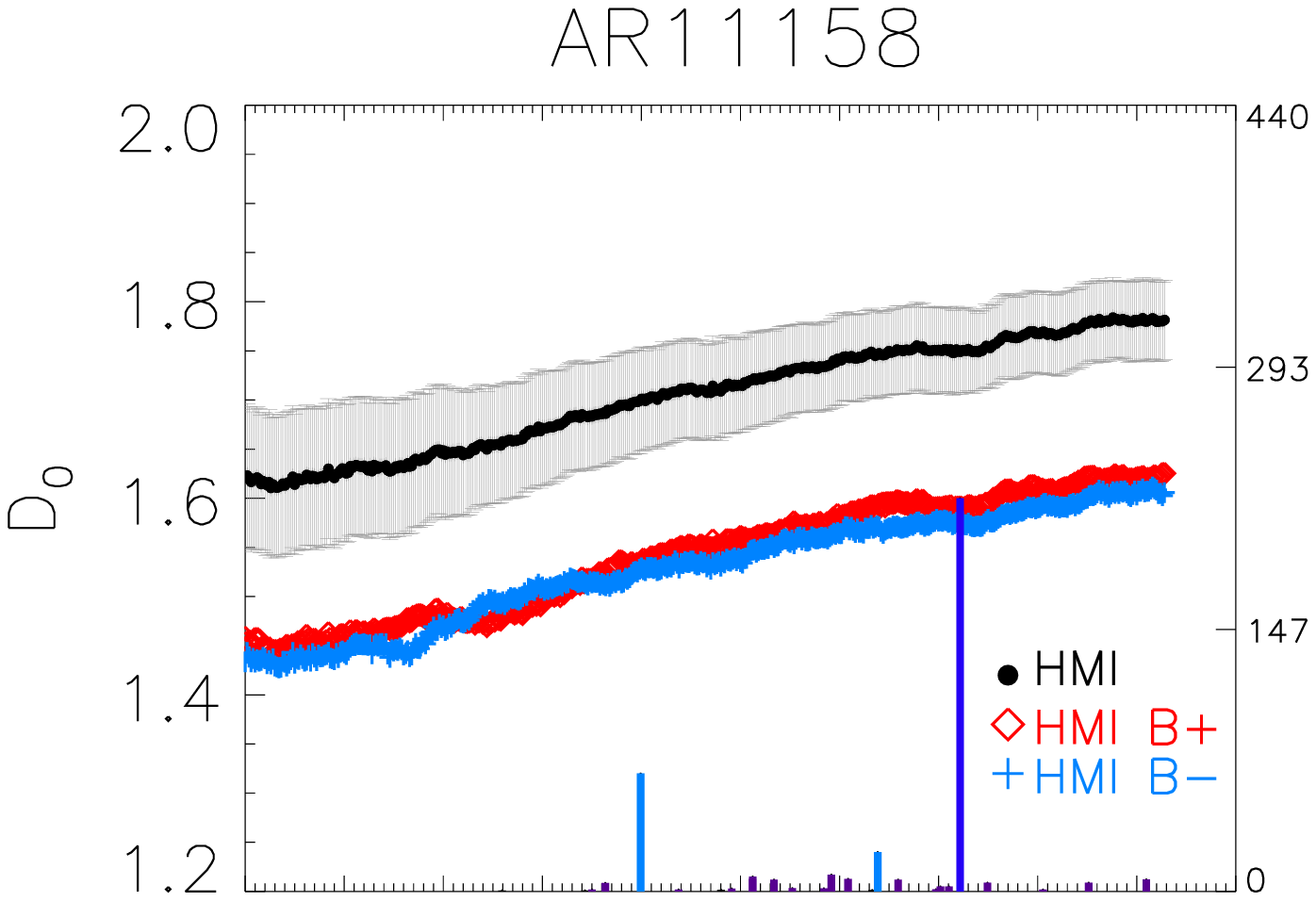}\includegraphics[trim={0.5cm 2cm  0.5cm 0.2cm},clip,width=5.2cm]{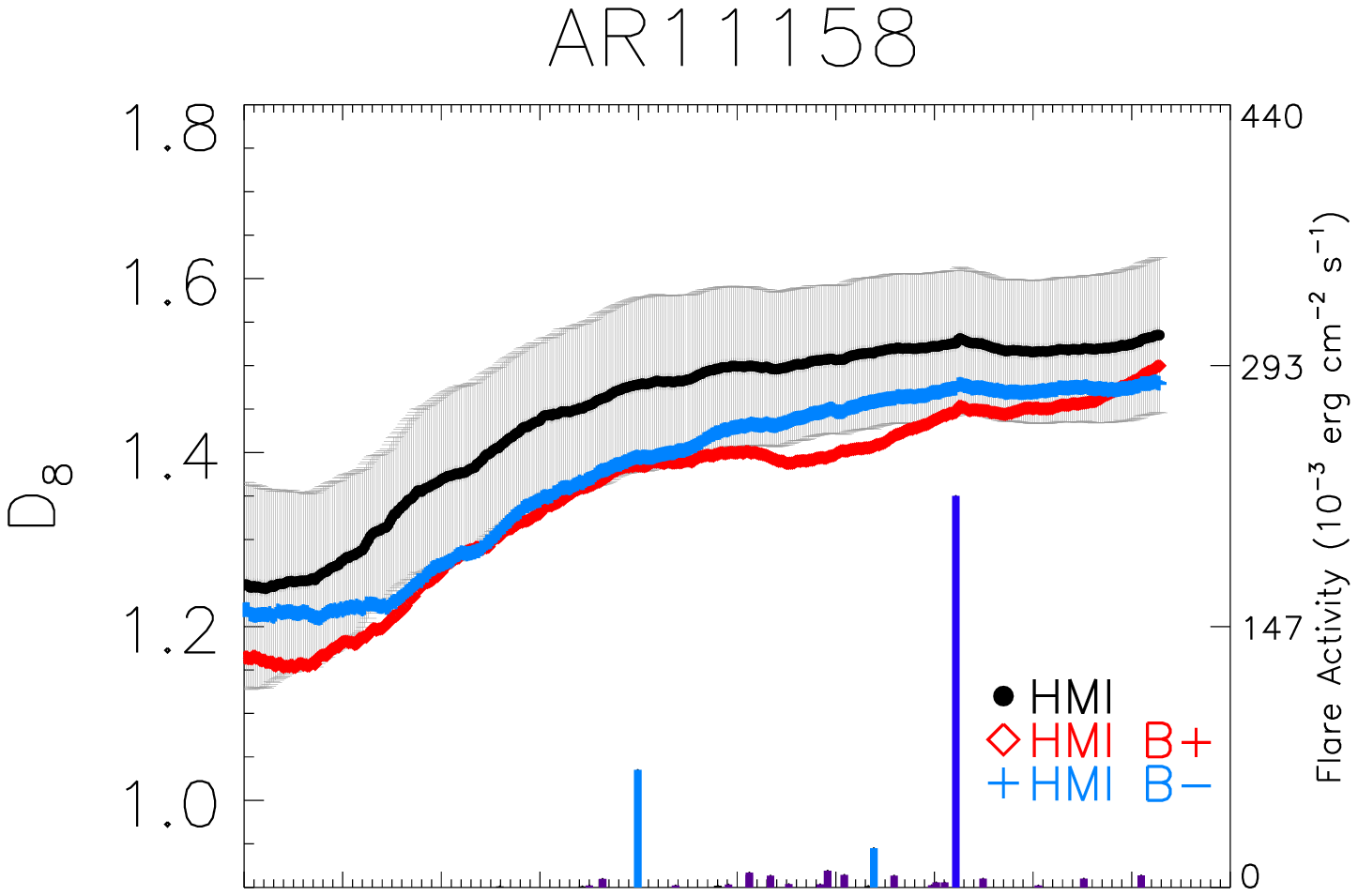}}
      \centerline{\includegraphics[trim={0.5cm 2cm  0.5cm  0.2cm},clip,width=5.2cm]{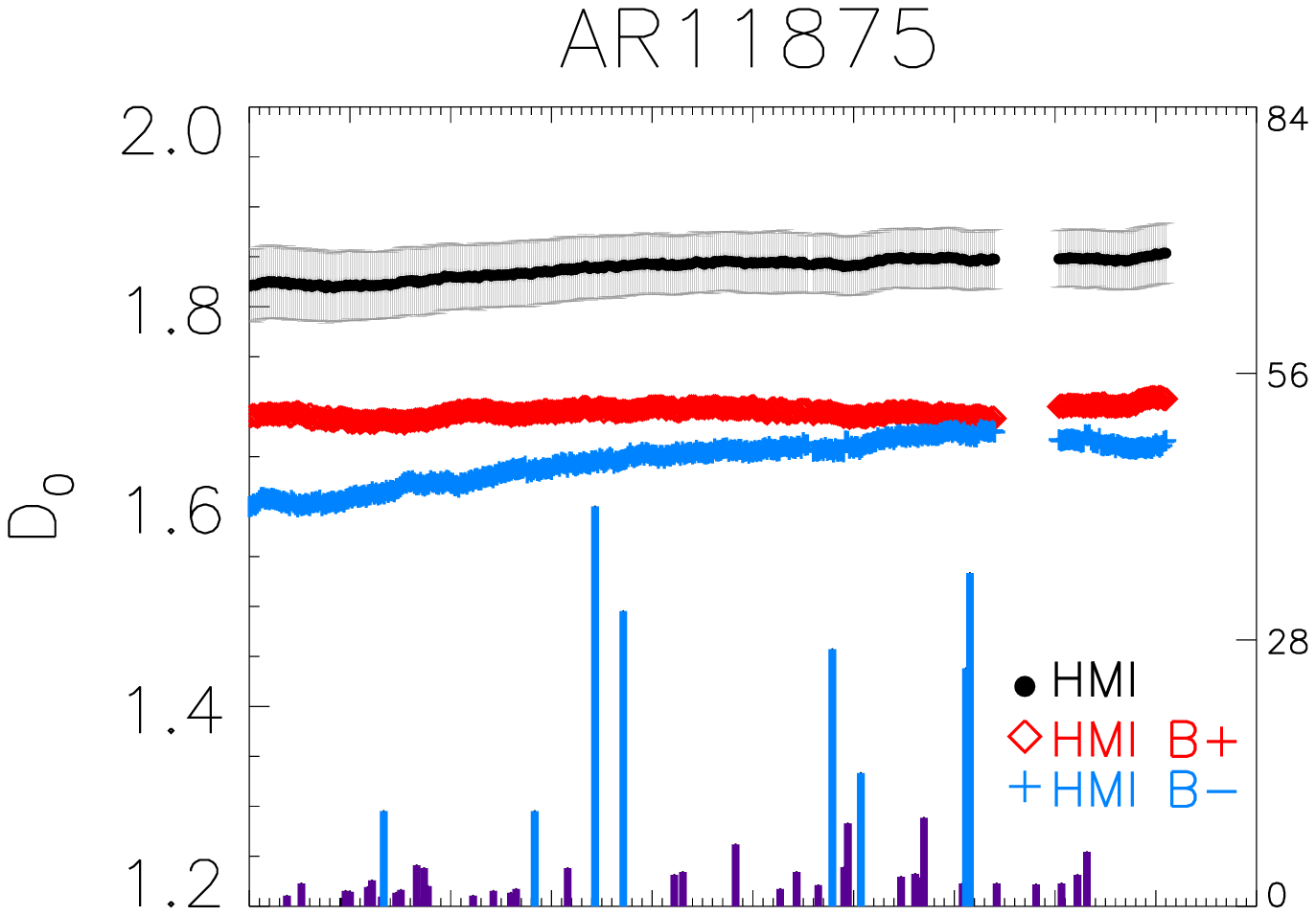}\includegraphics[trim={0.5cm 2cm  0.5cm  0.2cm},clip,width=5.2cm]{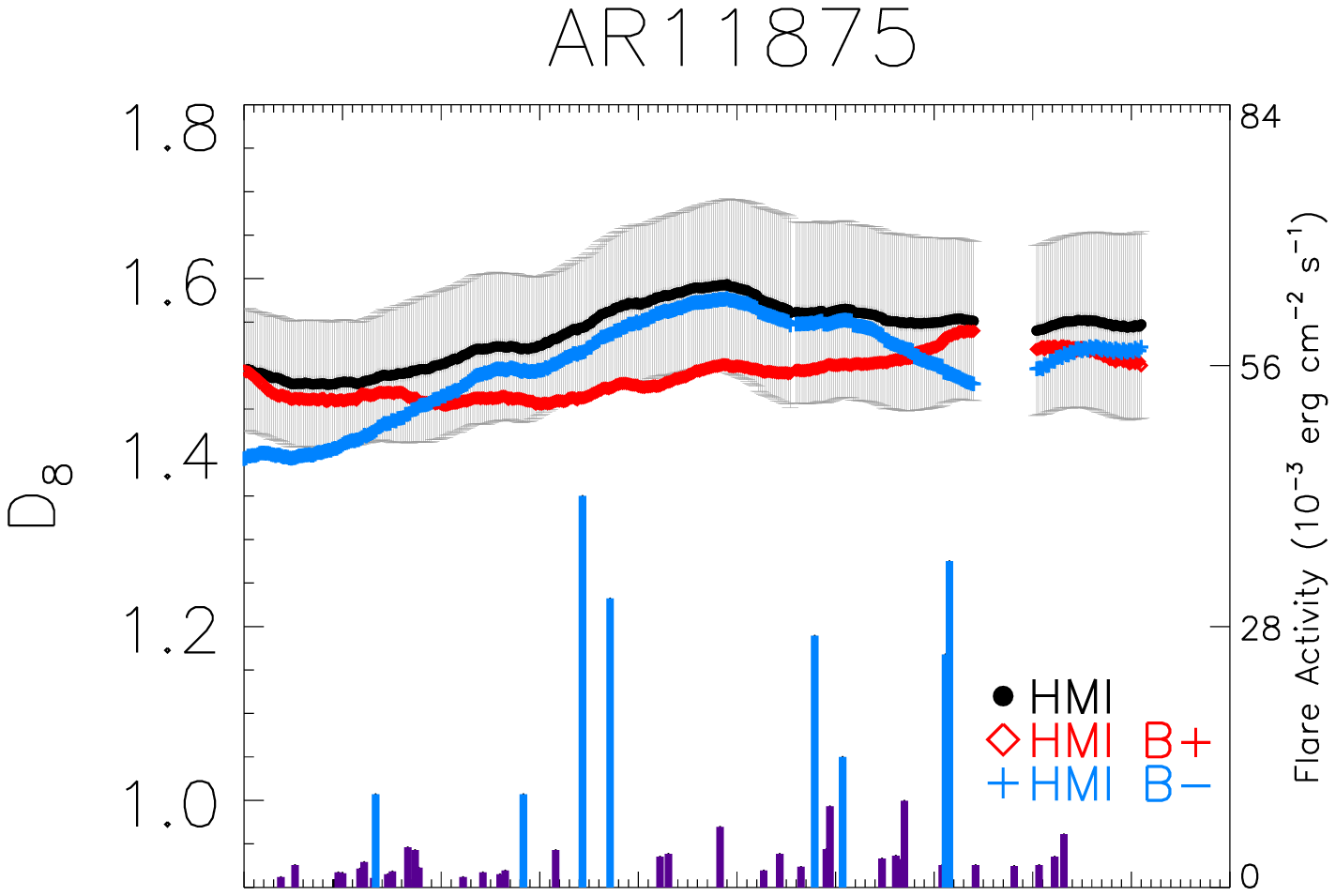}}
      \centerline{\includegraphics[trim={0.5cm 2cm  0.5cm  0.2cm},clip,width=5.2cm]{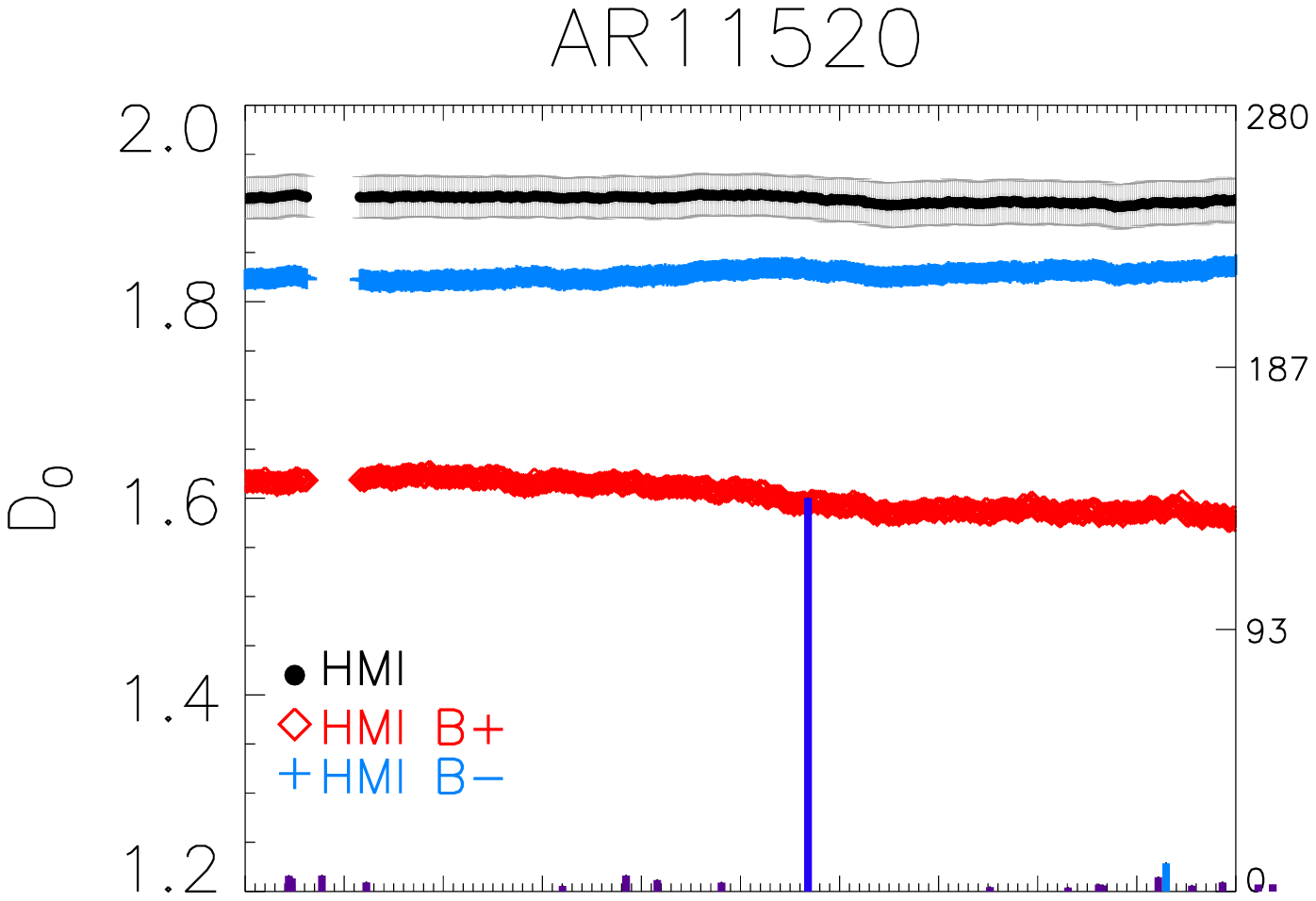}\includegraphics[trim={0.5cm 2cm  0.5cm  0.2cm},clip,width=5.2cm]{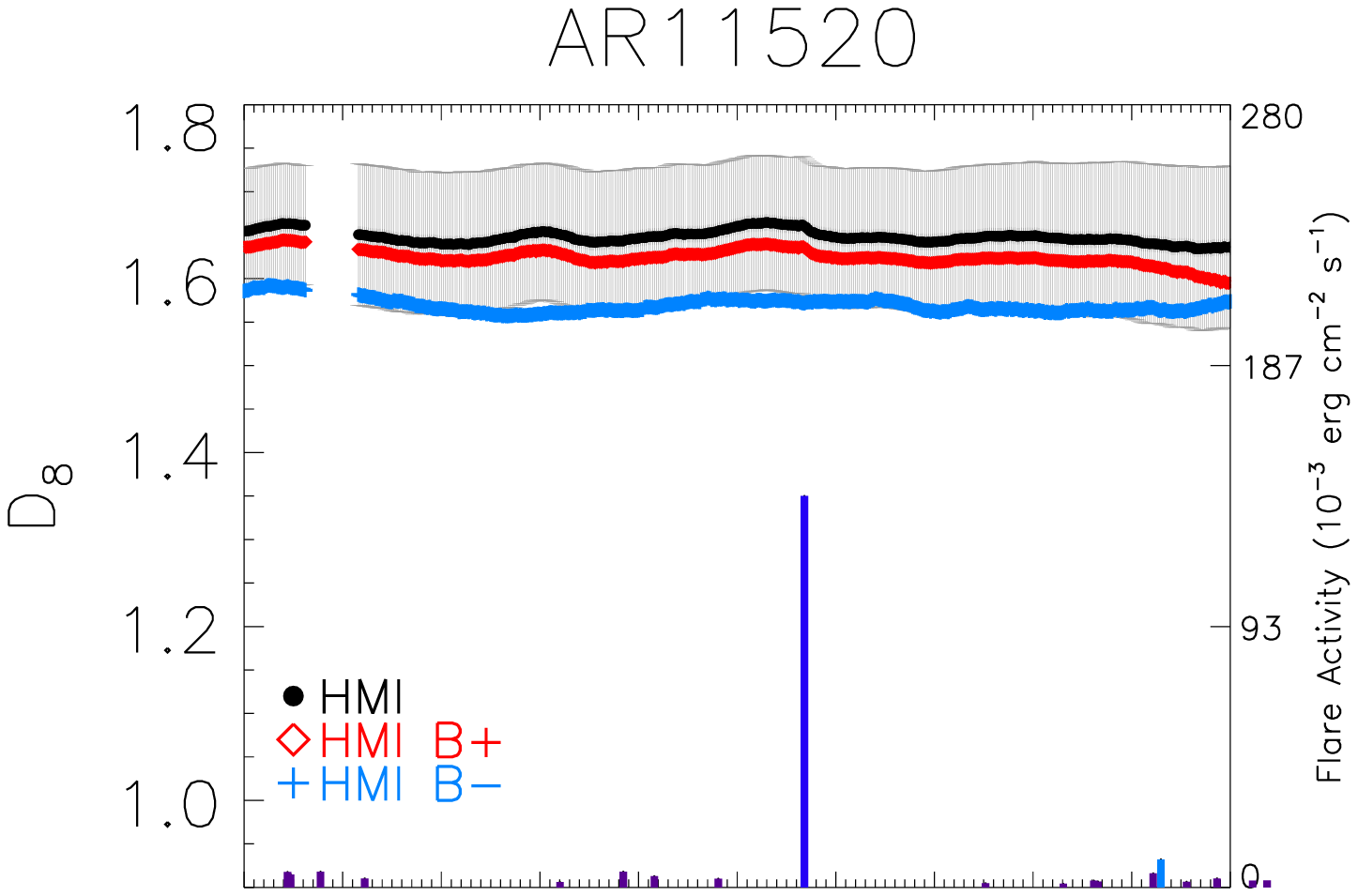}}
      \centerline{\includegraphics[trim={0.5cm 2cm  0.5cm  0.2cm},clip,width=5.2cm]{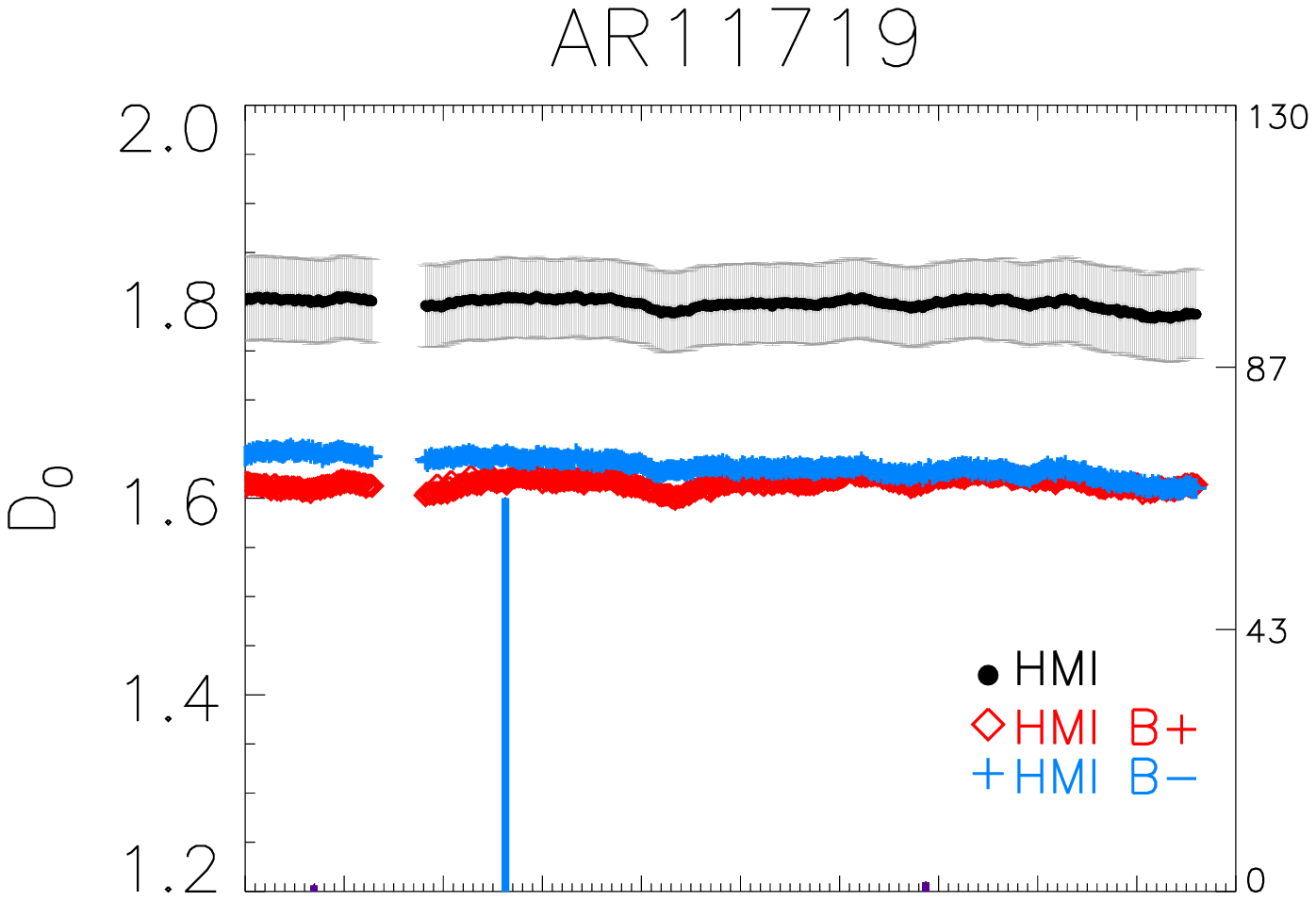}\includegraphics[trim={0.5cm 2cm  0.5cm  0.2cm},clip,width=5.2cm]{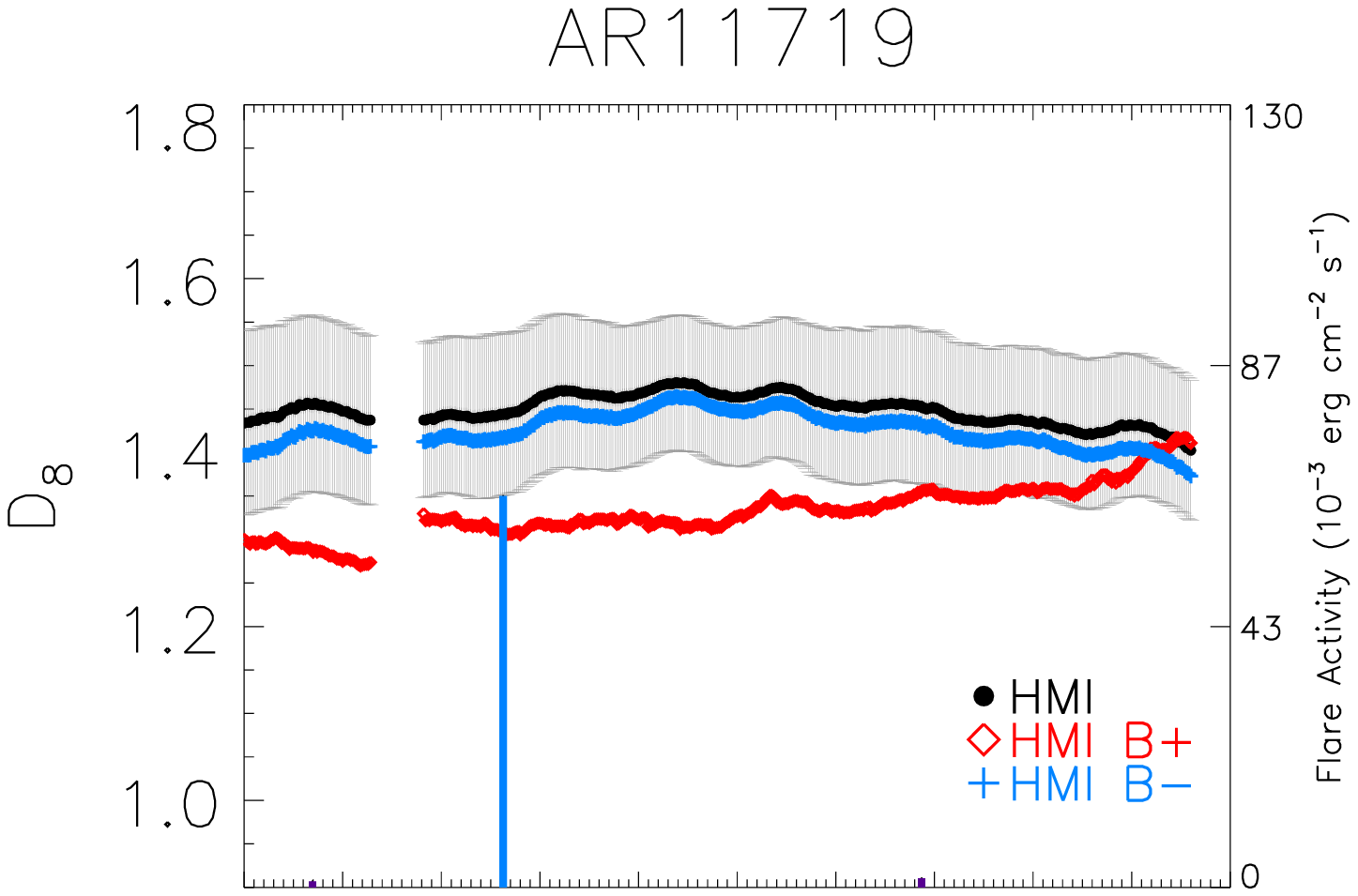}}
      \centerline{\includegraphics[trim={0.5cm 2cm  0.5cm  0.2cm},clip,width=5.2cm]{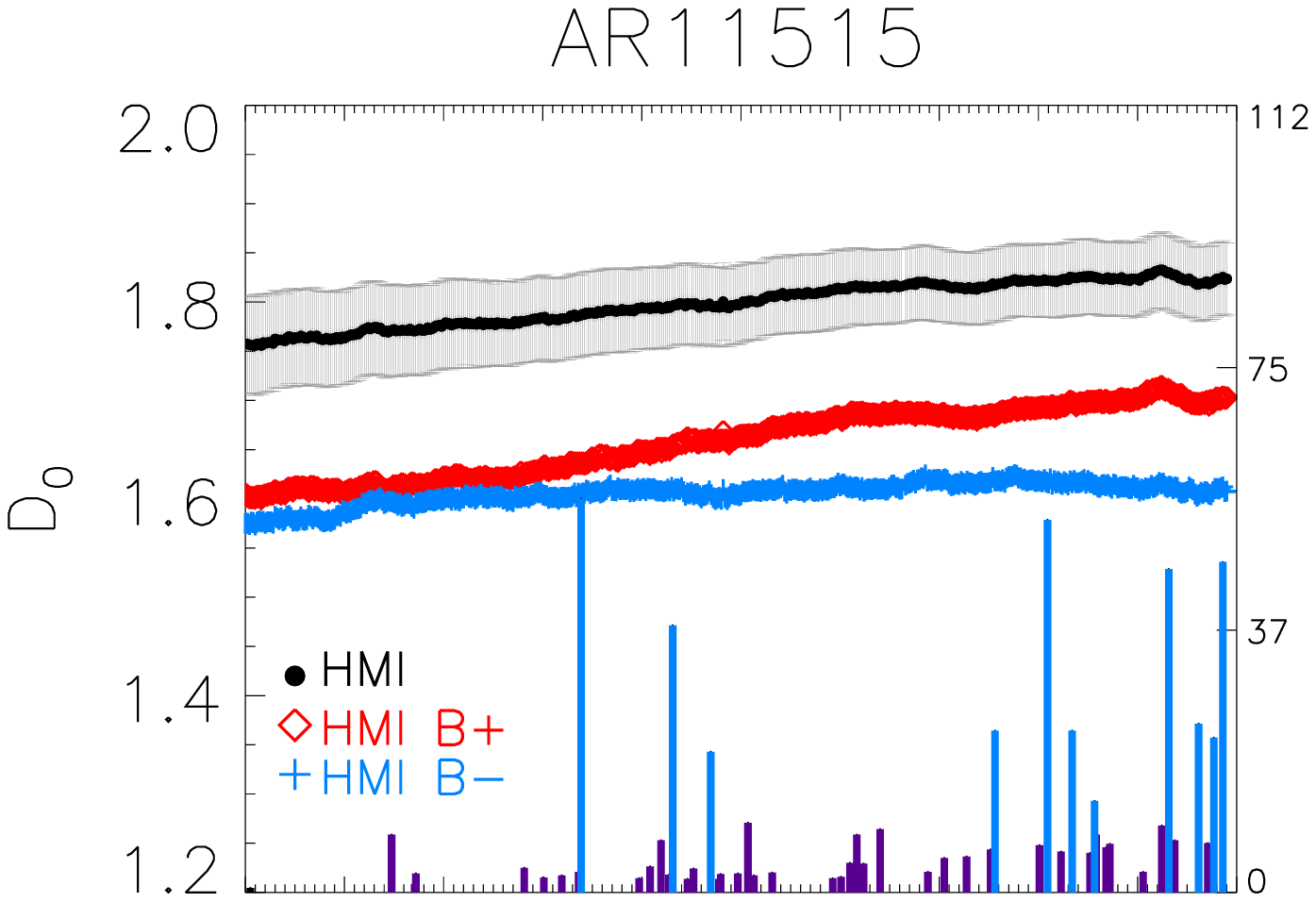}\includegraphics[trim={0.5cm 2cm  0.5cm  0.2cm},clip,width=5.2cm]{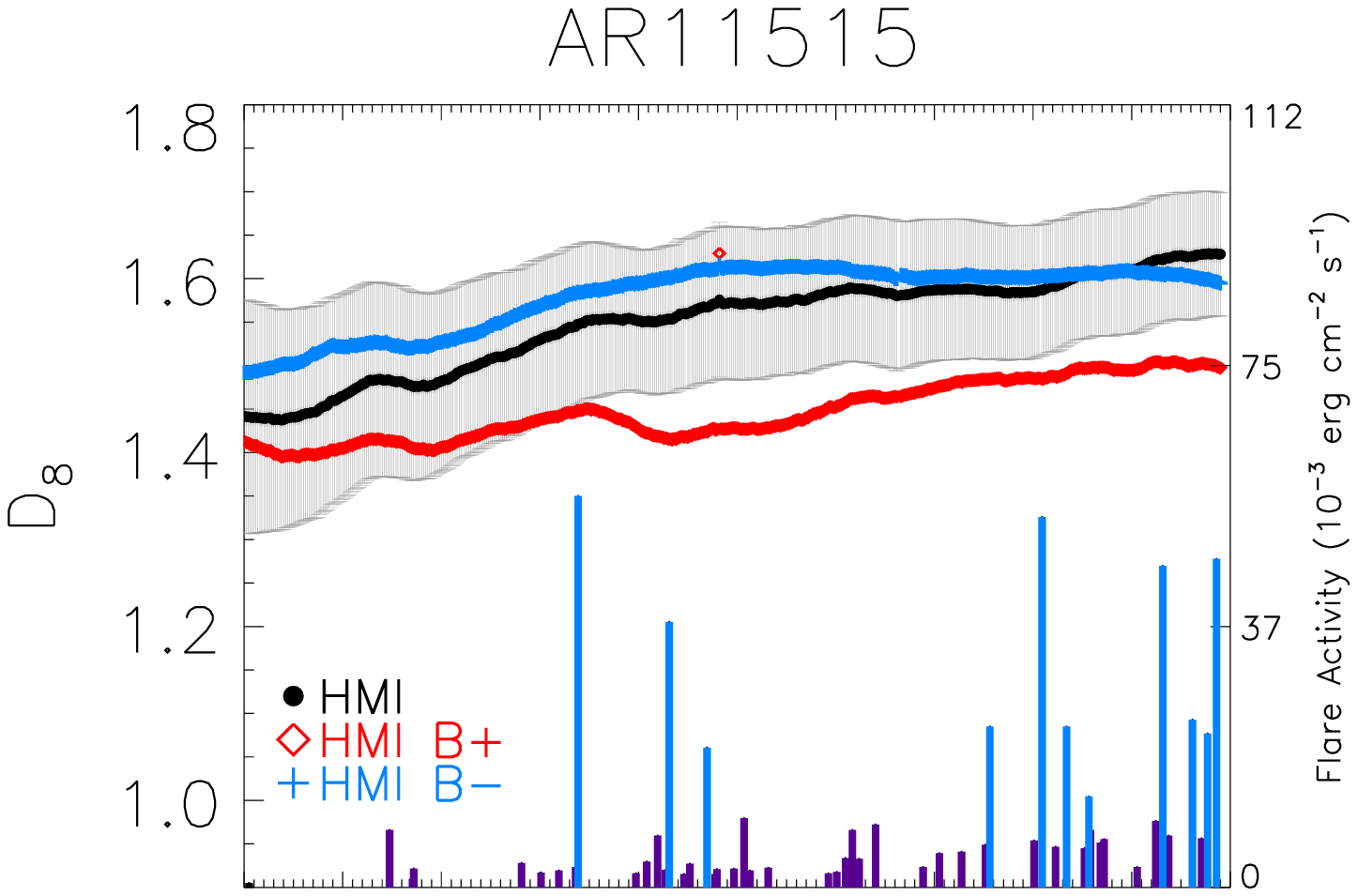}}
      \centerline{\includegraphics[trim={0.5cm 0.5cm  0.5cm  0.2cm},clip,width=5.2cm]{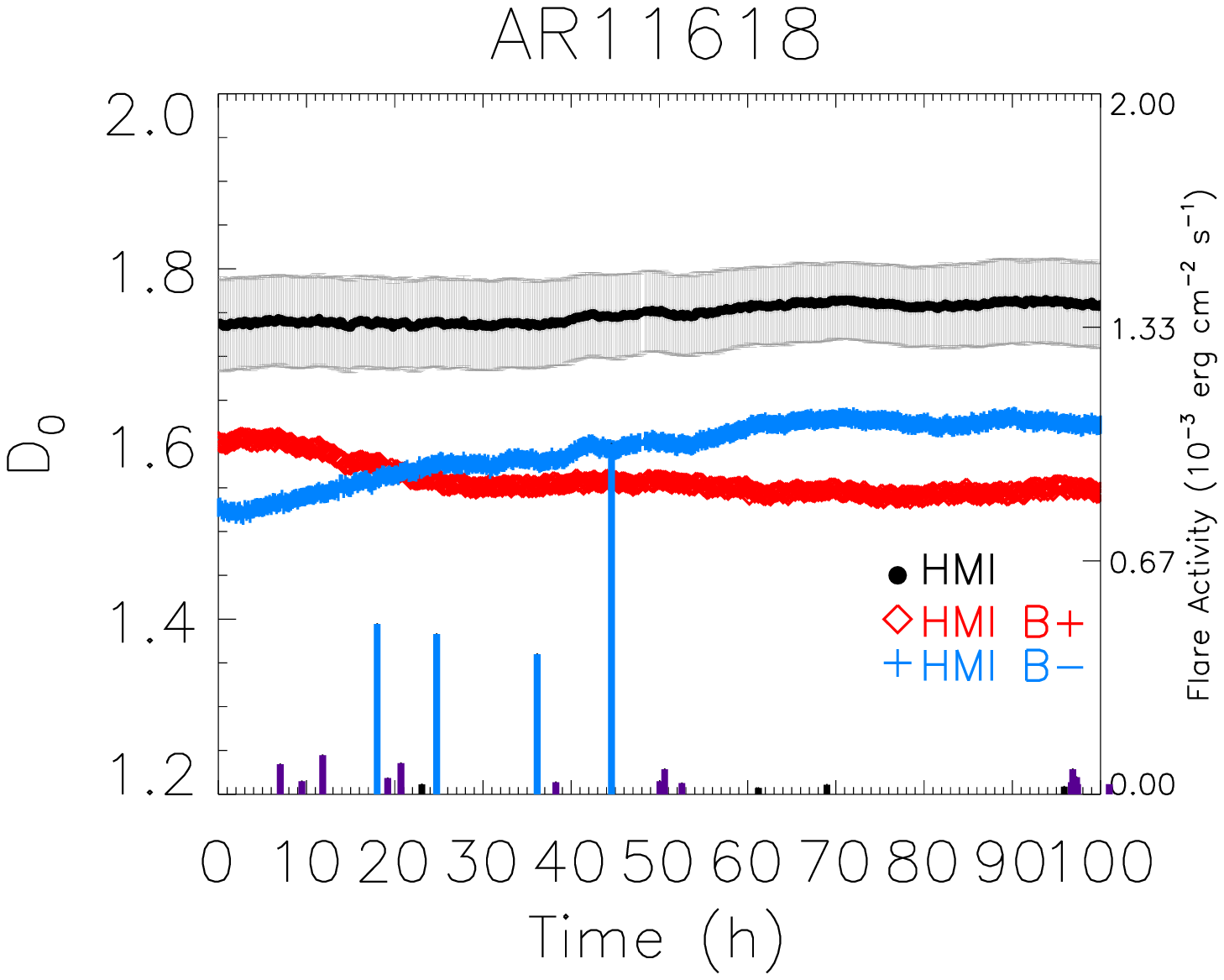}\includegraphics[trim={0.5cm 0.5cm  0.5cm  0.2cm},clip,width=5.2cm]{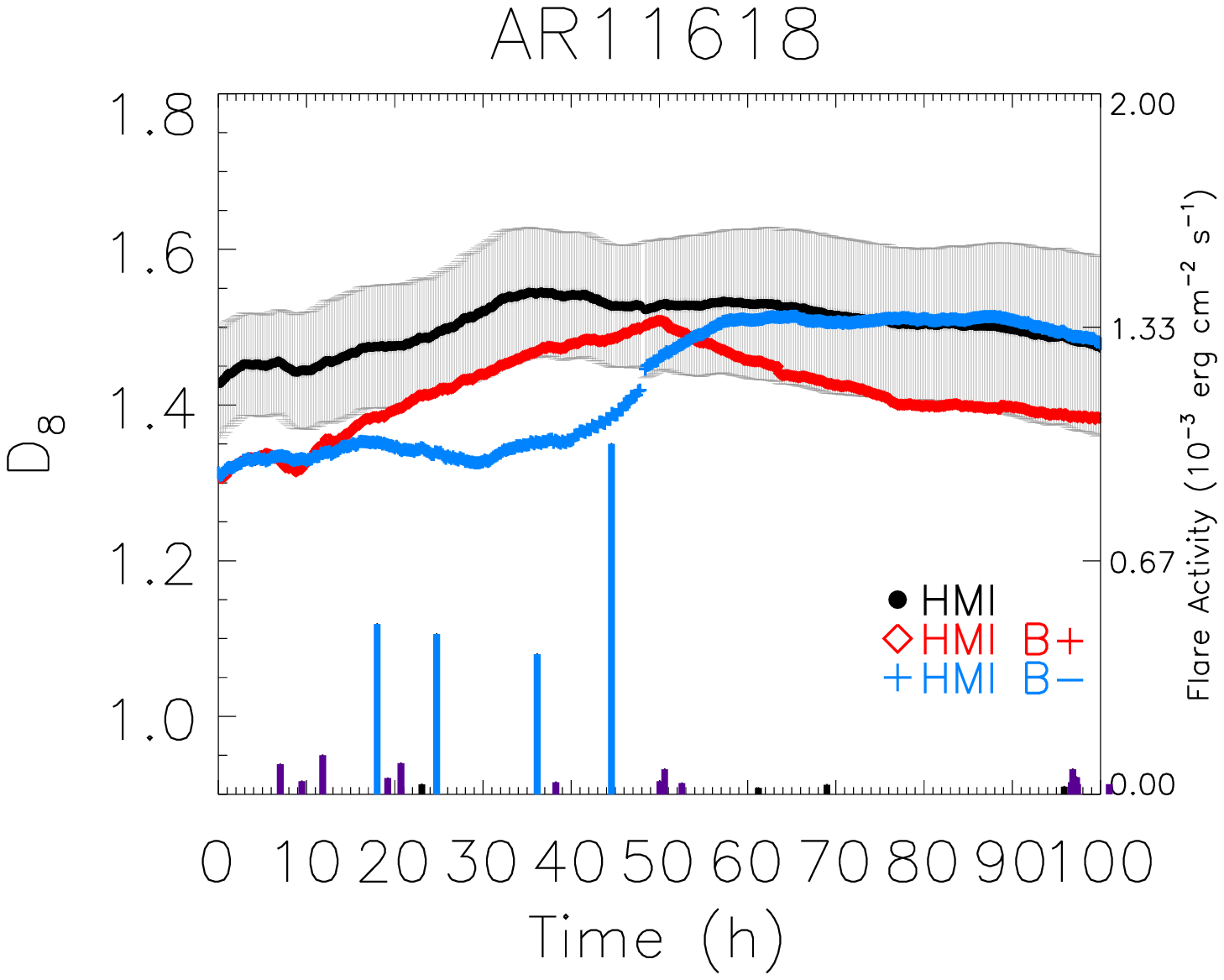}}
              \caption{Time series of the fractal parameters $D_0$ (left) and $D_8$ (right)  measured on six ARs, by considering both unsigned (HMI, black circles) and signed (HMI B+ and HMI B-, red diamonds and blue crosses, respectively) flux data in the analyzed regions. ARs NOAA 11158, 11520, and 11515 appeared in the southern hemisphere, while ARs NOAA 11875, 11719, and 11618 in the northern hemisphere.  { The leading polarity in the ARs is B+ and B- for the regions in the southern and northern hemisphere, respectively. Vertical bars indicate the flare activity of the AR; each single flare produced by the AR is reported with a bar whose length is proportional to the X-ray flux observed by GOES. Reading for these values are given in the right ordinates. Error bars show the uncertainty associated with the measured values, details are given in the text.} For clarity, the error bars are only shown for the results from unsigned flux data.  The gaps in the time series are due to the lack of SDO/HMI observations.}
   \label{f8}
   \end{figure*}

     \begin{figure*}
      \centerline{\includegraphics[trim={0.5cm 2cm  0.5cm 0.2cm},clip,width=5.2cm]{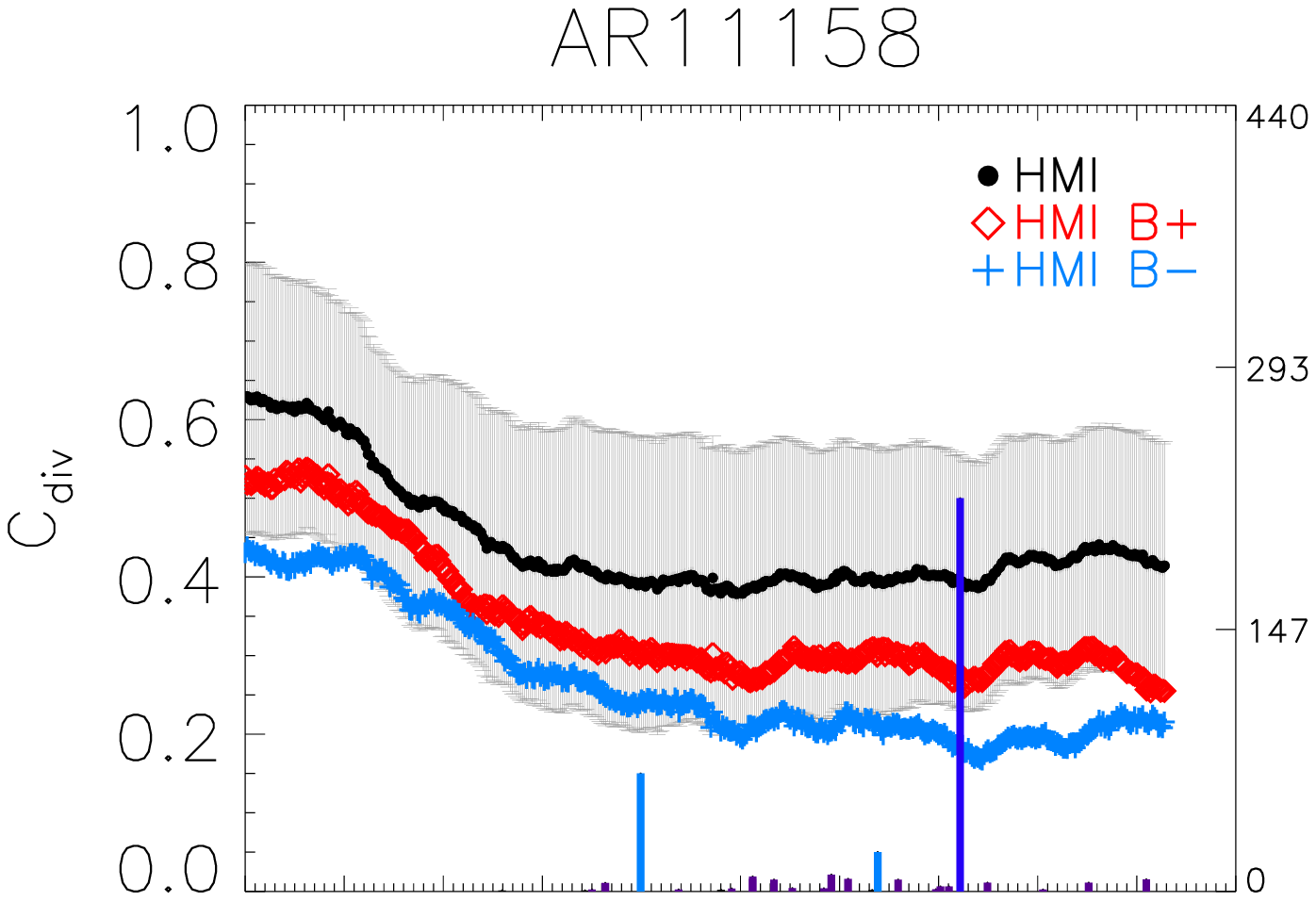}\includegraphics[trim={0.5cm 2cm  0.5cm 0.2cm},clip,width=5.2cm]{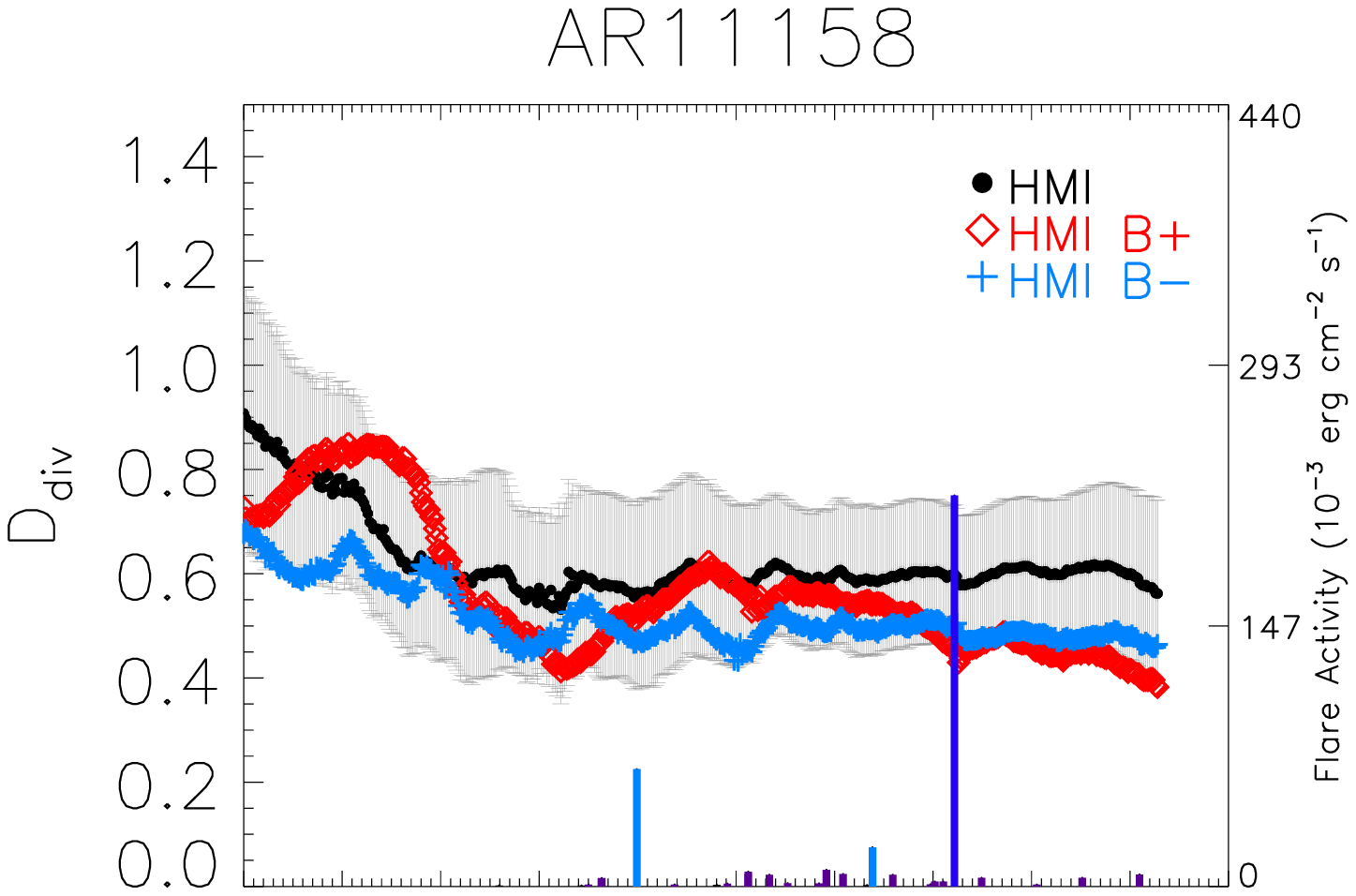}}
      \centerline{\includegraphics[trim={0.5cm 2cm  0.5cm  0.2cm},clip,width=5.2cm]{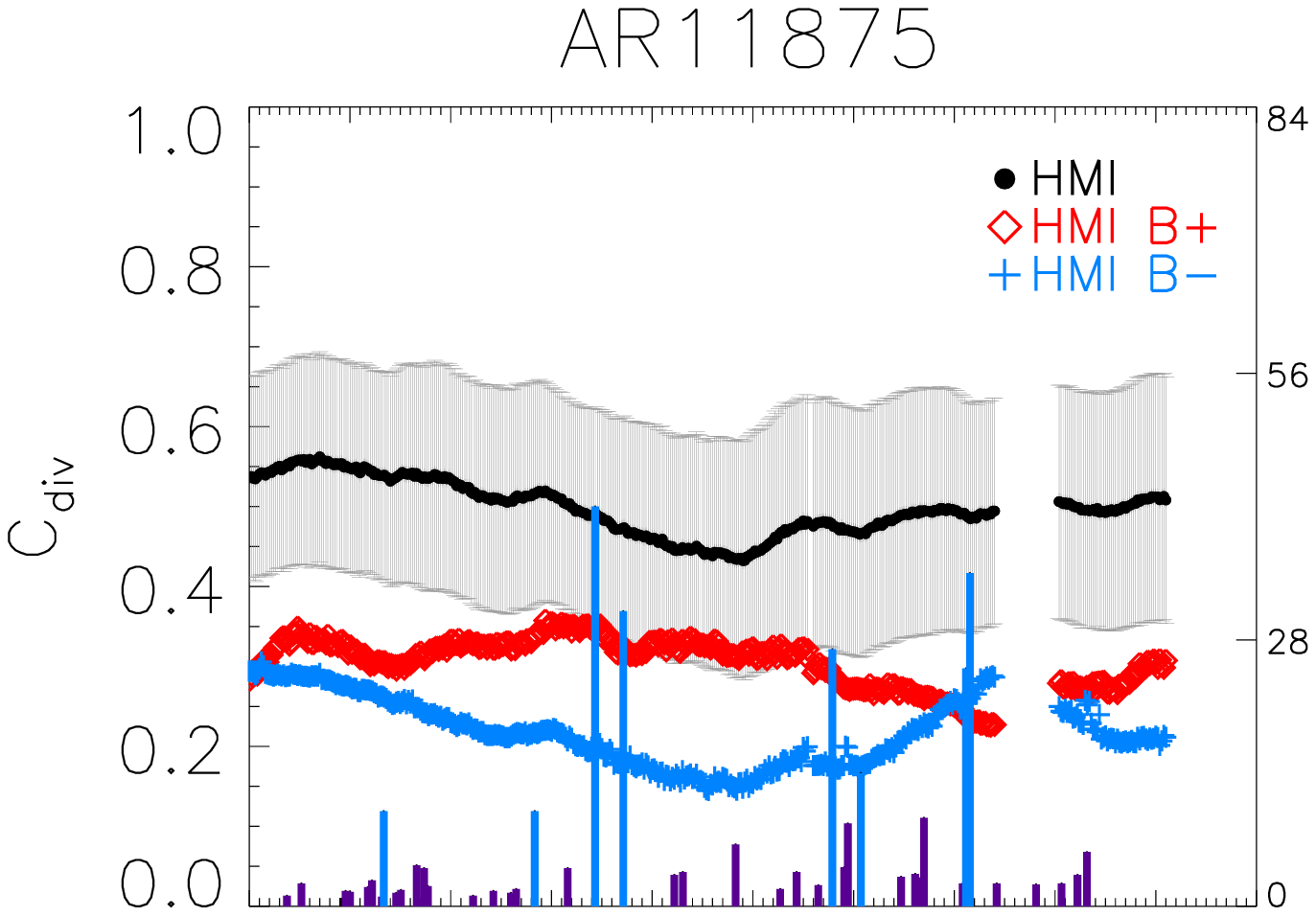}\includegraphics[trim={0.5cm 2cm  0.5cm  0.2cm},clip,width=5.2cm]{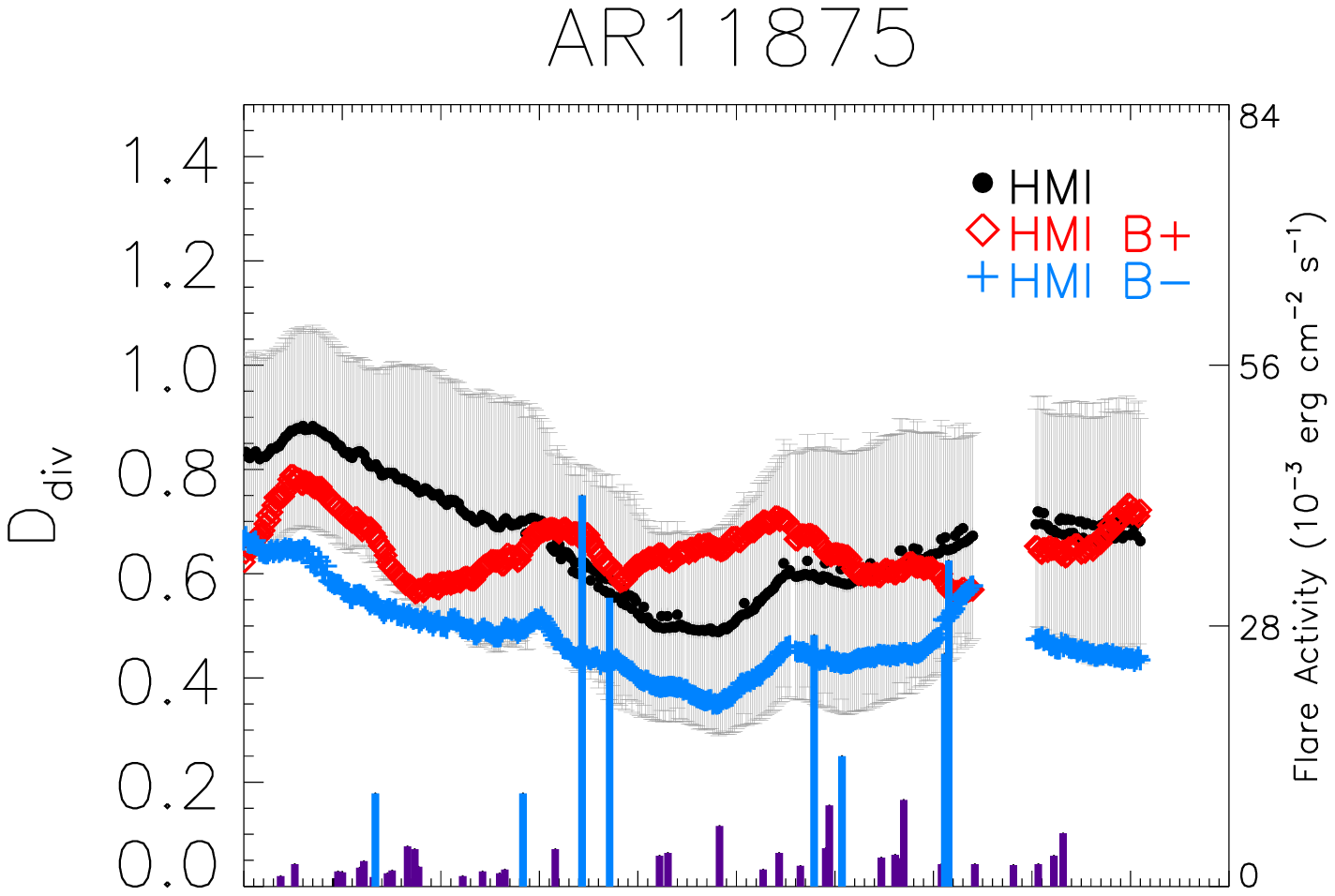}}
      \centerline{\includegraphics[trim={0.5cm 2cm  0.5cm  0.2cm},clip,width=5.2cm]{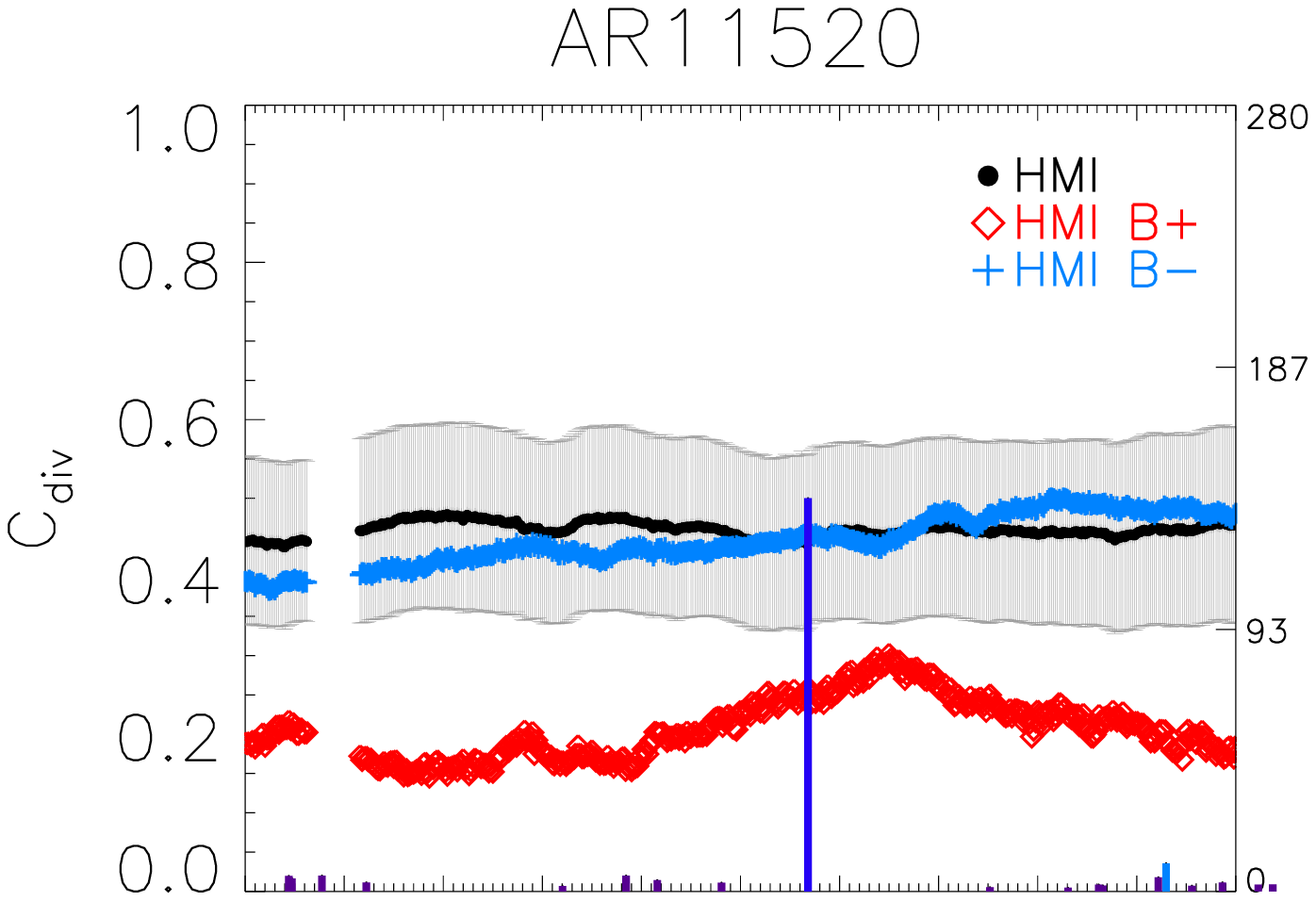}\includegraphics[trim={0.5cm 2cm  0.5cm  0.2cm},clip,width=5.2cm]{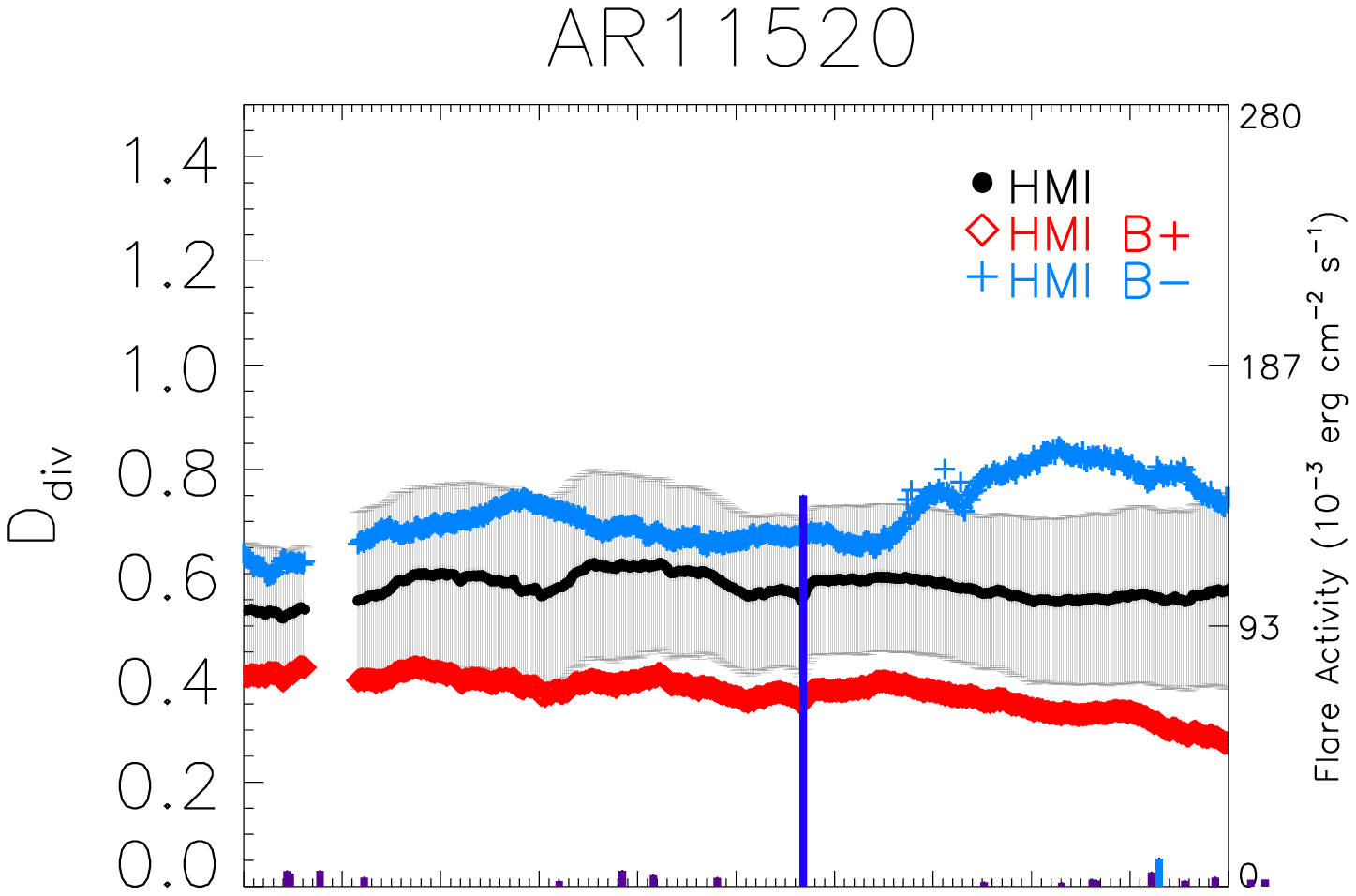}}
      \centerline{\includegraphics[trim={0.5cm 2cm  0.5cm  0.2cm},clip,width=5.2cm]{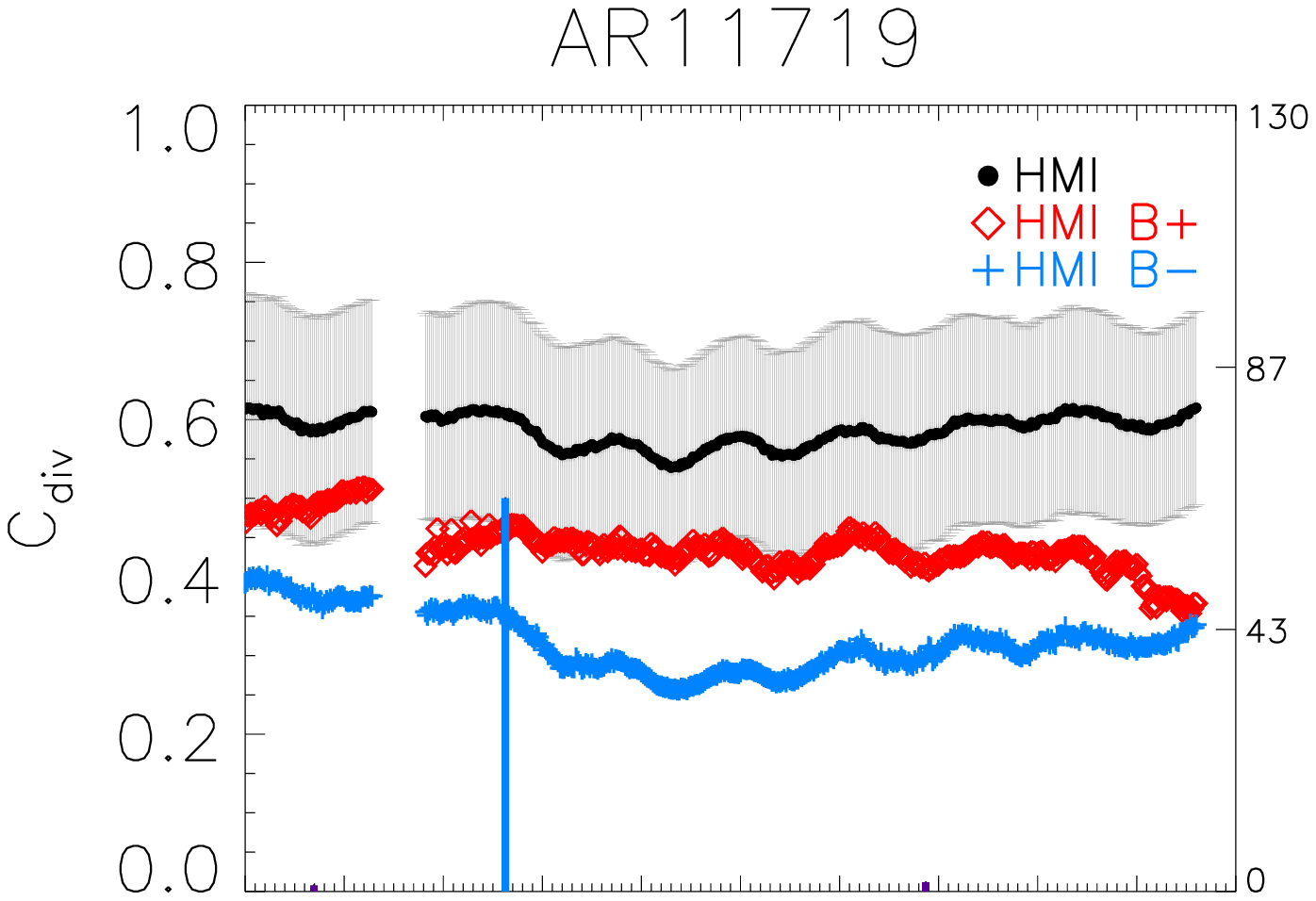}\includegraphics[trim={0.5cm 2cm  0.5cm  0.2cm},clip,width=5.2cm]{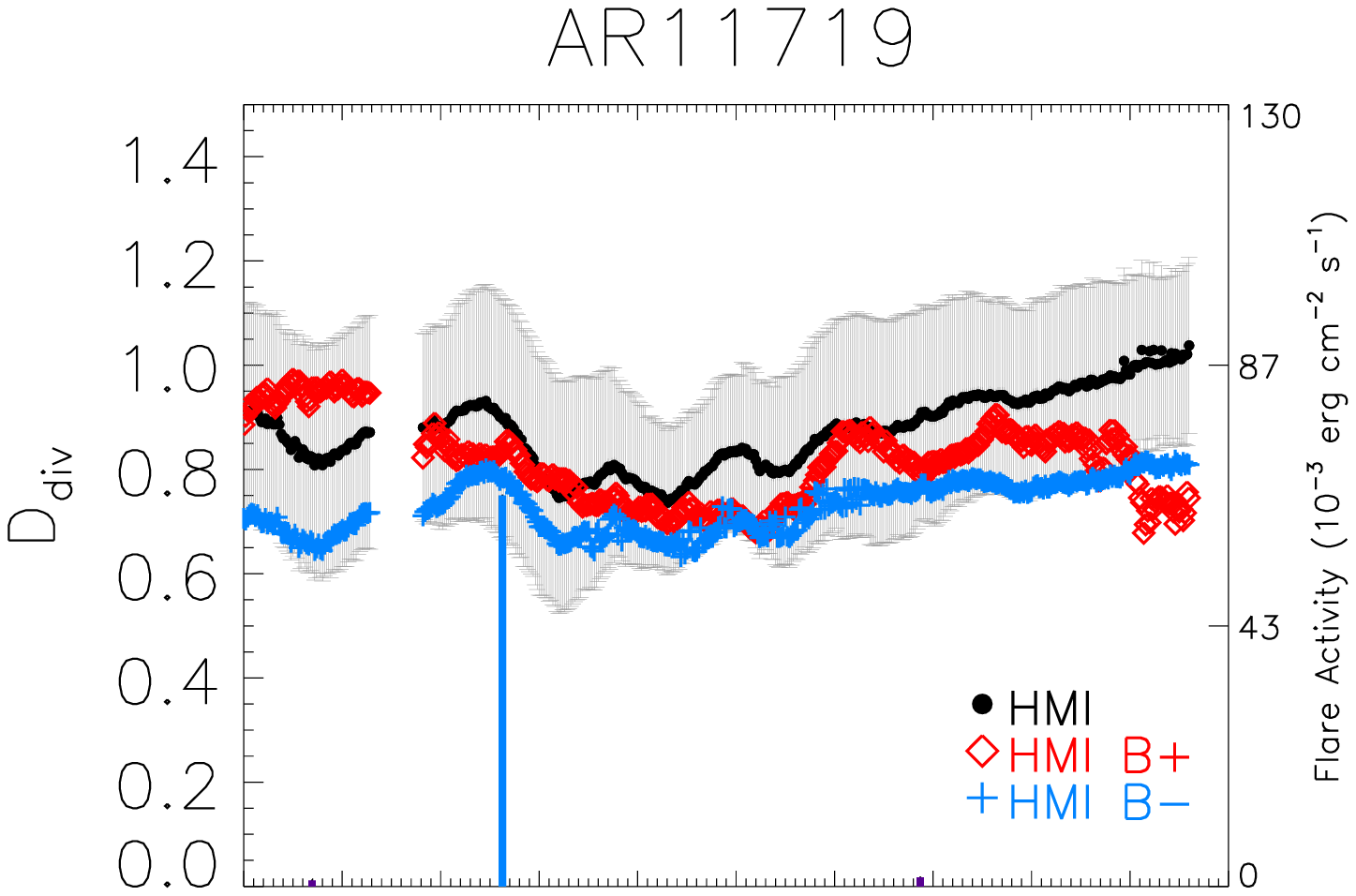}}
      \centerline{\includegraphics[trim={0.5cm 2cm  0.5cm  0.2cm},clip,width=5.2cm]{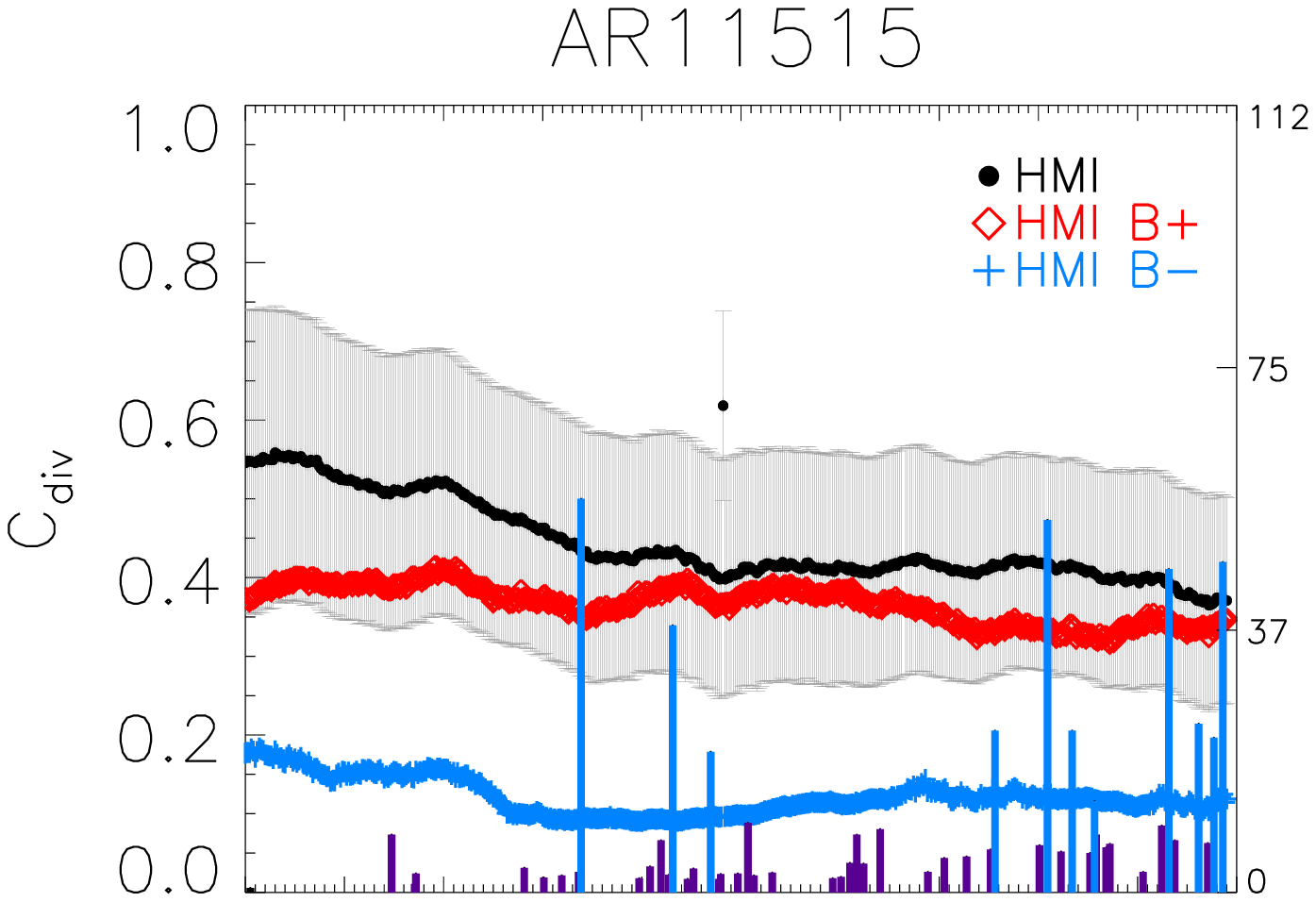}\includegraphics[trim={0.5cm 2cm  0.5cm  0.2cm},clip,width=5.2cm]{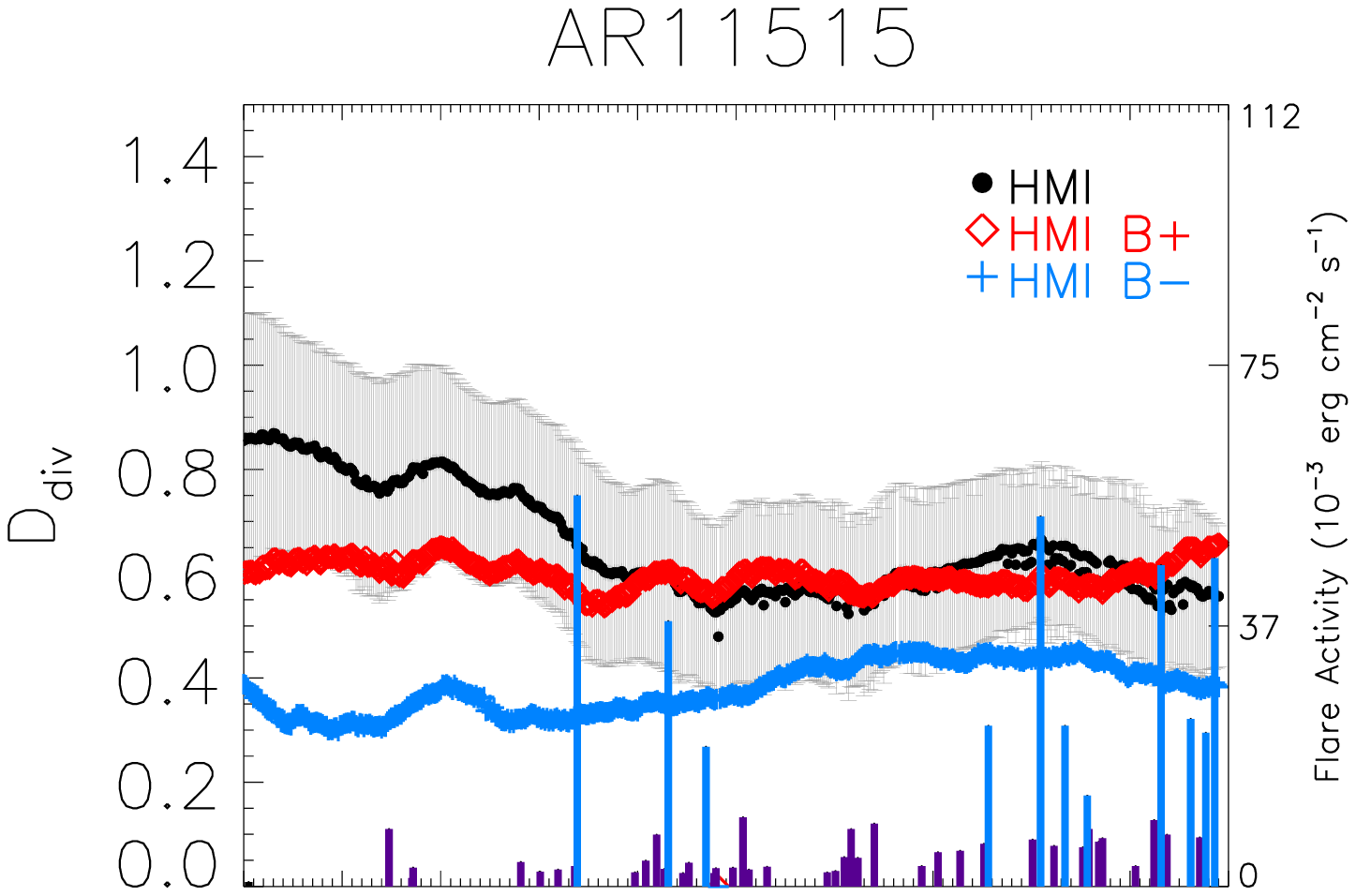}}
      \centerline{\includegraphics[trim={0.5cm 0.5cm  0.5cm  0.2cm},clip,width=5.2cm]{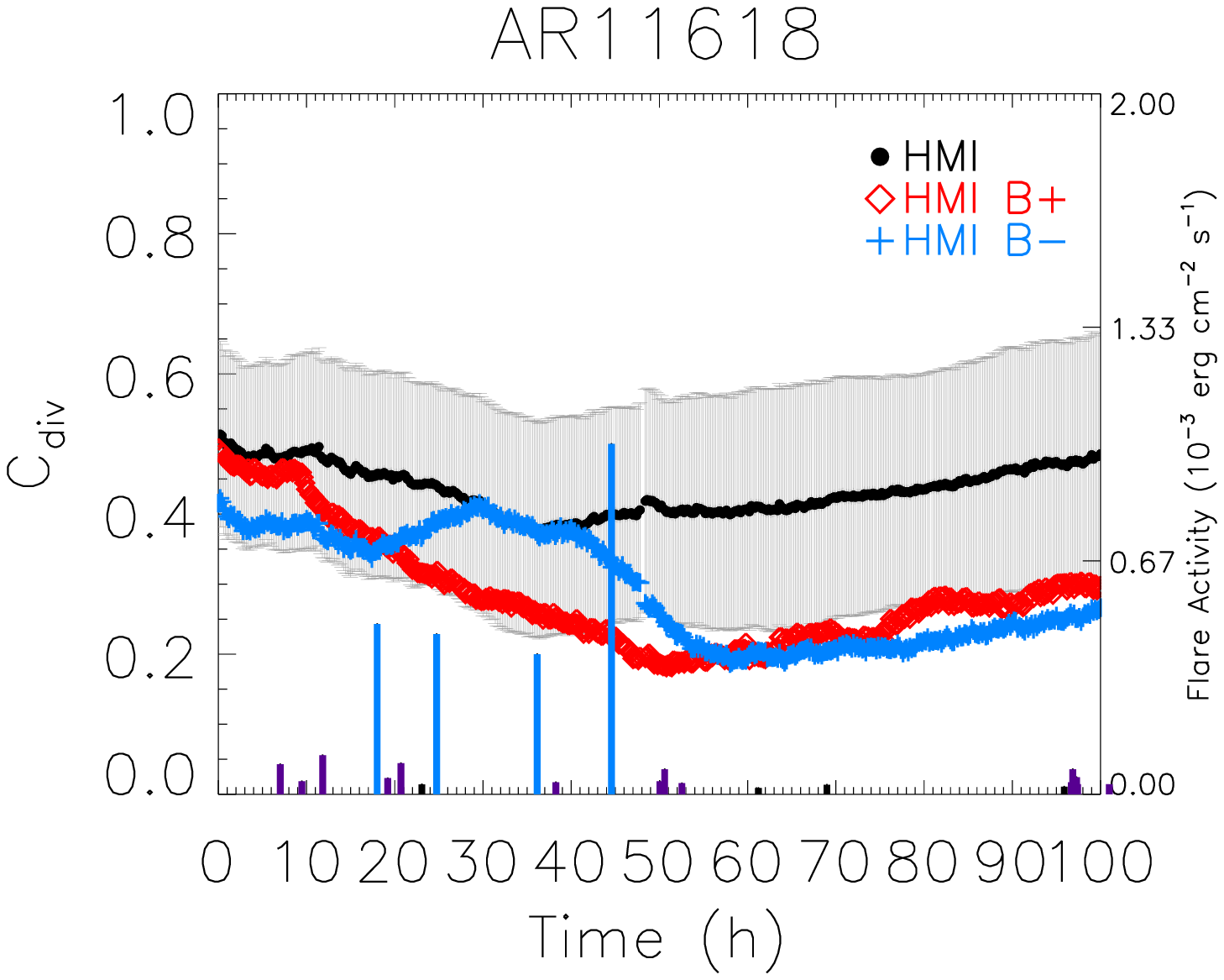}\includegraphics[trim={0.5cm 0.5cm  0.5cm  0.2cm},clip,width=5.2cm]{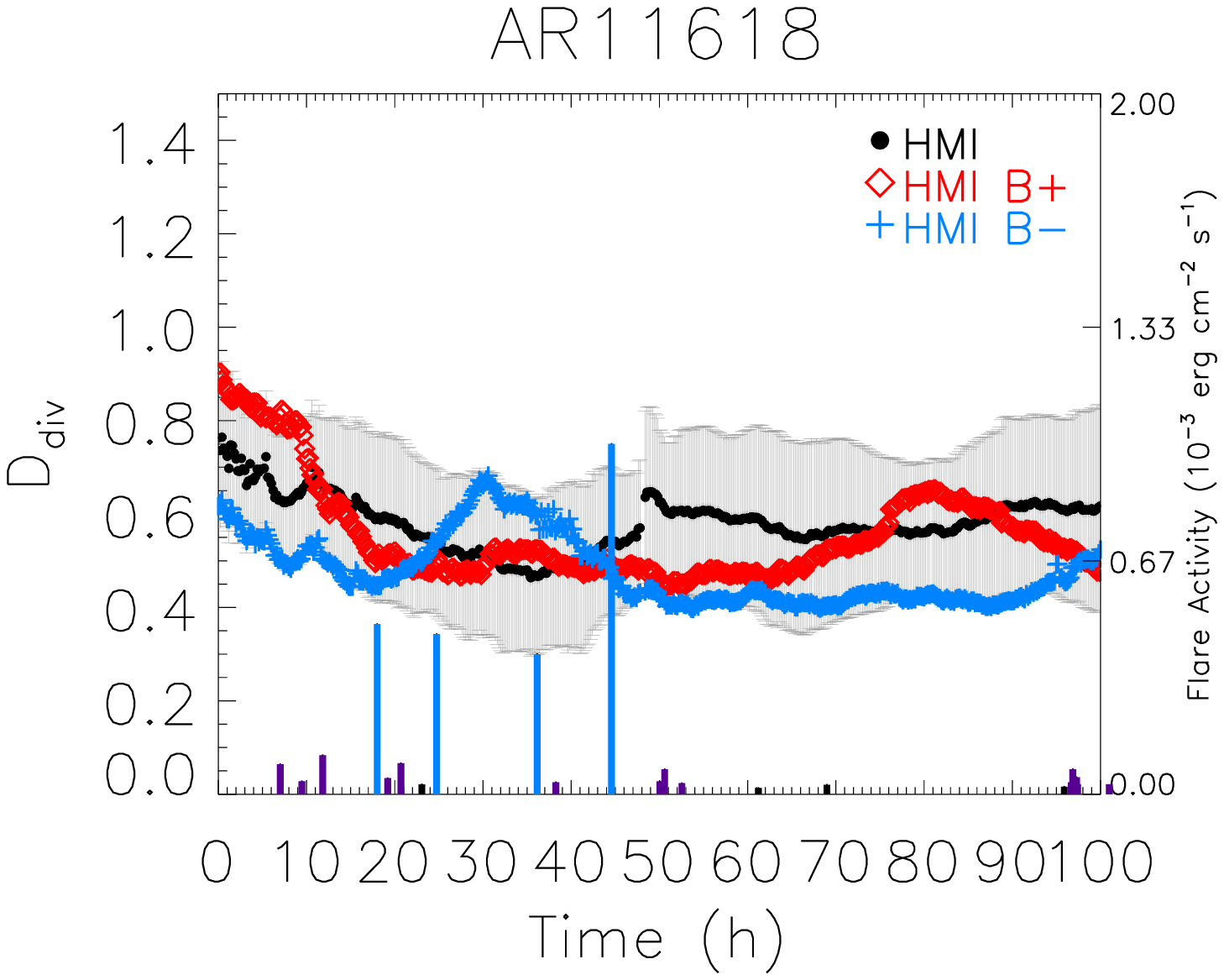}}
              \caption{Time series of the  multifractal parameters  $C_{\mathrm{div}}$ (left) and $D_{\mathrm{div}}$ (right)  measured on six ARs, by considering both unsigned (HMI, black circles) and signed (HMI B+ and HMI B-, red diamonds and blue crosses, respectively) flux data in the analyzed regions. ARs NOAA 11158, 11520, and 11515 appeared in the southern hemisphere, while ARs NOAA 11875, 11719, and 11618 in the northern hemisphere.  { The leading polarity in the ARs is B+ and B- for the regions in the southern and northern hemisphere, respectively. } Error bars and vertical bars as in Figure \ref{f8}. The gaps in the time series are due to the lack of SDO/HMI observations.}
   \label{f9}
   \end{figure*}

 We focussed our attention on the time series of  the various parameters measured on the analyzed ARs. Figures \ref{f8} and \ref{f9}  show a sample of  the analyzed data, which correspond to the time series of some among  the most flaring ARs in the data set. 
 We first studied the main features of these series, aiming at investigating  the possibility that a reduced dispersion of the parameters measured on SDO/HMI observations   may improve the efficiency of these measurements to discriminate  flaring from flare-quiet regions with respect to previous reports in  the literature.
 Section 3.1 presents  the average values of the parameters derived from unsigned flux data of the analyzed ARs, in close analogy to previous results  in the literature, whereas Section 3.2 shows the findings from the analysis of the signed flux data of the analyzed regions, aiming  at singling out effects of flux discrimination on measurement results. 
 Section 3.3 presents the main features of the temporal evolution of the parameters measured from both unsigned and signed flux data of the studied regions, aiming at highlighting any distinctive patterns in the time series that can be associated with the flaring activity.

\subsection{Average Values of the  Parameters from Unsigned Flux Data}

     \begin{figure*}
    \centerline{\includegraphics[width=6.5cm]{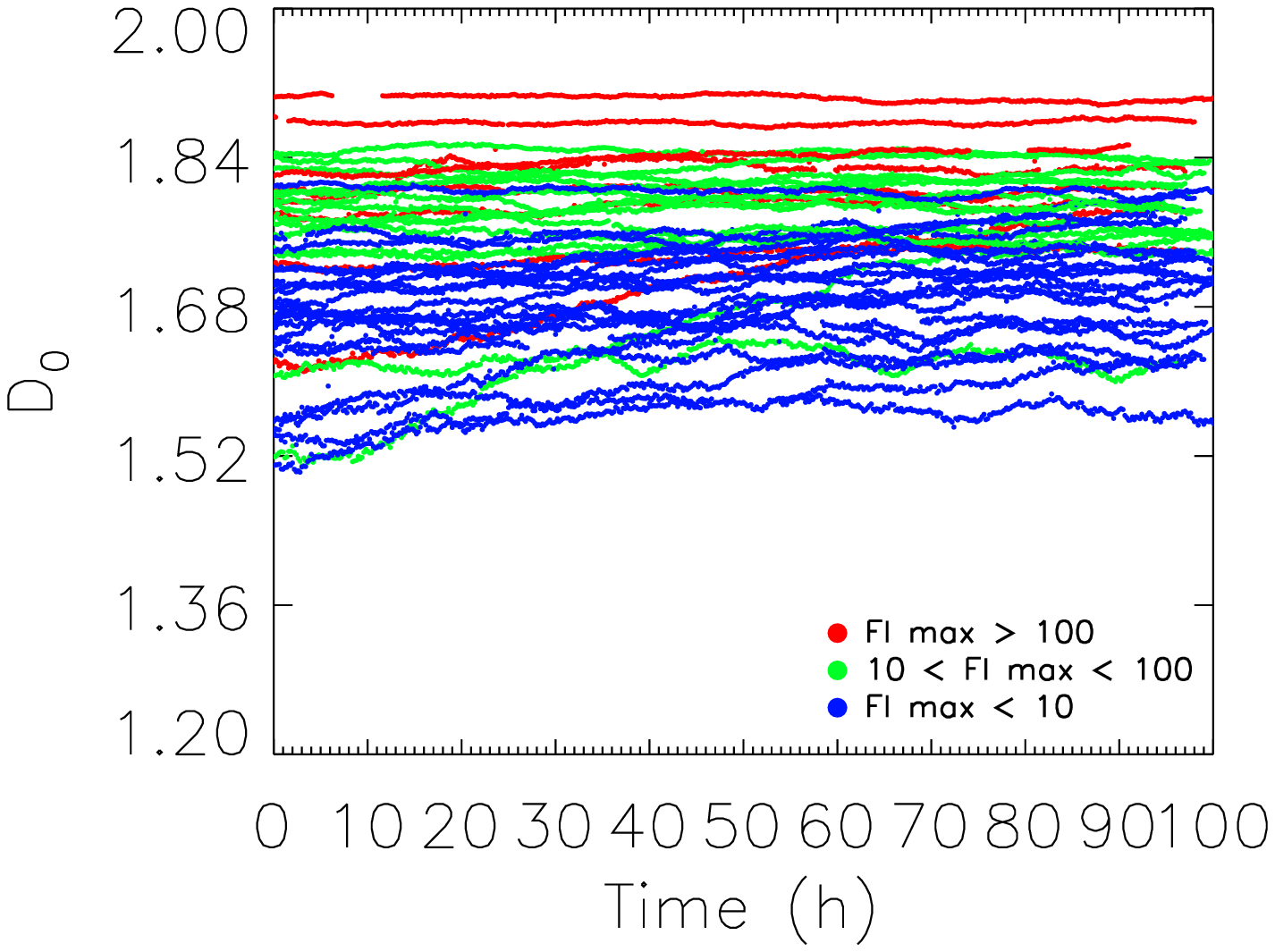}\includegraphics[width=6.5cm]{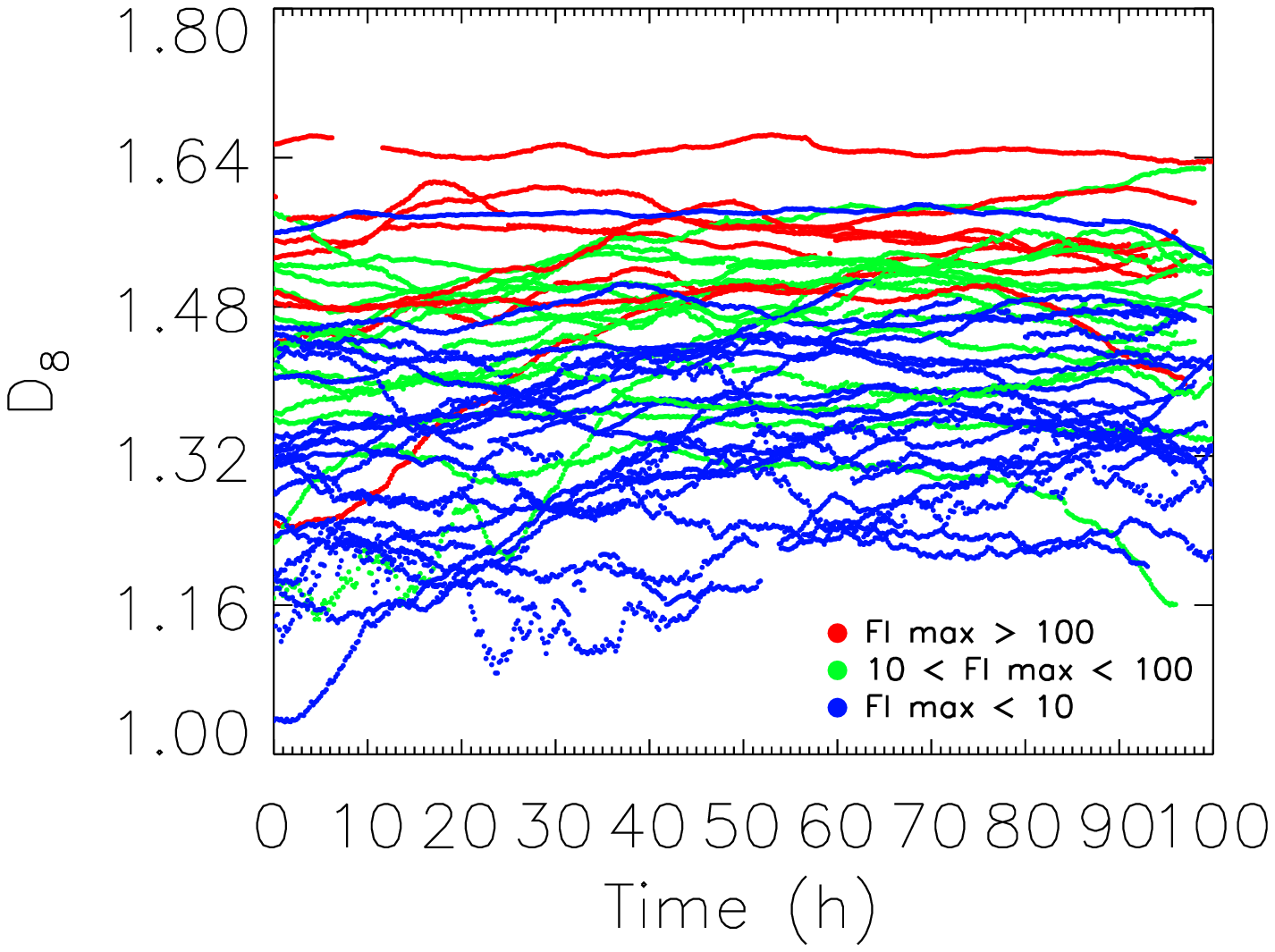}}
              \centerline{\includegraphics[width=6.5cm]{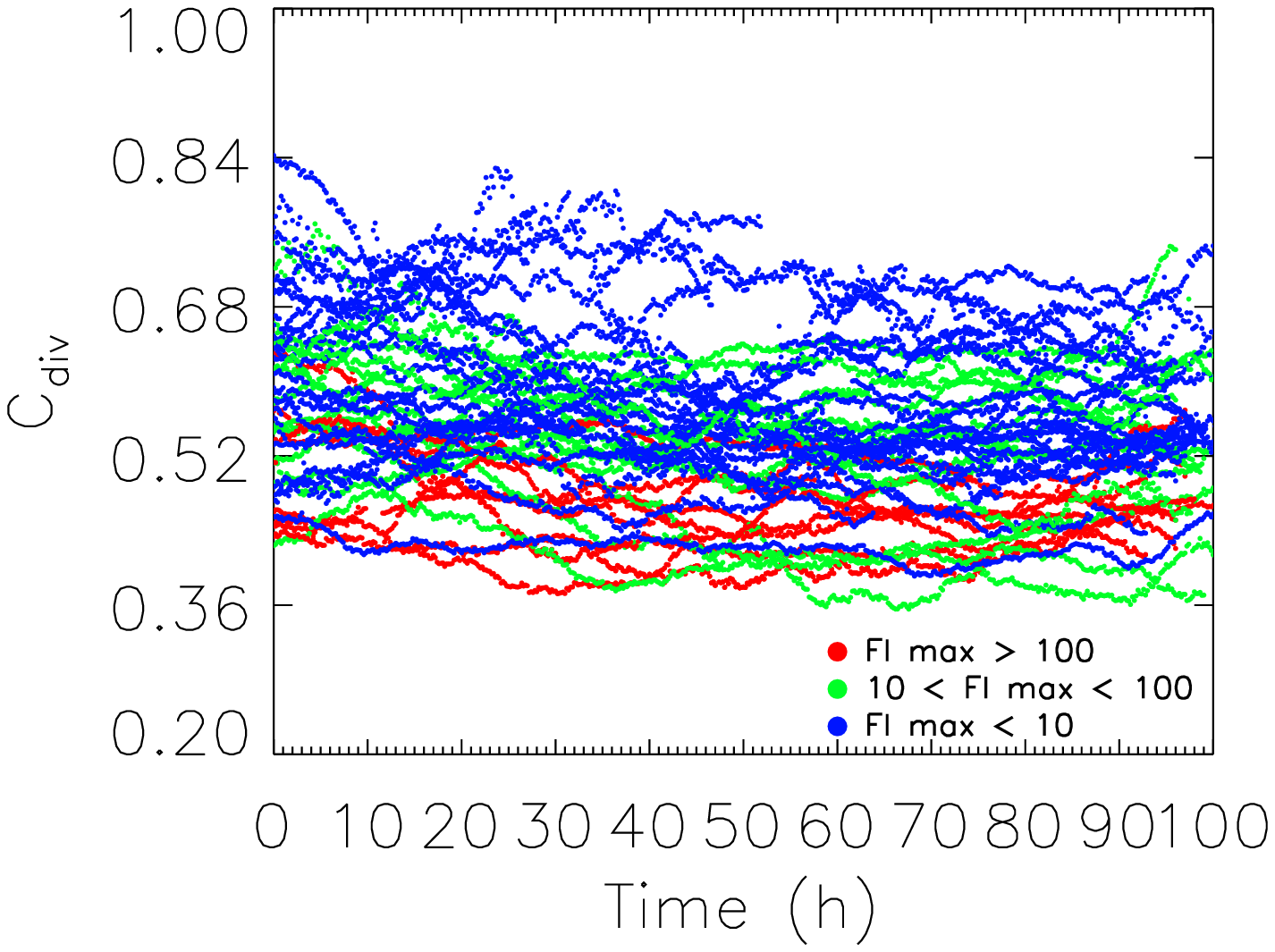}\includegraphics[width=6.5cm]{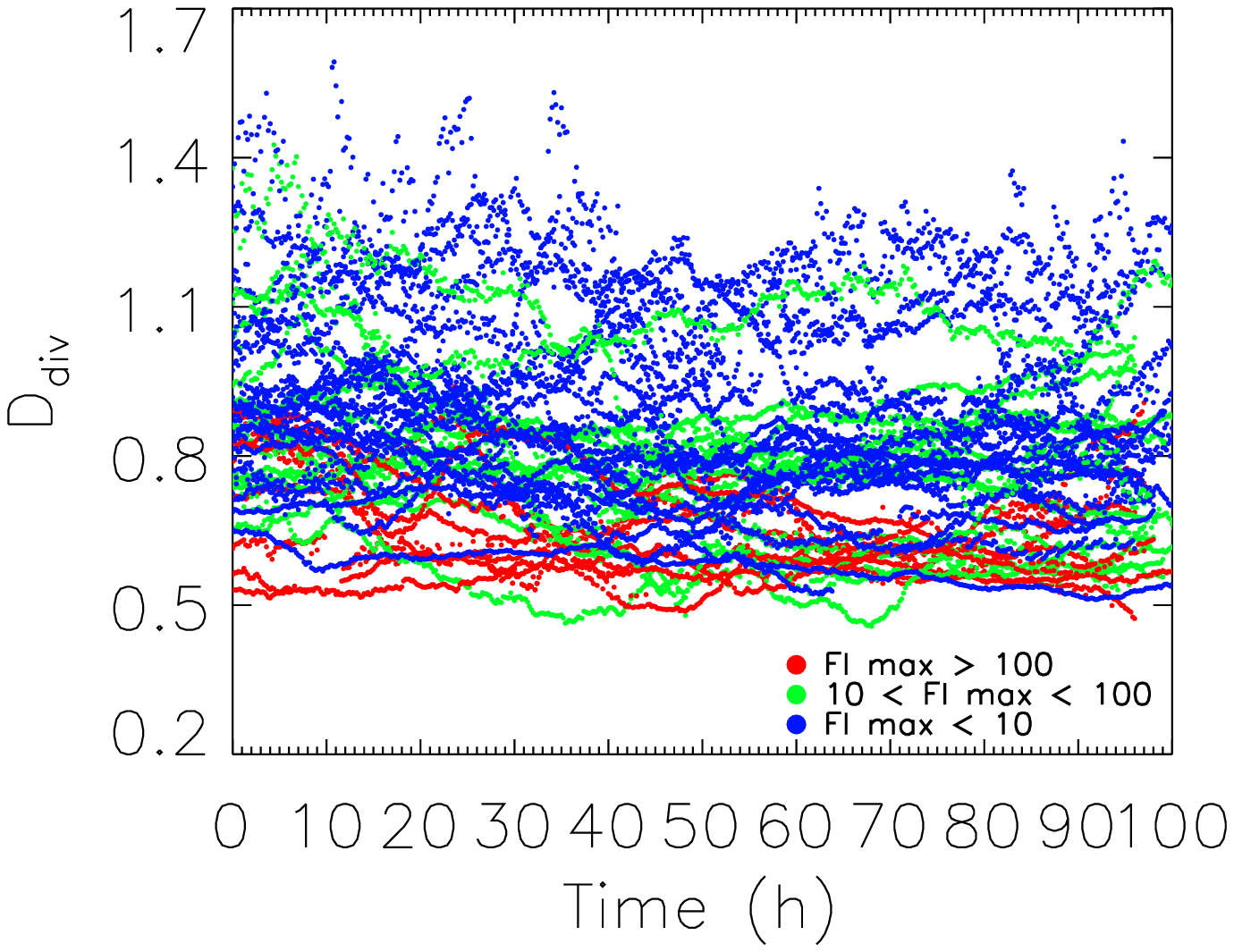}}

              \caption{Comparison among results of the fractal and multifractal parameters measured on the whole sample of analyzed ARs by taking into account unsigned magnetic flux data  in the studied regions.
The time series of fractal parameters $D_0$ and $D_8$ (top panels) and multifractal parameters $C_{\mathrm{div}}$ and $D_{\mathrm{div}}$ (bottom panels) were divided according to the flaring level of the analyzed AR. The results derived from  regions that have hosted B- or C- (FI max $<$10), M- (10$<$ FI max$<$100), and X- class (FI max$>$100) flares are shown with blue, green, and red symbols, respectively.
 }
   \label{f5}
   \end{figure*}

      \begin{figure*}
     \centerline{\includegraphics[width=6.5cm]{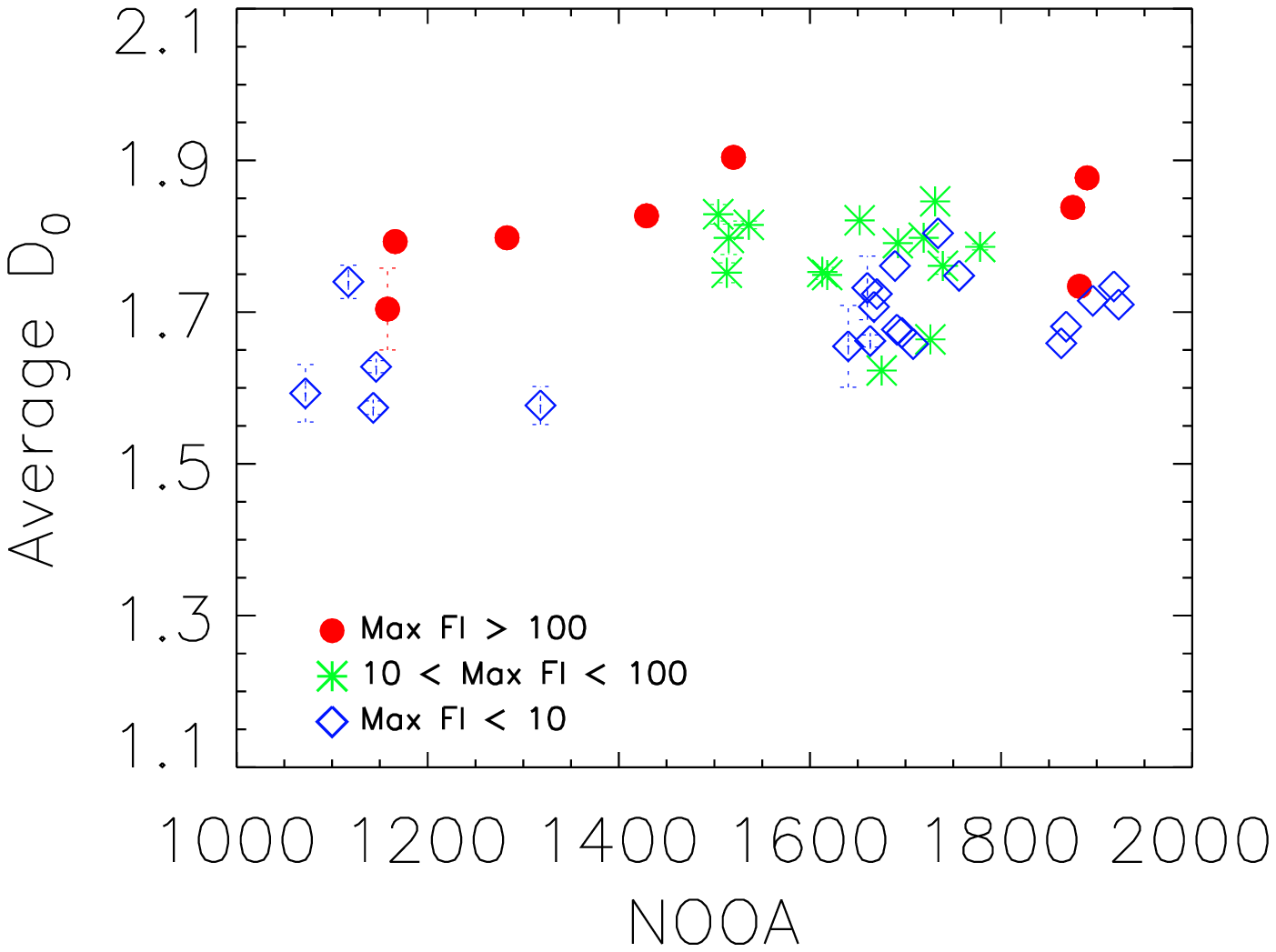}\includegraphics[width=6.5cm]{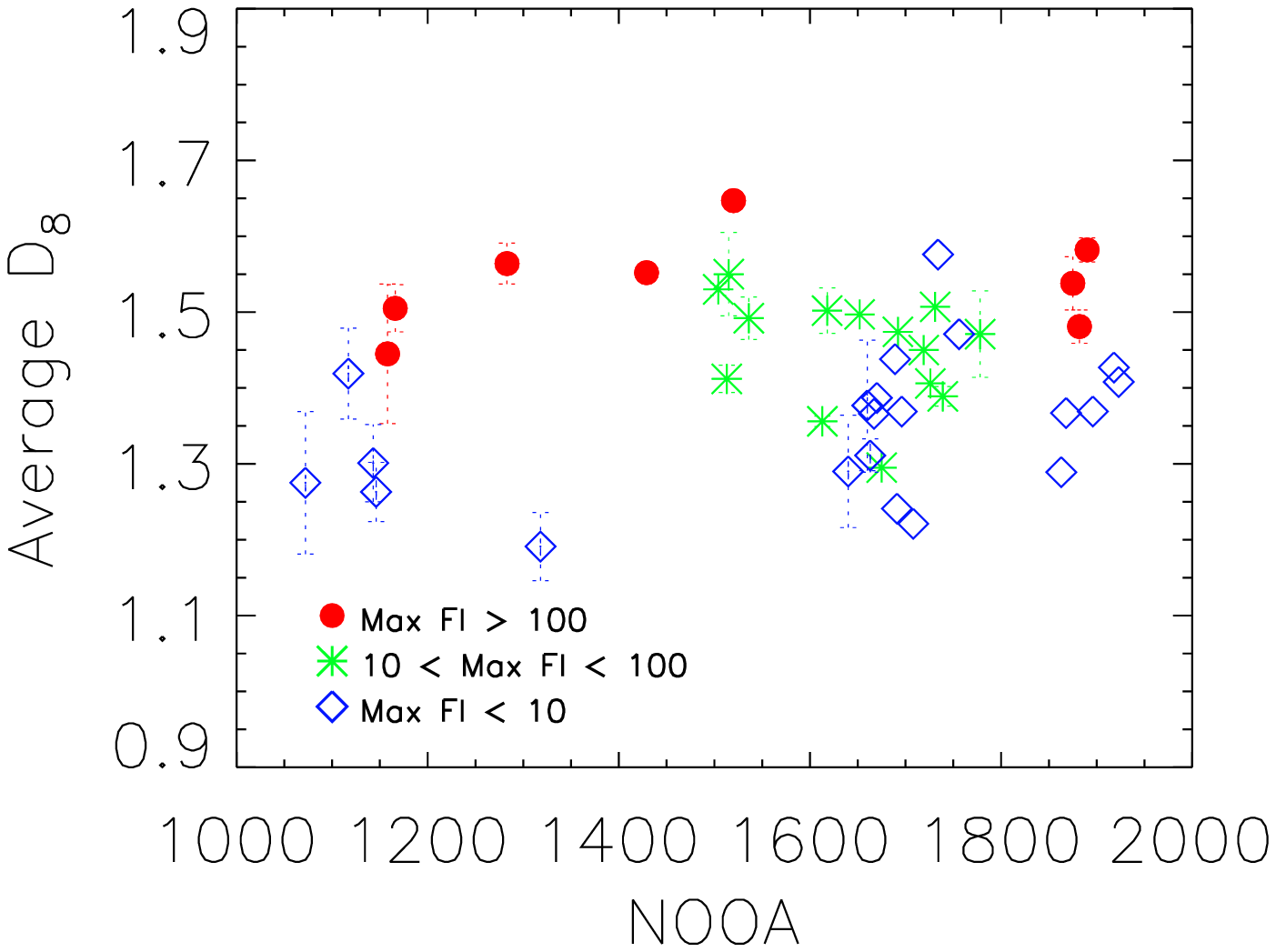}}
              \centerline{\includegraphics[width=6.5cm]{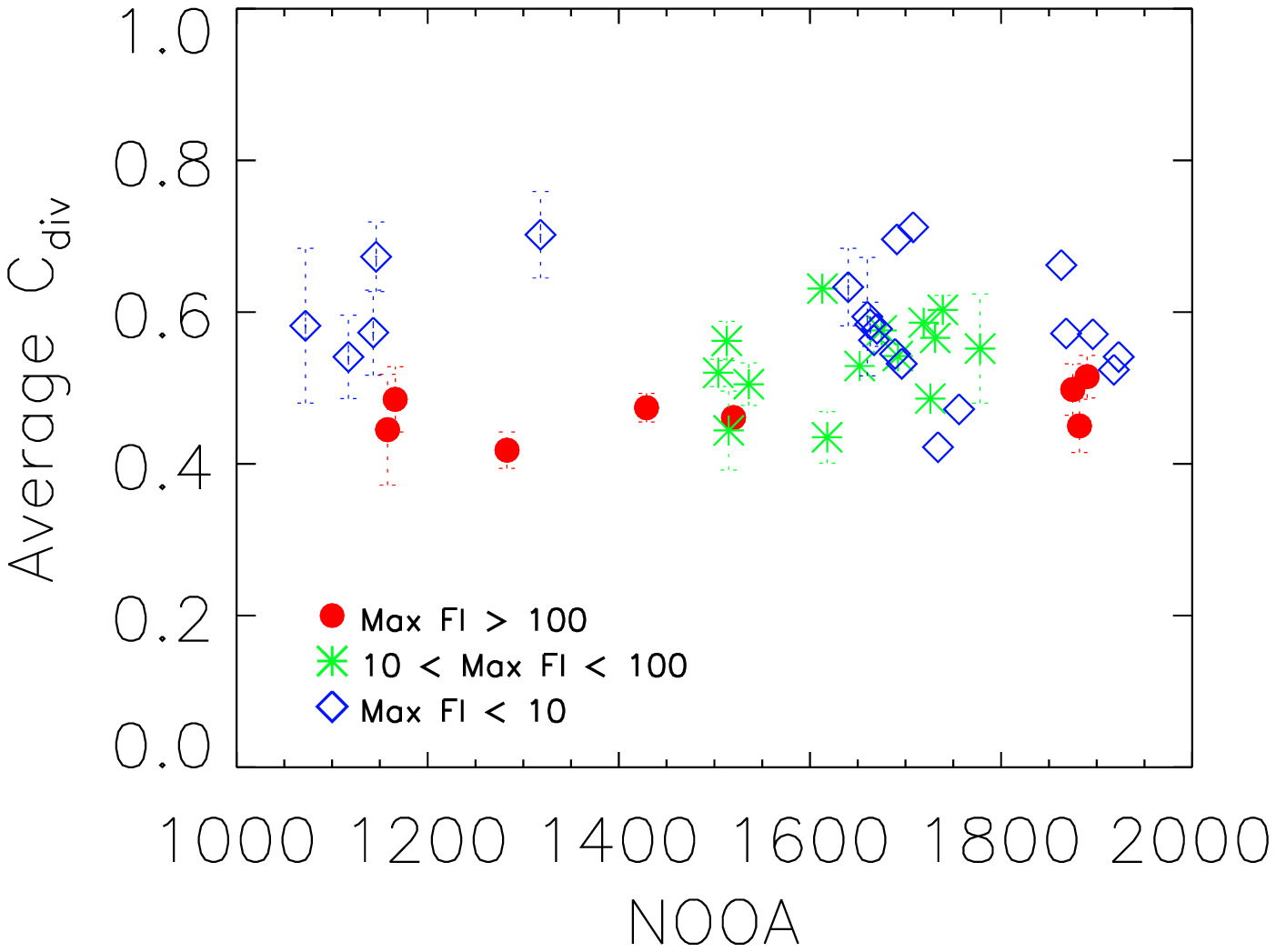}\includegraphics[width=6.5cm]{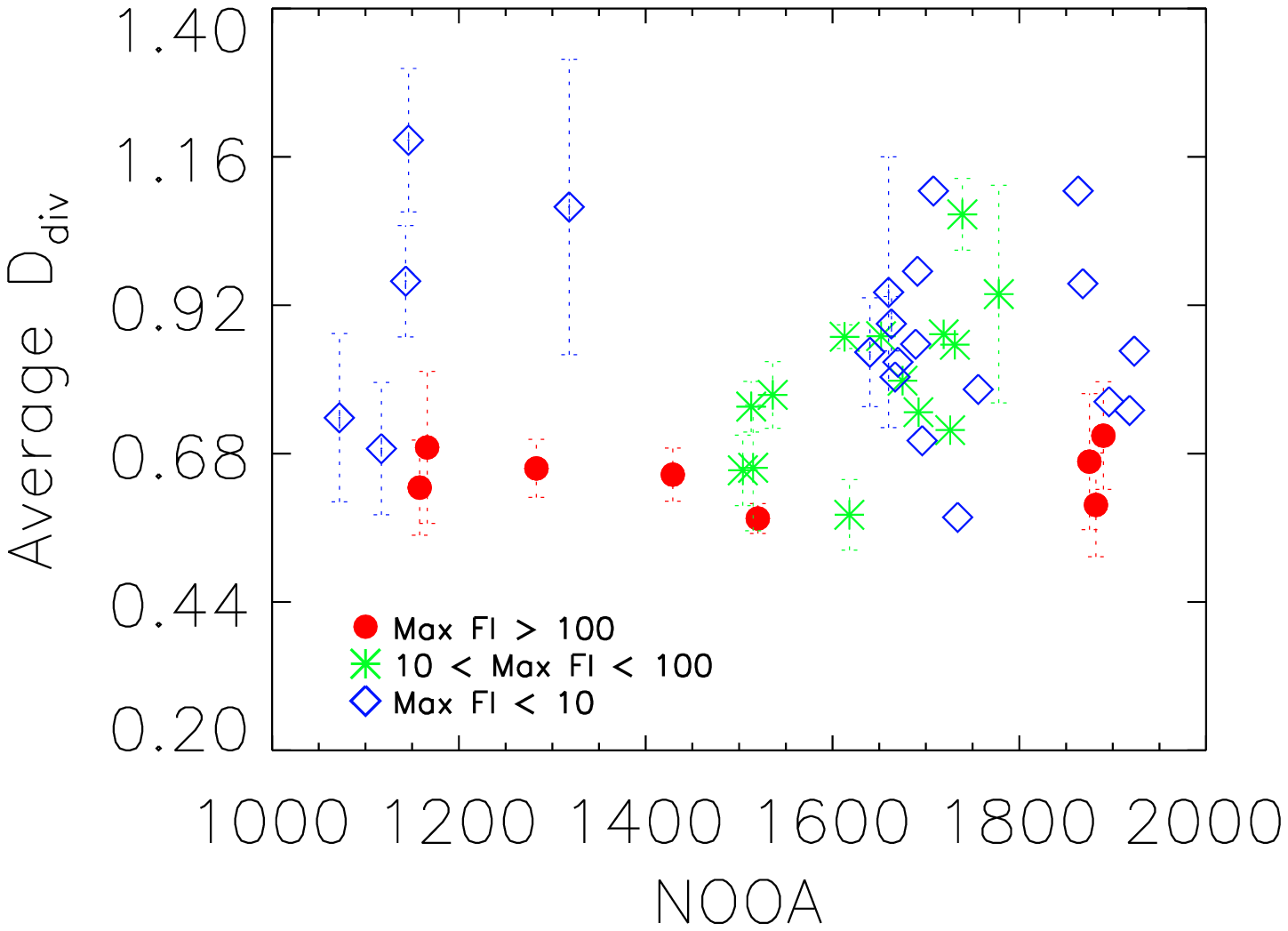}}
              \caption{Comparison between the average values of the fractal  and multifractal  parameters measured on the sample of { 43} ARs.  
              Shown are  results from unsigned magnetic flux measurements in the AR of the generalized fractal dimension  $D_0$ and $D_8$ (top panels), and multifractal parameters $C_{\mathrm{div}}$ and $D_{\mathrm{div}}$  (bottom panels).
              The regions were divided into three groups according to the flaring level of the studied AR. Details are given in Section 2 and  caption of Figure \ref{f5}. 
}
   \label{f6}
   \end{figure*}

      \begin{figure*}
             \centerline{\includegraphics[width=6.5cm]{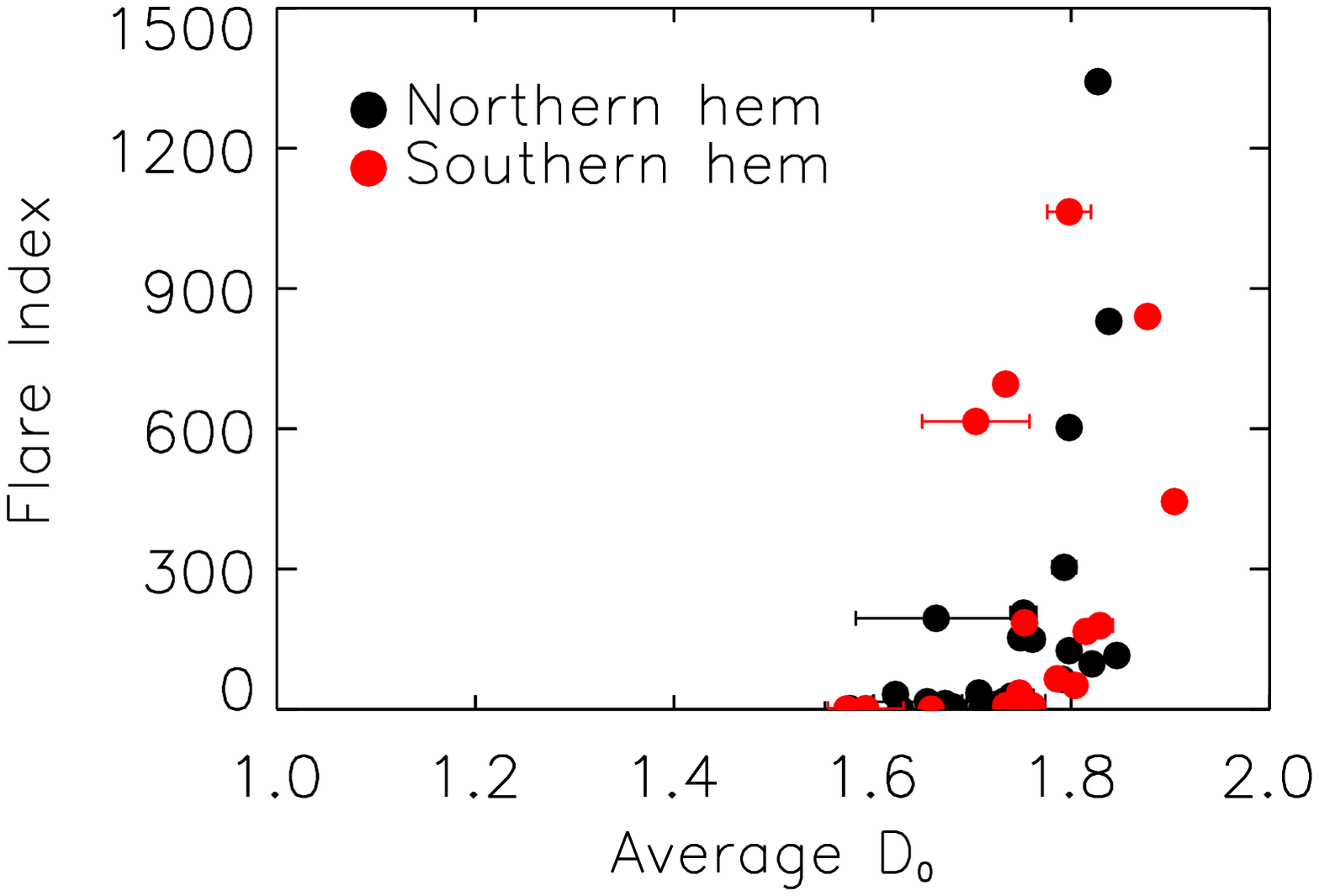}\includegraphics[width=6.5cm]{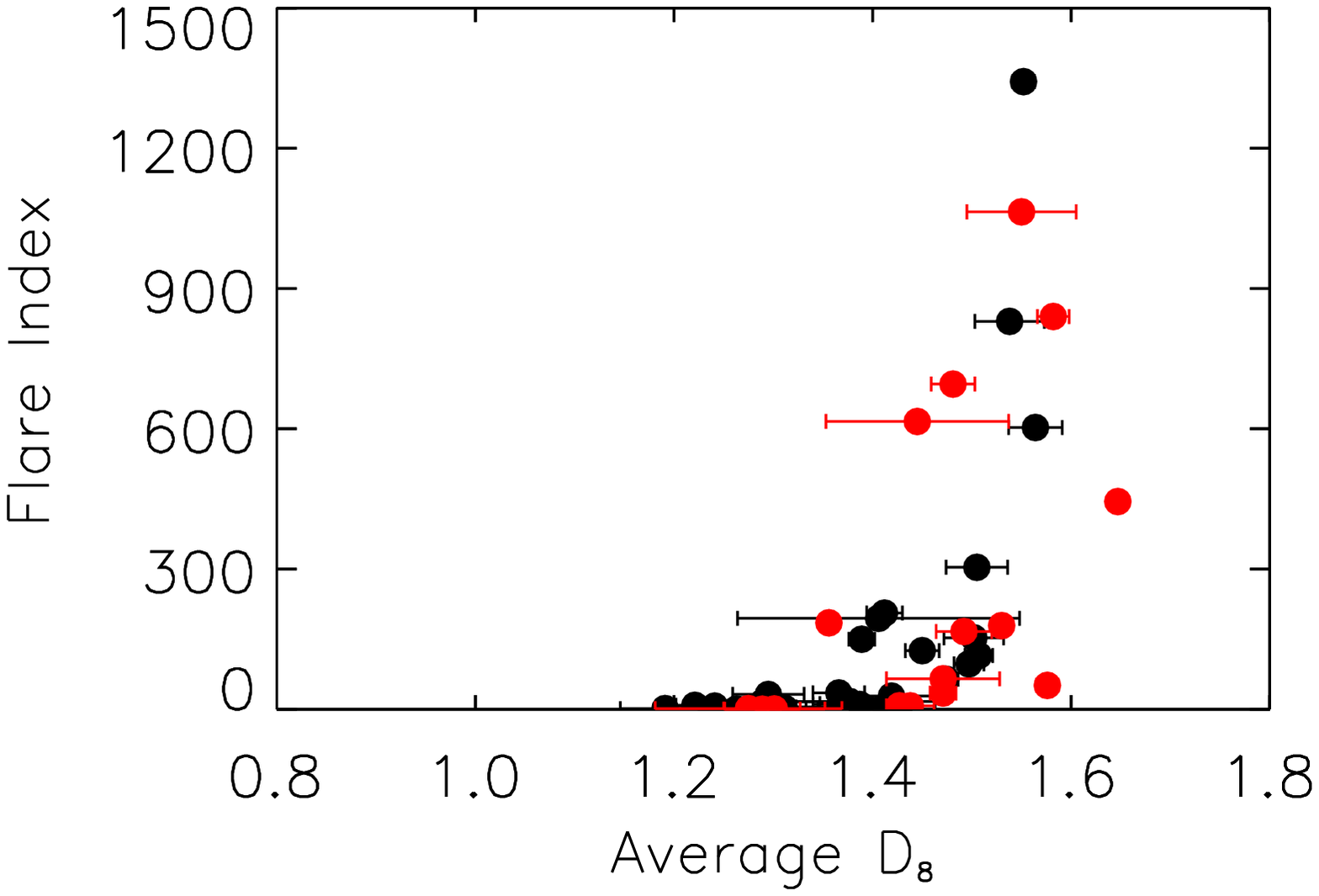}}
              \centerline{\includegraphics[width=6.5cm]{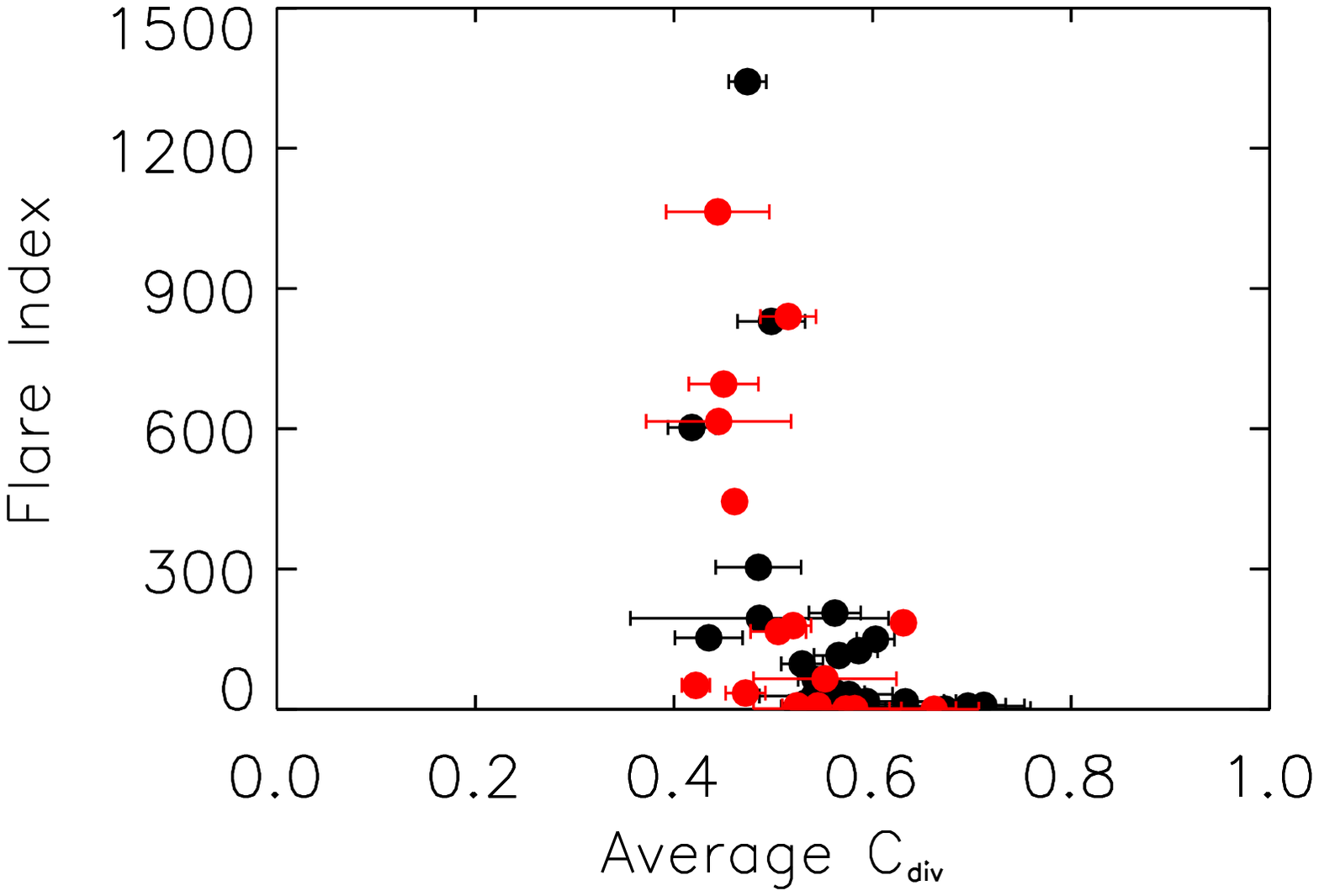}\includegraphics[width=6.5cm]{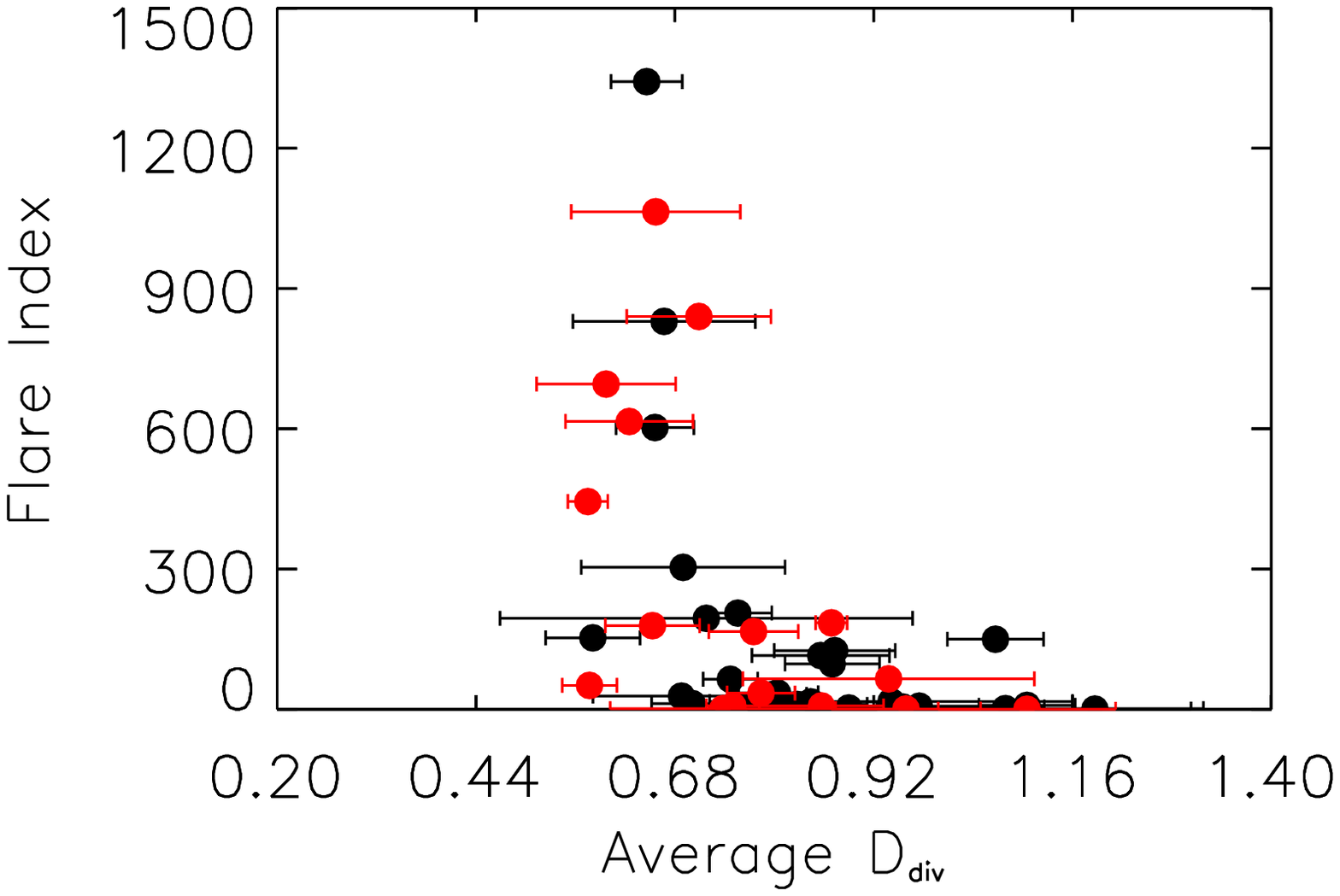}}
              \caption{{ Flare index of the analyzed ARs versus average values of the fractal 
  and multifractal  parameters measured on our sample of { 43}  ARs, which include  B- or C-,  M-, and X-class flaring regions. Shown are  results from unsigned magnetic flux measurements in the ARs of the generalized fractal dimension  $D_0$ and $D_8$ (top panels), and multifractal parameters $C_{\mathrm{div}}$ and $D_{\mathrm{div}}$  (bottom panels). 
  Black and red symbols show results for the ARs in the northern and southern solar hemispheres, respectively. }}
   \label{f7}
   \end{figure*}


{ 
Figure \ref{f5} shows the time series of fractal and multifractal parameters  derived from total unsigned flux measurements of the analyzed ARs. In order  to point out the dependence of measured values on the AR flaring activity, the results were divided into three groups, according to the three classes of flaring  ARs introduced in Section 2. The three classes concern the  ARs that have hosted B- or C-class, M-class, and X-class events. 

Figure \ref{f5} shows that the  $D_0$ values measured on all ARs  range  between $\approx$ 1.50 and $\approx$ 1.90, in agreement with previous reports in the literature on a significant fractality of the morphology of the magnetic flux concentration in both flaring and flare-quiet ARs (see {\it e.g.} \inlinecite{Mcateer_etal2005}, \inlinecite{Conlon_etal2008}, \inlinecite{Criscuoli_etal2009},  and \inlinecite{Georgoulis_2012} for results of the fractal dimension estimate of ARs derived from the box-counting method and {\it e.g.} \inlinecite{Criscuoli_etal2007}  and references therein for the findings from other methods). 
Figure \ref{f5} also shows that 
higher $D_0$ and $D_8$ values, as well as lower $C_{\mathrm{div}}$ and $D_{\mathrm{div}}$, are estimated, on average,  for the most flaring regions in the AR sample, than derived from  the  more  quiet ARs. 
We found that the difference between the values derived from the extreme classes of flaring ARs in our sample  is   larger for the  $D_0$ and $D_8$ values 
than for the $C_{\mathrm{div}}$ and $D_{\mathrm{div}}$  parameters, whose trends obtained from the  three classes of analyzed  ARs  overlap more than for the fractal parameters.

Figure \ref{f6} shows  the average   of the parameters  measured on  each AR.  We found that  the   parameters derived from the three classes  of analyzed ARs are  distributed between a lower and an upper limit, but a systematic difference between the trends obtained from distinct  classes of flaring ARs seems also to occur.
Indeed, 
Figure   \ref{f7}  shows the relation between the average value of the parameters estimated for each AR in the sample and its  FI. The measurements were divided according to the hemisphere hosting the AR, to test any hemispheric signature on them. In agreement with previous works ({\it e.g.} \opencite{Mcateer_etal2005}; \opencite{Abramenko_2005}; \opencite{Criscuoli_etal2009}), we found that higher values of fractal $D_0$ and $D_8$ parameters, as well as a lower value of $C_{\mathrm{div}}$ and $D_{\mathrm{div}}$,  imply  a higher flare activity of the  ARs, in terms of their  FI that account for the flare size and event frequency. The  scatter of data  does not support the assumption of a linear relation between the compared quantities, although a non-linear relation may hold between them.  
Figure  \ref{f7}  shows that this result holds likewise for the ARs of both hemispheres.}

\begin{table}
\caption{Summary of the average value and standard deviation  of the fractal ($D_0$ and $D_8$) and multifractal ($C_{\mathrm{div}}$ and $D_{\mathrm{div}}$) parameters measured  in our sample of { 43} ARs, which includes  { 8, 14, 16, and  5 regions that have produced at least one X-, M-, C, and B-class} flare, respectively, by considering unsigned and signed  flux data in the analyzed ARs, { with distinction of the leading and following flux polarities in each AR.}
}
\label{tbl:2}
\begin{tabular}{llllll}     
  \hline                   
Parameter & flux data & B and C flaring  & M flaring & X flaring & All regions\\
\hline
$D_0$     &          unsigned       &    1.69$\pm$ 0.06      &         1.77$\pm$ 0.06    &     1.81$\pm$0.07 & 1.74$\pm$0.08\\
$D_8$      &          unsigned      &     1.35$\pm$0.09     &          1.45$\pm$0.07    &     1.54$\pm$0.06 & 1.42$\pm$0.11\\
$C_{\mathrm{div}}$ &            unsigned    &       0.58$\pm$0.07   &            0.54$\pm$0.06   &      0.47$\pm$0.03 & 0.55$\pm$0.08\\
$D_{\mathrm{div}}$   &           unsigned   &        0.88$\pm$0.16  &             0.80$\pm$0.13    &     0.65$\pm$0.05 & 0.81$\pm$0.15\\
\hline
$D_0$       &              leading polarity      &        1.51$\pm$0.08        &          1.61$\pm$0.07    &        1.61$\pm$0.04 & 1.57$\pm$0.10\\
$D_8$       &          leading polarity       &          1.30$\pm$0.10     &             1.41$\pm$0.07     &       1.50$\pm$0.08 & 1.37$\pm$0.12\\
$C_{\mathrm{div}}$ &        leading polarity     &      0.40$\pm$0.13         &         0.34$\pm$0.10     &       0.23$\pm$0.05 & 0.35$\pm$0.12\\
$D_{\mathrm{div}}$ &        leading polarity       &        0.71$\pm$0.18       &           0.64$\pm$0.12      &      0.49$\pm$0.06 & 0.66$\pm$0.17\\
\hline
$D_0$       &       trailing polarity   &     1.49$\pm$0.11        &          1.58$\pm$0.09    &        1.66$\pm$0.11& 1.54$\pm$0.14\\
$D_8$       &        trailing polarity    &           1.31$\pm$0.13        &          1.40$\pm$0.14     &       1.47$\pm$0.07 & 1.37$\pm$0.14\\
$C_{\mathrm{div}}$ &        trailing polarity      &        0.38$\pm$0.11     &             0.33$\pm$0.11    &        0.31$\pm$0.06 & 0.36$\pm$0.11\\
$D_{\mathrm{div}}$ &       trailing polarity        &       0.78$\pm$0.13      &            0.71$\pm$0.14   &         0.64$\pm$0.14 & 0.74$\pm$0.14\\
\hline
\end{tabular}
\end{table}

\begin{table}
\caption{ The linear Pearson correlation coefficients between the measured  parameters and flaring level of the analyzed ARs, as expressed by their FI and FI max. The coefficients were computed by assuming measurement results from both unsigned and signed flux data in the analyzed ARs, { with distinction of the leading and following flux polarities in each AR.}
}
\label{tbl:3}
\begin{tabular}{llllll}     
\hline
Flare index & flux data   &   $D_0$ &     $D_8$ &    $C_{\mathrm{div}}$ & $D_{\mathrm{div}}$\\
\hline
FI      & unsigned  & 0.51   & 0.60  & -0.55  & -0.52\\
FI  & leading polarity  & 0.39 &   0.51 &  -0.39 &  -0.45\\
FI  & trailing polarity   & 0.48  &  0.46  & -0.33 &  -0.49\\
\hline
FI max  &    unsigned &  0.48   & 0.53 &  -0.45  & -0.44\\
FI max  &   leading polarity & 0.32 &   0.48  & -0.42 &  -0.43\\
FI max  &   trailing polarity & 0.47  &  0.38  & -0.20 &  -0.35\\
\hline
\end{tabular}
\end{table}

{
The results shown in Figures \ref{f5} to \ref{f7} are quantified in Table \ref{tbl:2}, which  summarizes the  average and standard deviation  of the parameters measured for each class of analyzed ARs. We found 
distinct average values of  the parameters measured on ARs that have hosted flares of different class. However,   the dispersion of  values measured on  ARs that have produced same class events  is such that the parameters deduced from distinct classes  of flaring regions  can also largely overlap.
Based on the results of our measurements, C- and M-class flaring ARs are practically indistinguishable, as well as M- and X-class flaring ARs. However,  C- and X-class regions can be either discerned or not depending on the analyzed ARs. 
  These findings  give  reasons to some  conflicting conclusions presented in the literature on the efficiency of the fractal and multifractal  measurements to discriminate flaring ARs. In fact, they confirm  the statistical trend reported by \inlinecite{Mcateer_etal2005}. They also show that  depending on the analyzed ARs, the fractal and multifractal  measurements derived from regions hosting distinct  classes of events may have or may have not similar values, as reported  from the analysis of two flaring and flare-quiet  ARs by {\it e.g.} \inlinecite{Georgoulis_2013} and  by considering the results from the most flaring and most flare-quiet ARs in the sample analyzed in this study, respectively.  
}


{ 
Table \ref{tbl:3} lists  the linear Pearson correlation coefficients between the measured  parameters and FI. The absolute values of the  coefficients derived by taking into account the parameters deduced from unsigned flux data of the analyzed ARs range from 0.51 to 0.60, being  0.51 ($D_0$), 0.60 ($D_8$), -0.55 ($C_{\mathrm{div}}$), and -0.52 ($D_{\mathrm{div}}$). These values  confirm the   weakly significant relation among the compared quantities  already reported in the literature. 
}

\subsection{Average Values of the  Parameters from Signed Flux Data}

      \begin{figure*}
     \centerline{\includegraphics[width=6.5cm]{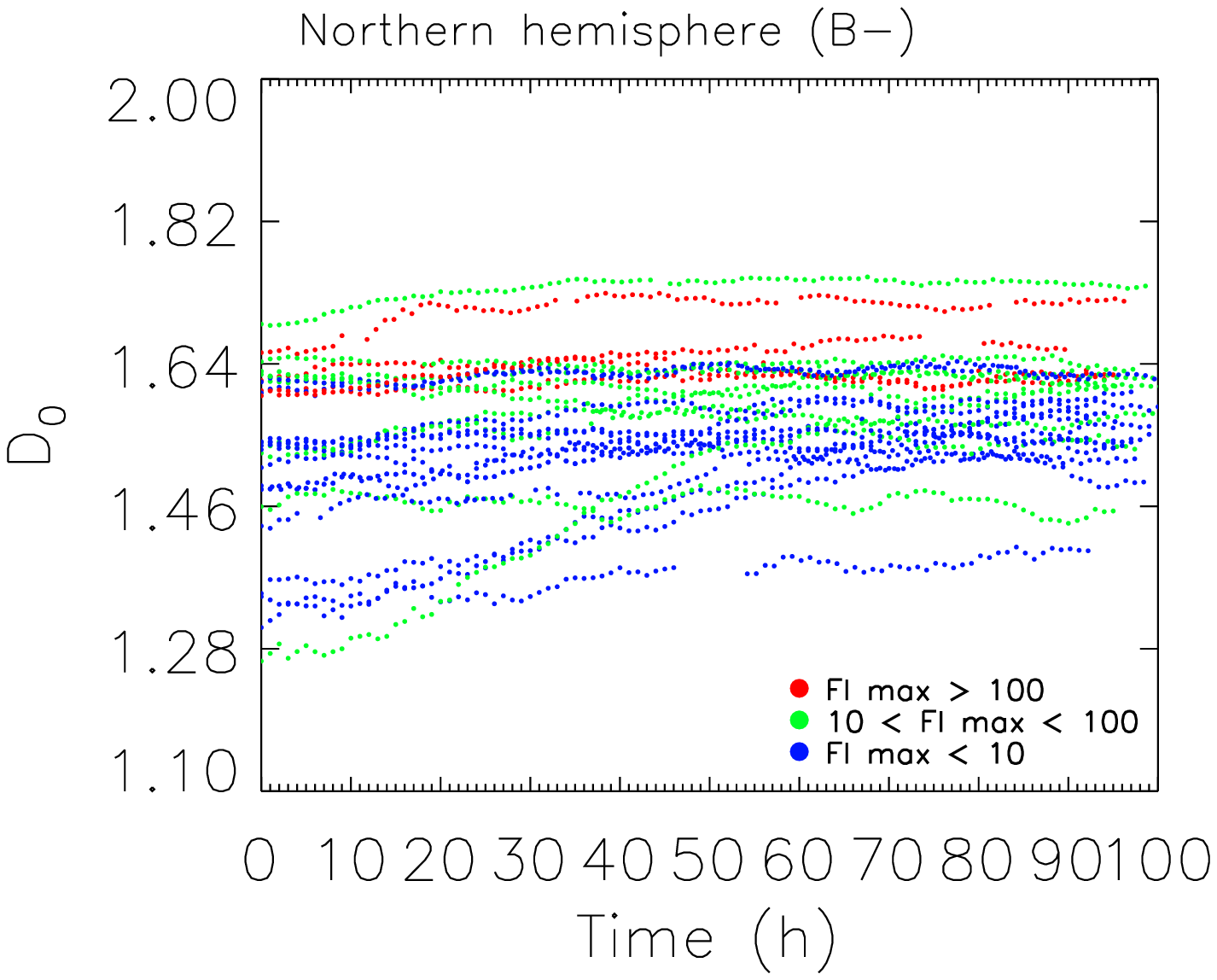}\includegraphics[width=6.5cm]{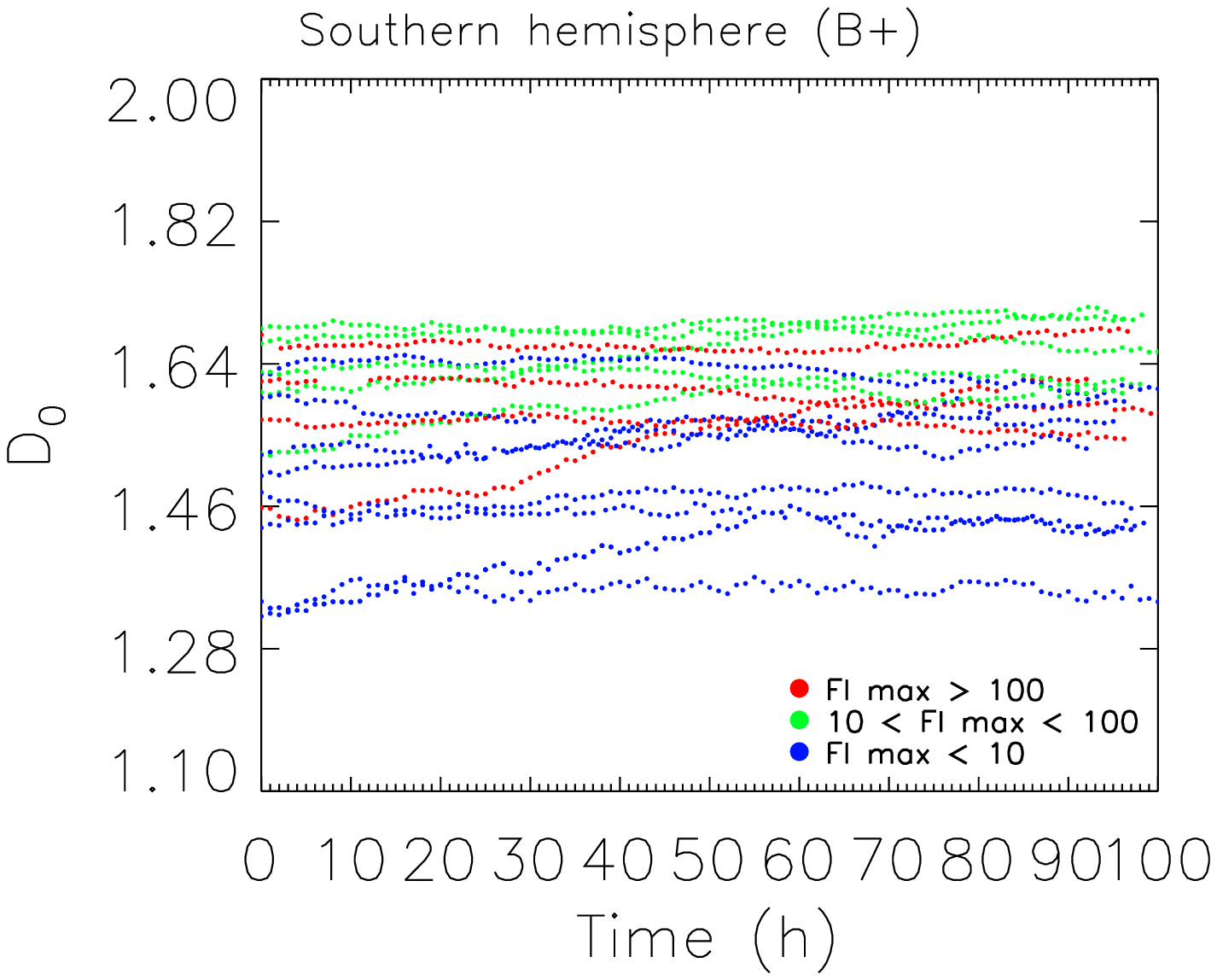}}
    \centerline{\includegraphics[width=6.5cm]{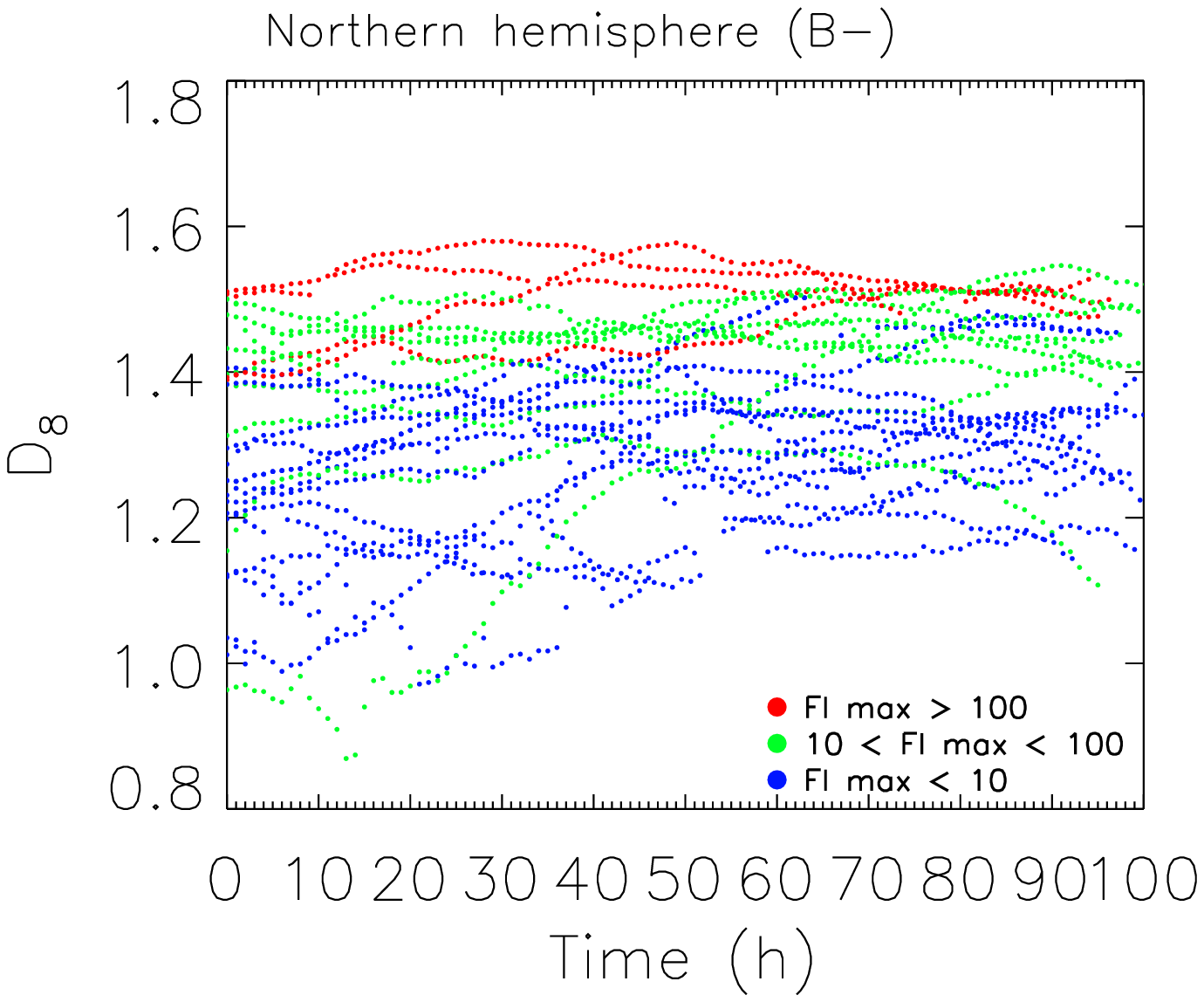}\includegraphics[width=6.5cm]{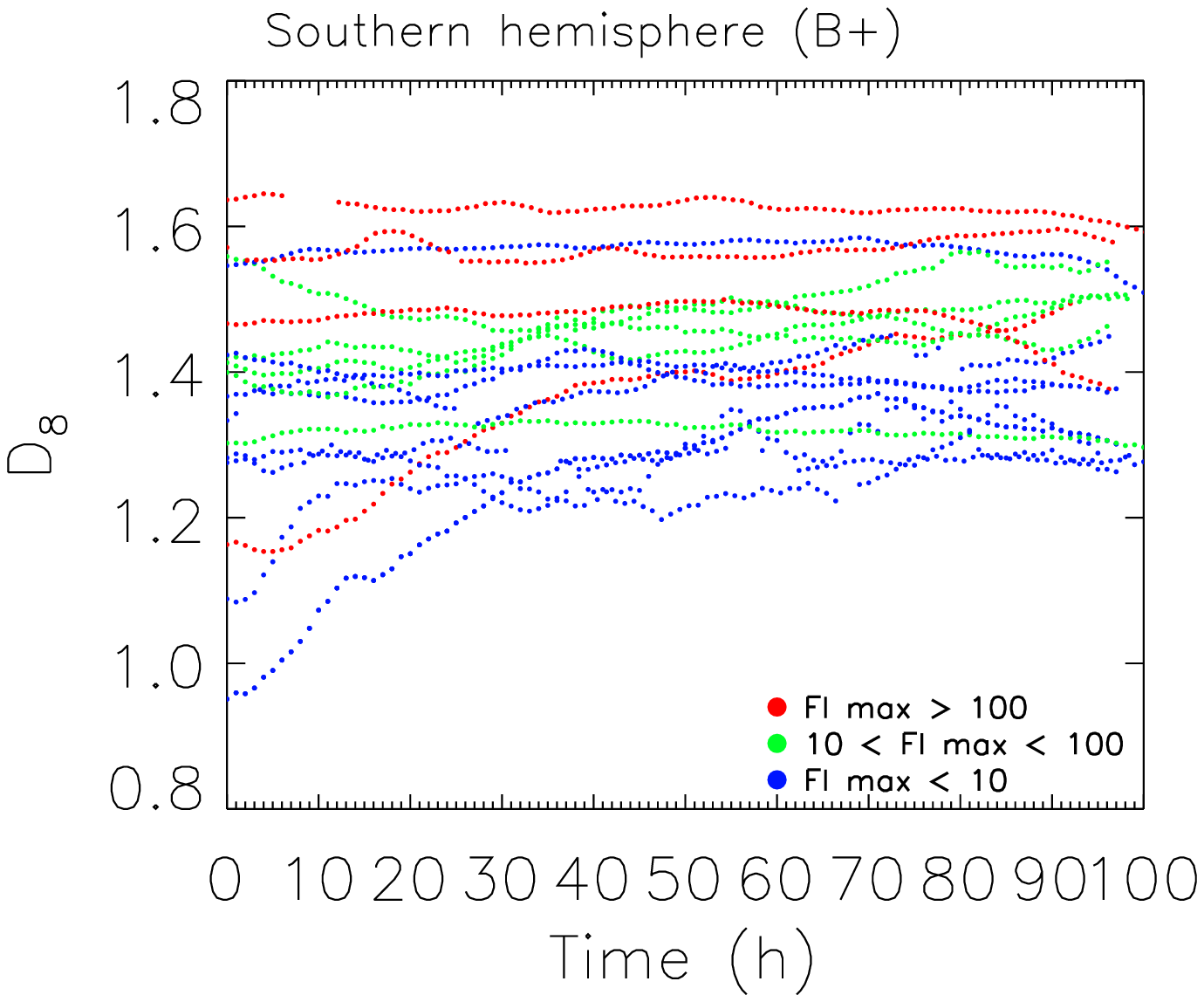}}
    \centerline{\includegraphics[width=6.5cm]{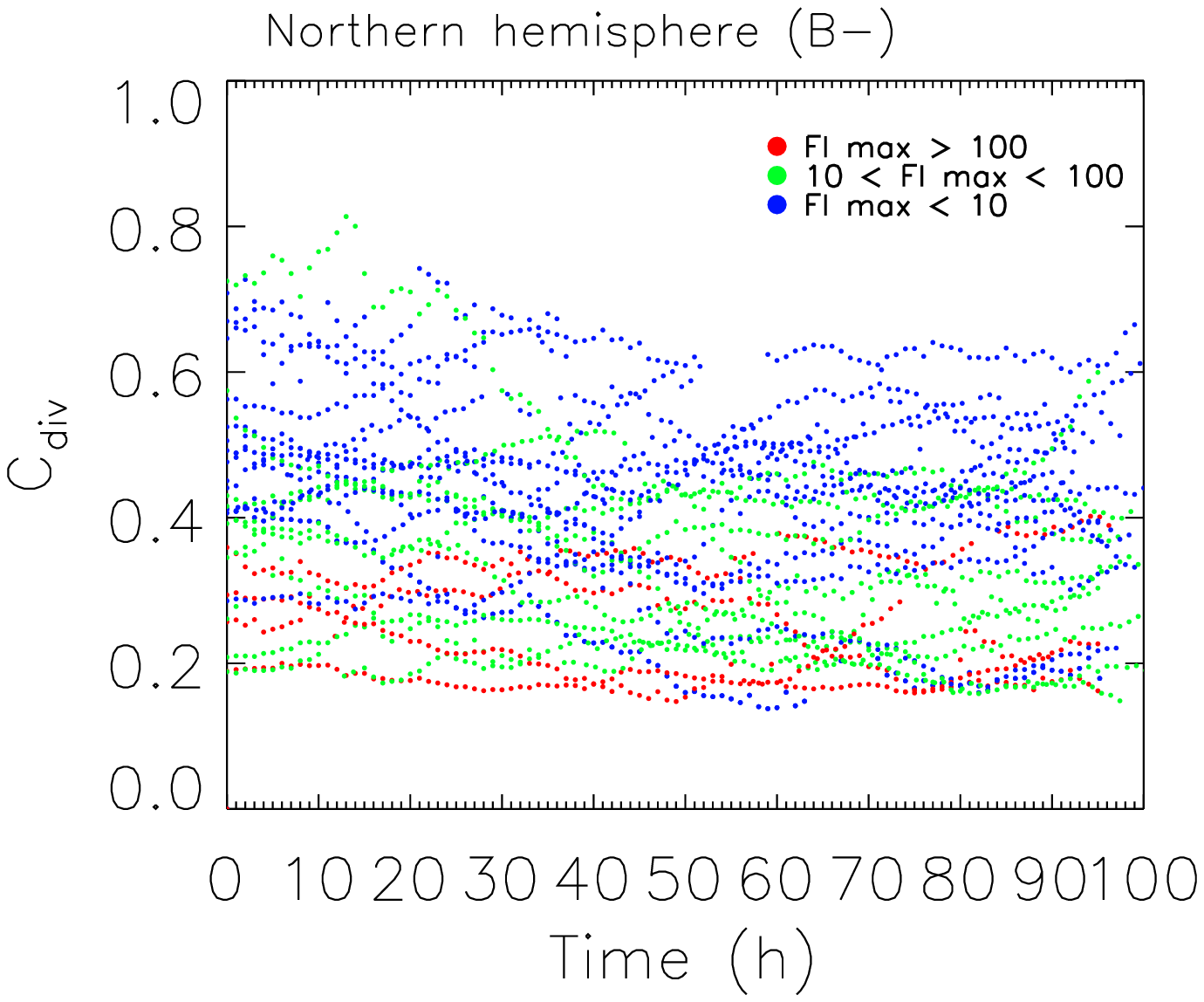}\includegraphics[width=6.5cm]{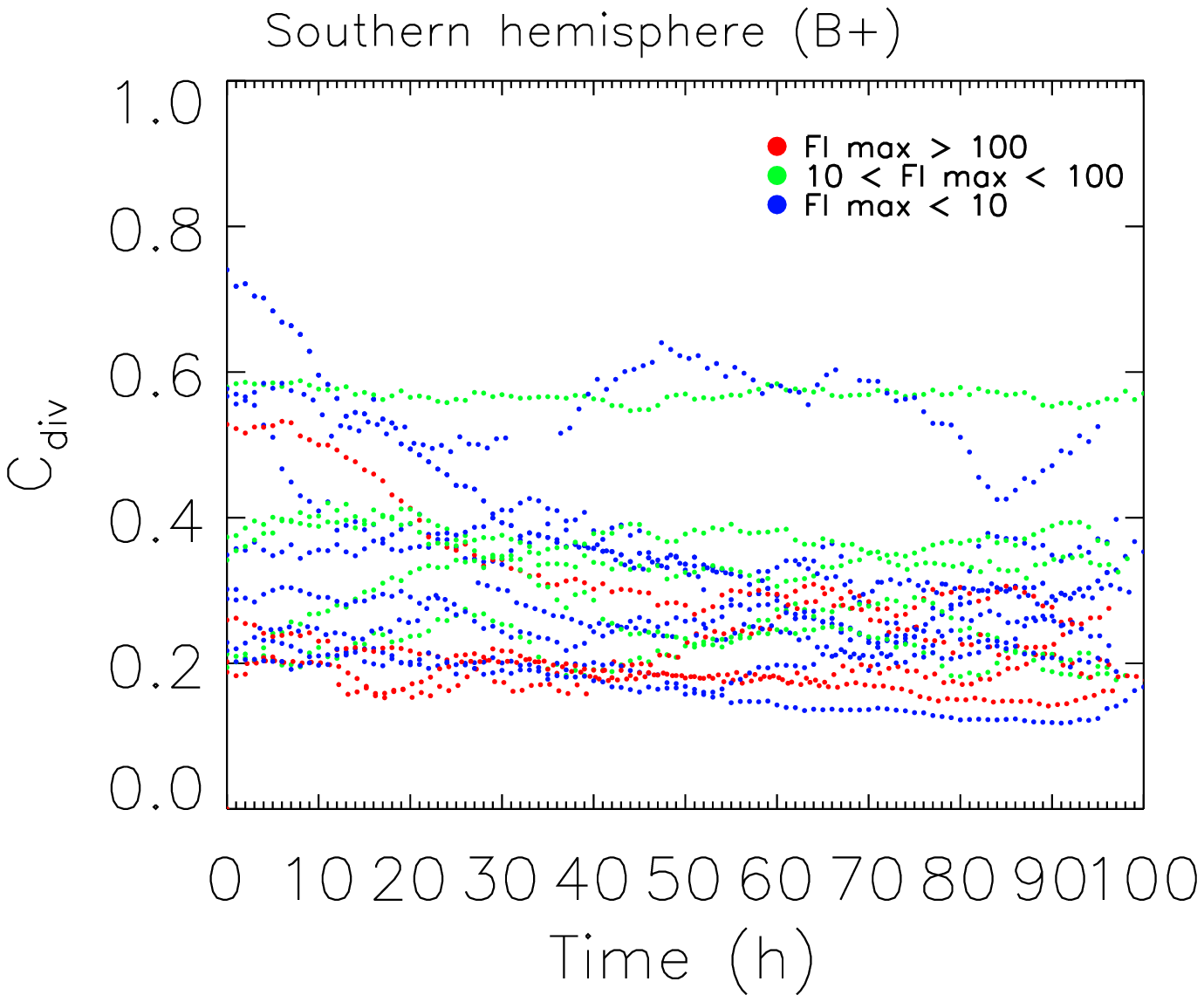}}
    \centerline{\includegraphics[width=6.5cm]{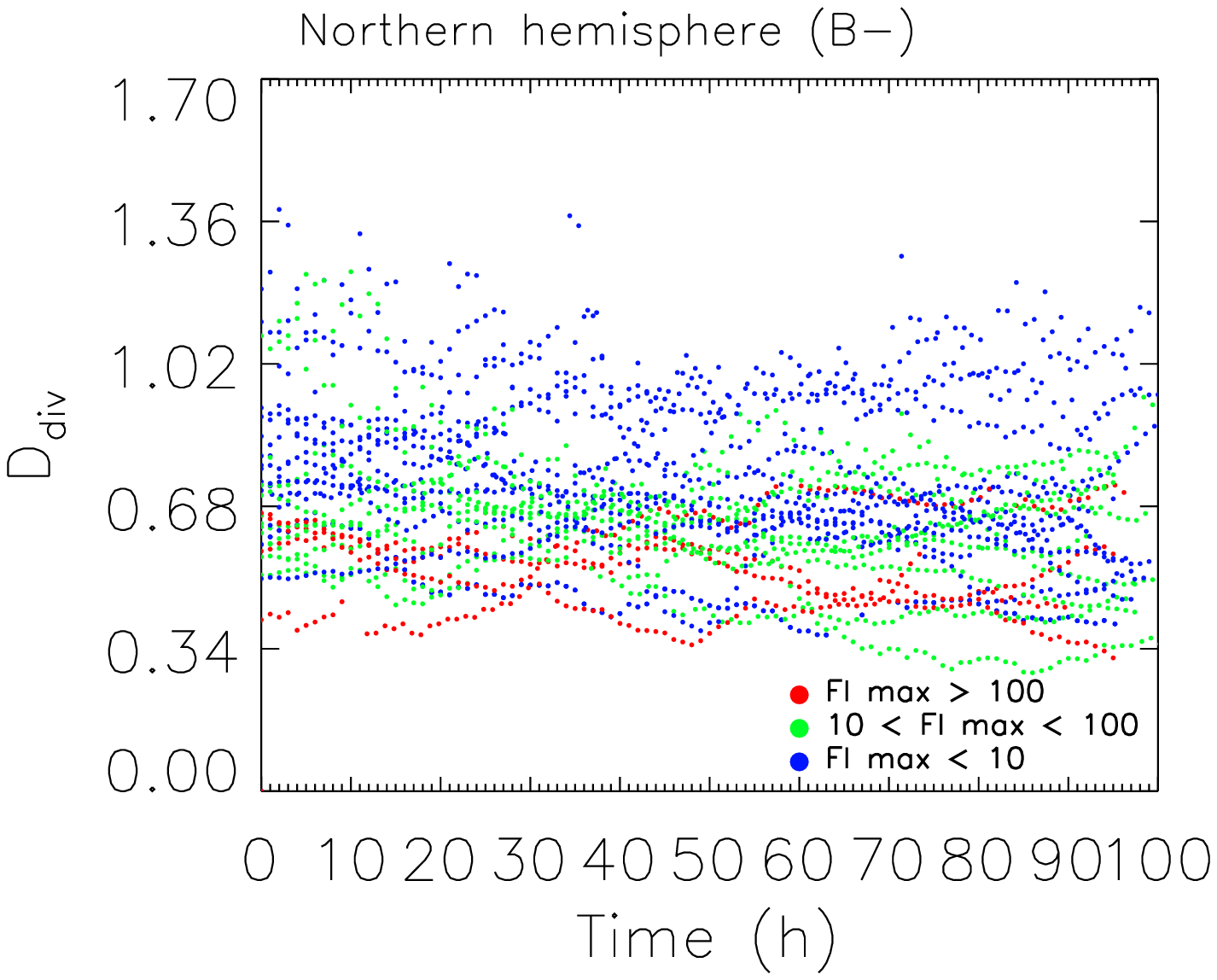}\includegraphics[width=6.5cm]{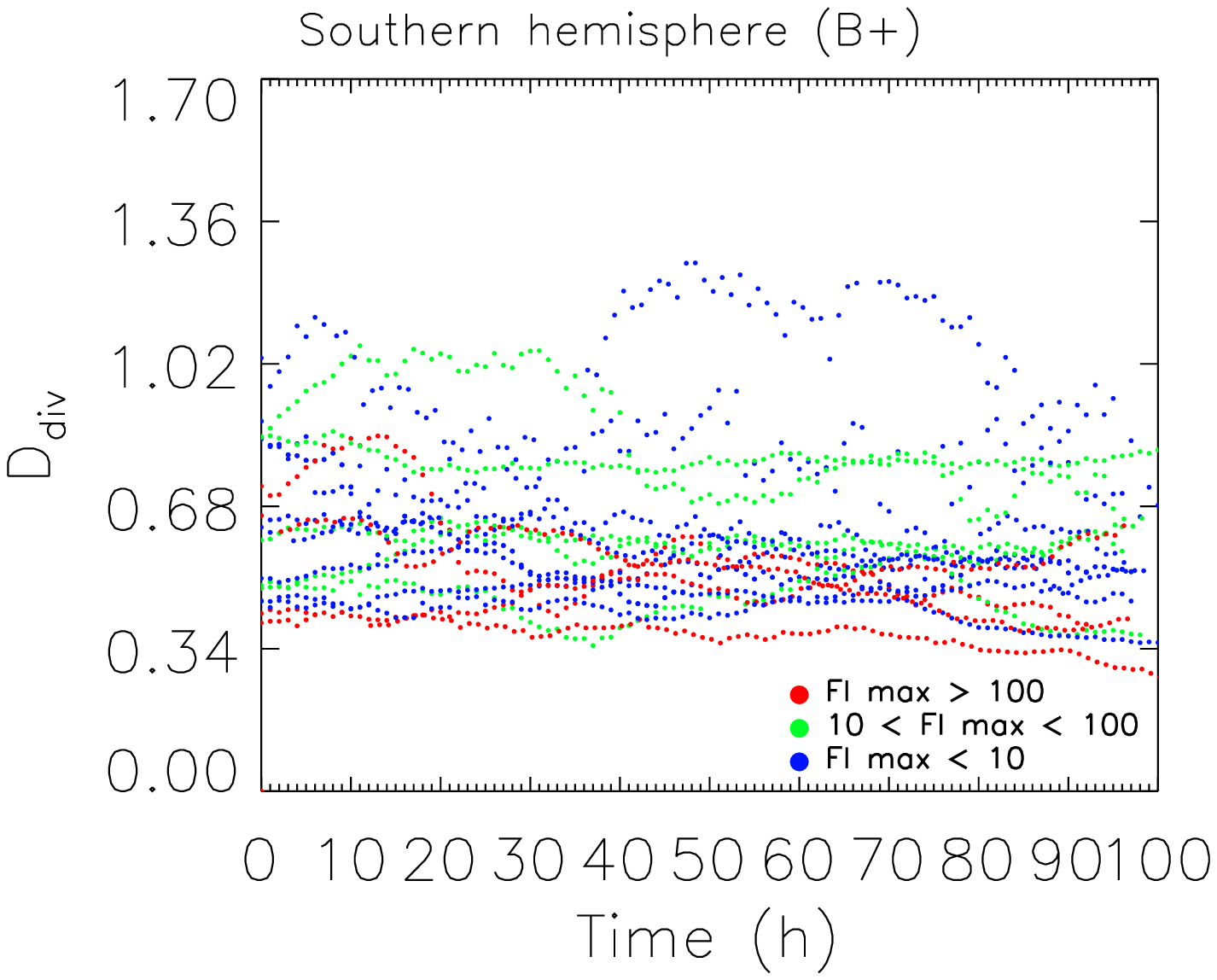}}
              \caption{{ Comparison among results of the fractal and multifractal parameters measured on the whole sample of analyzed ARs. Time series of fractal ($D_0$ and $D_8$) and multifractal ($C_{\mathrm{div}}$ and $D_{\mathrm{div}}$) parameters. Shown are  results from signed magnetic flux measurements of the leading polarity in the analyzed ARs that  were divided according to their flaring level and the hemisphere hosting them, northern (left panels) and southern (right panels). For clarity,  each symbol shows the average of results over 5 data points, which  correspond  to 1 h time interval.  Legend as in Figure \ref{f5}.}
 }
   \label{f5a}
   \end{figure*}
   
   \begin{figure*}
     \centerline{\includegraphics[width=6.5cm]{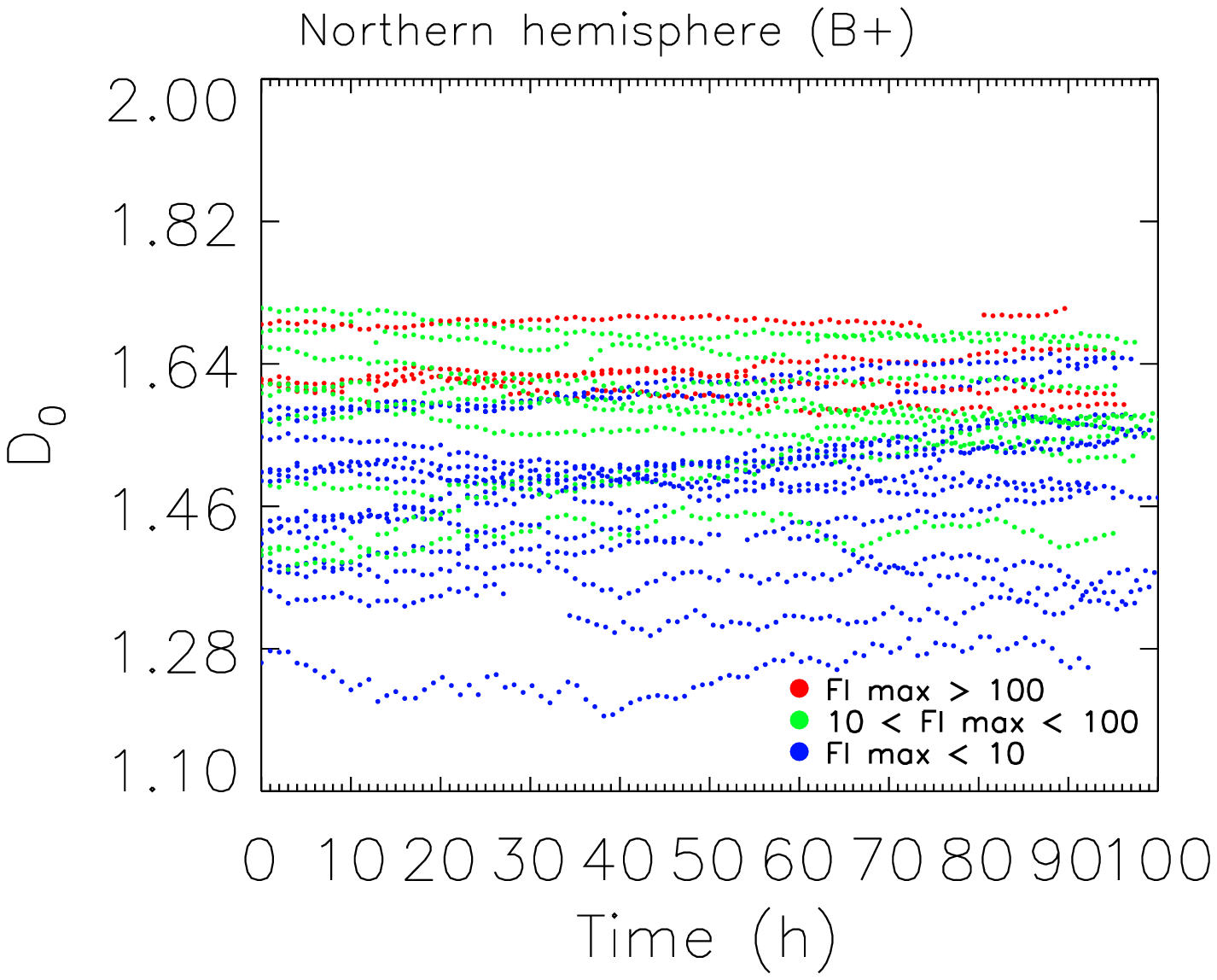}\includegraphics[width=6.5cm]{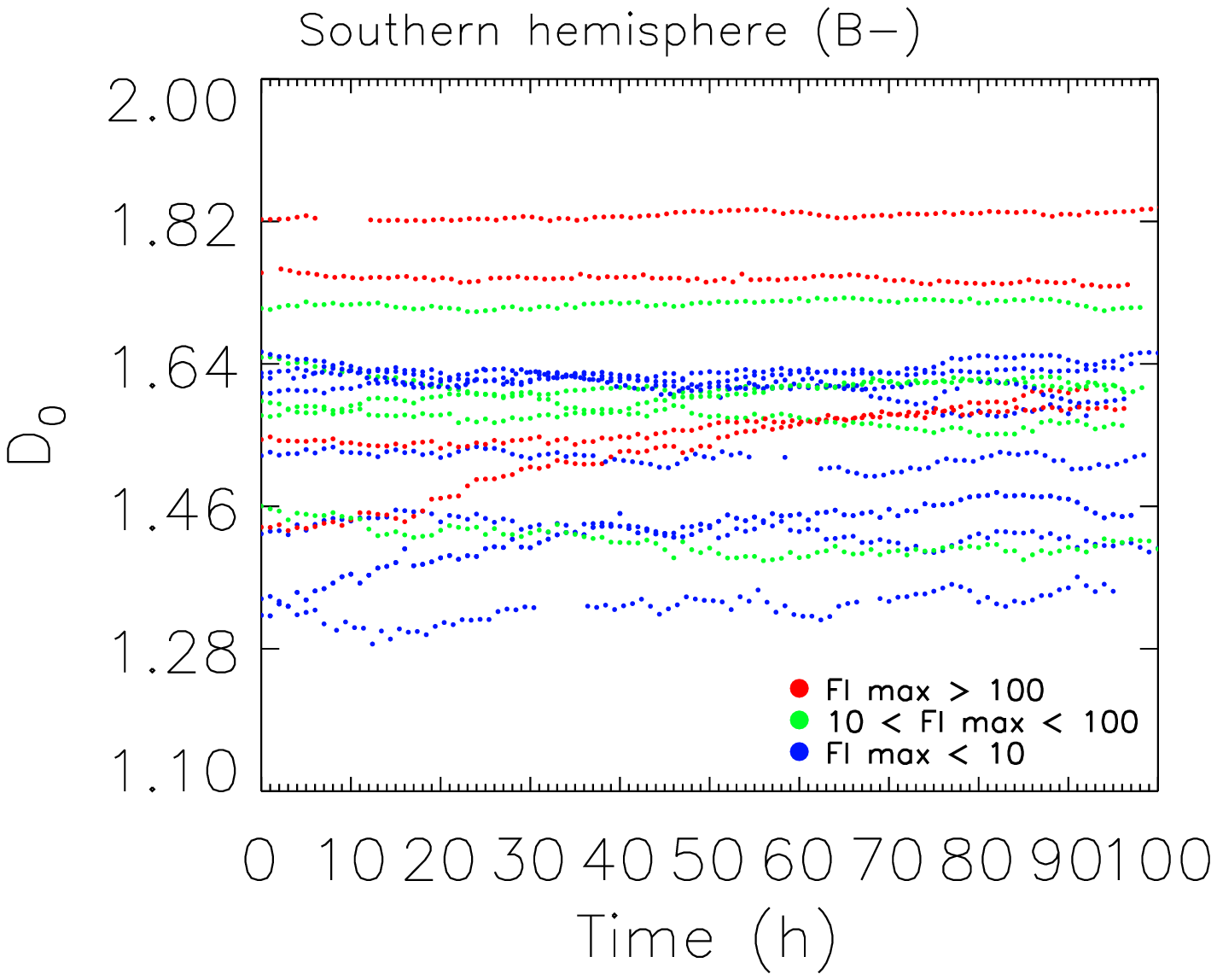}}
    \centerline{\includegraphics[width=6.5cm]{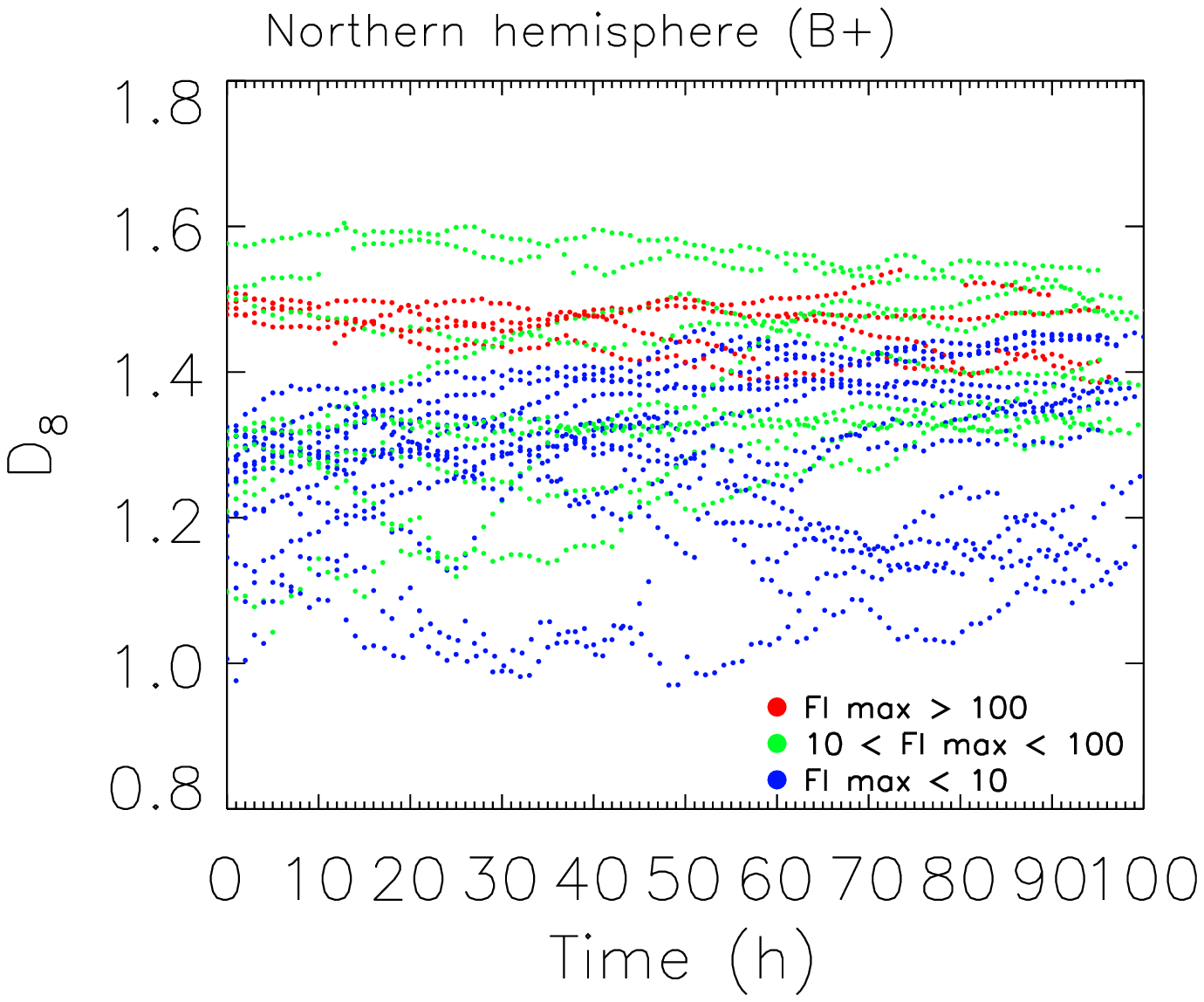}\includegraphics[width=6.5cm]{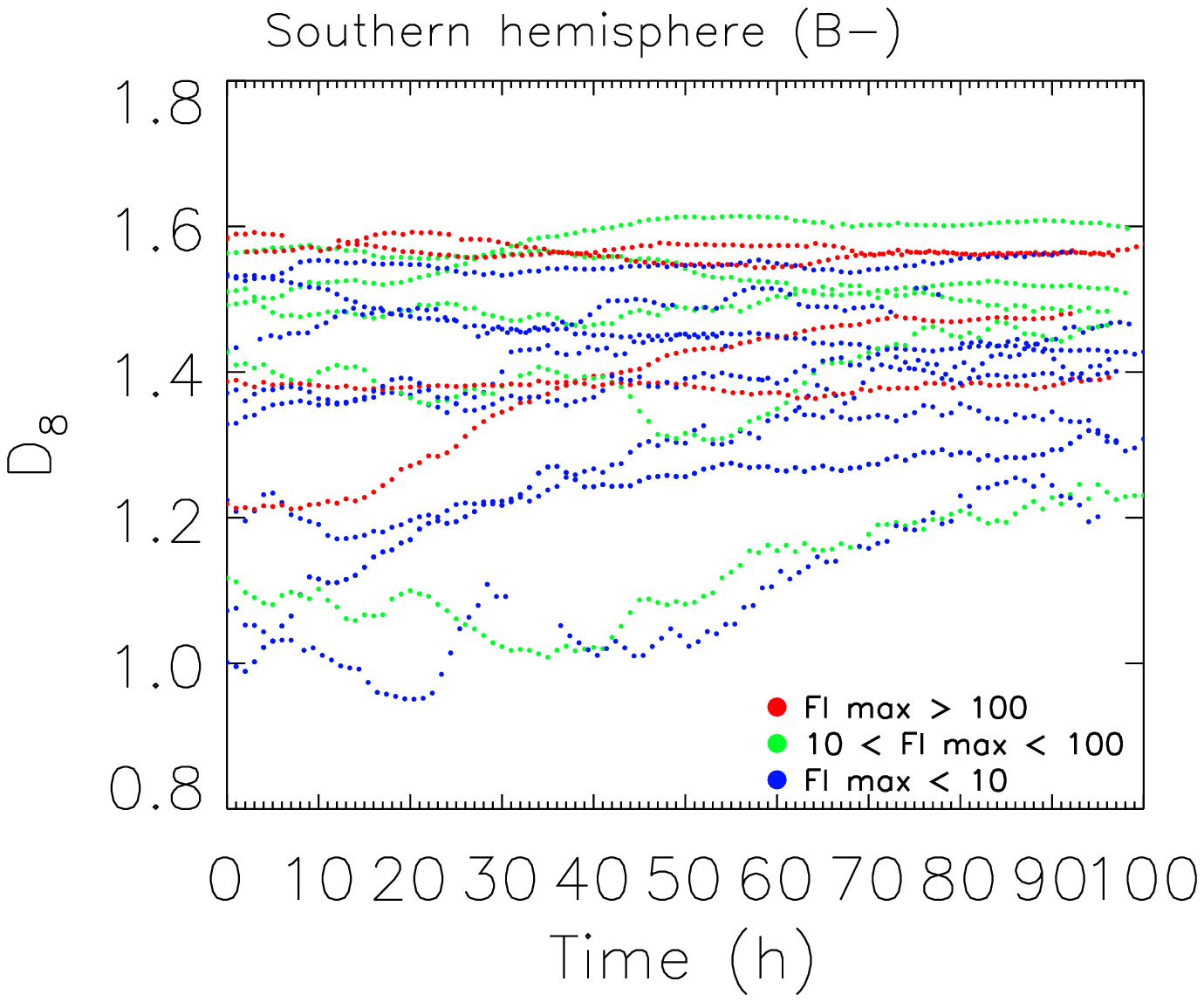}}
    \centerline{\includegraphics[width=6.5cm]{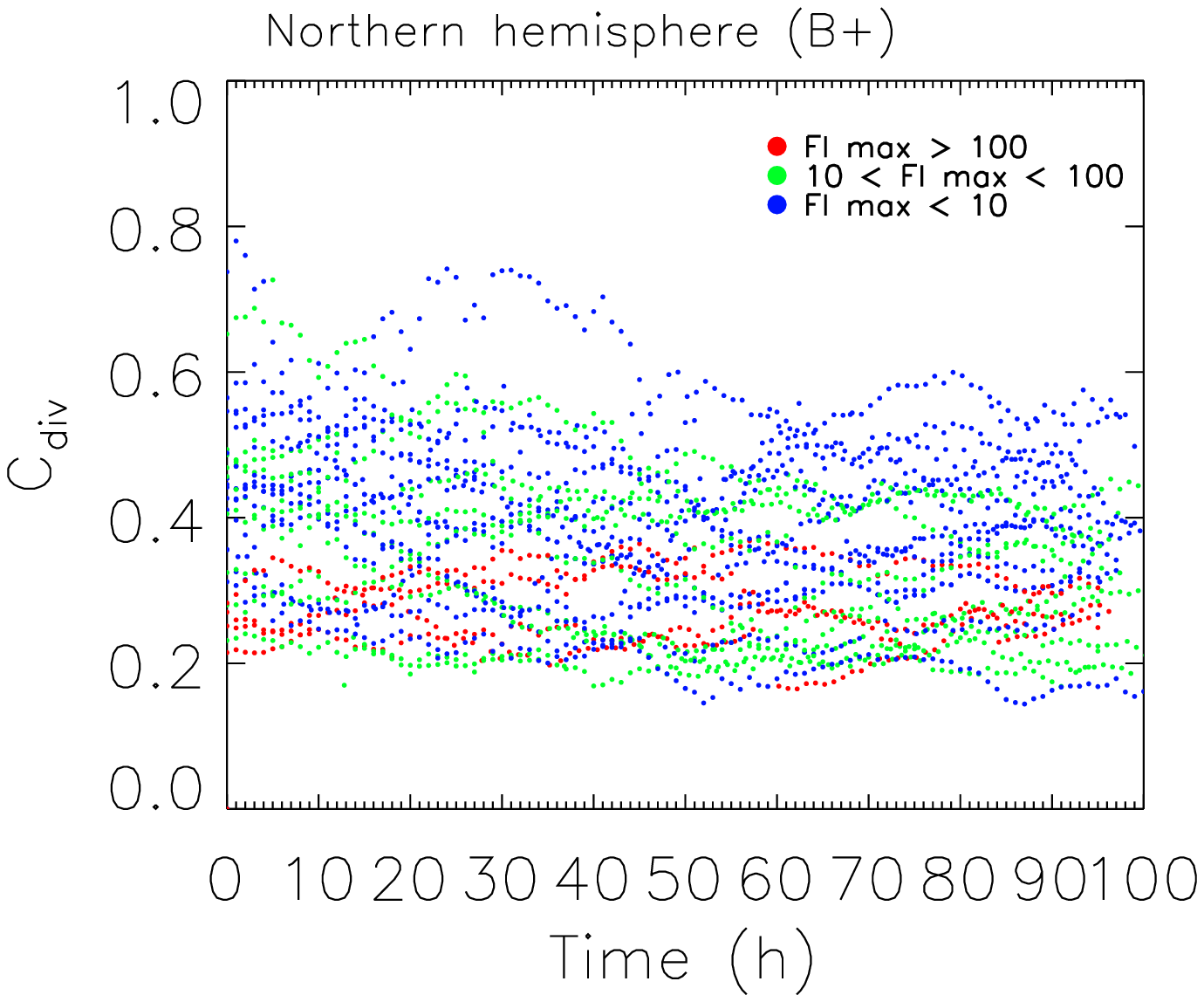}\includegraphics[width=6.5cm]{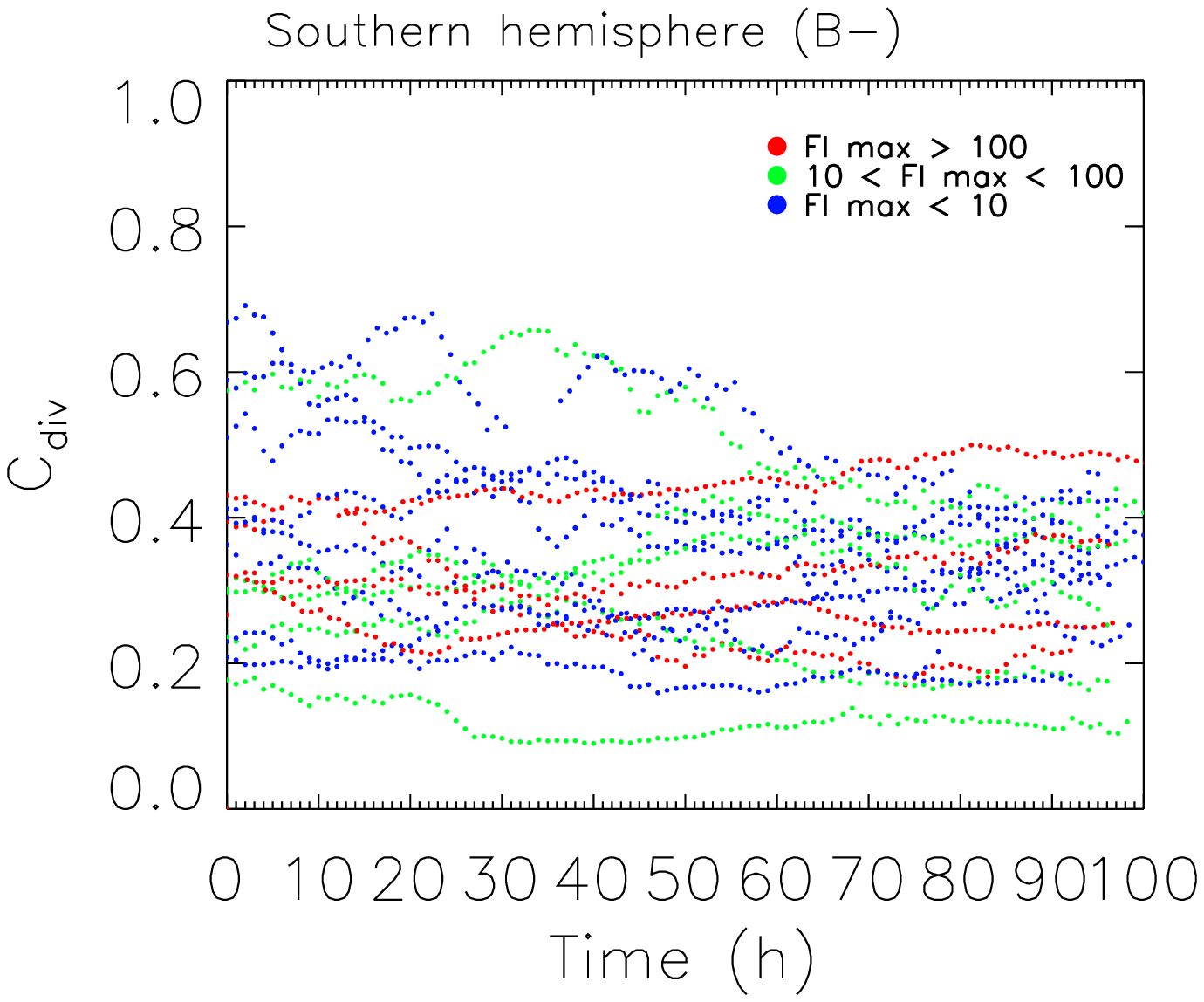}}
    \centerline{\includegraphics[width=6.5cm]{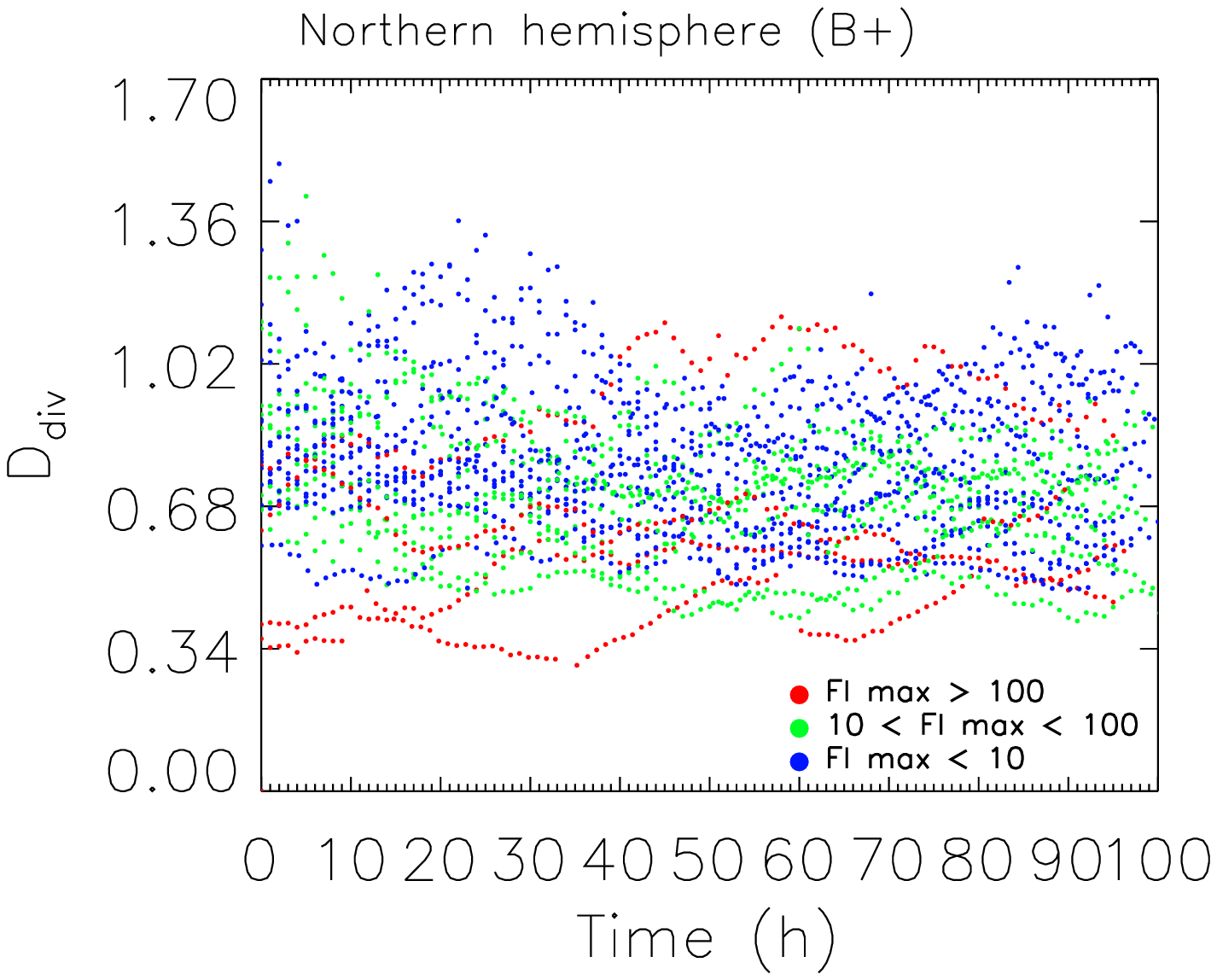}\includegraphics[width=6.5cm]{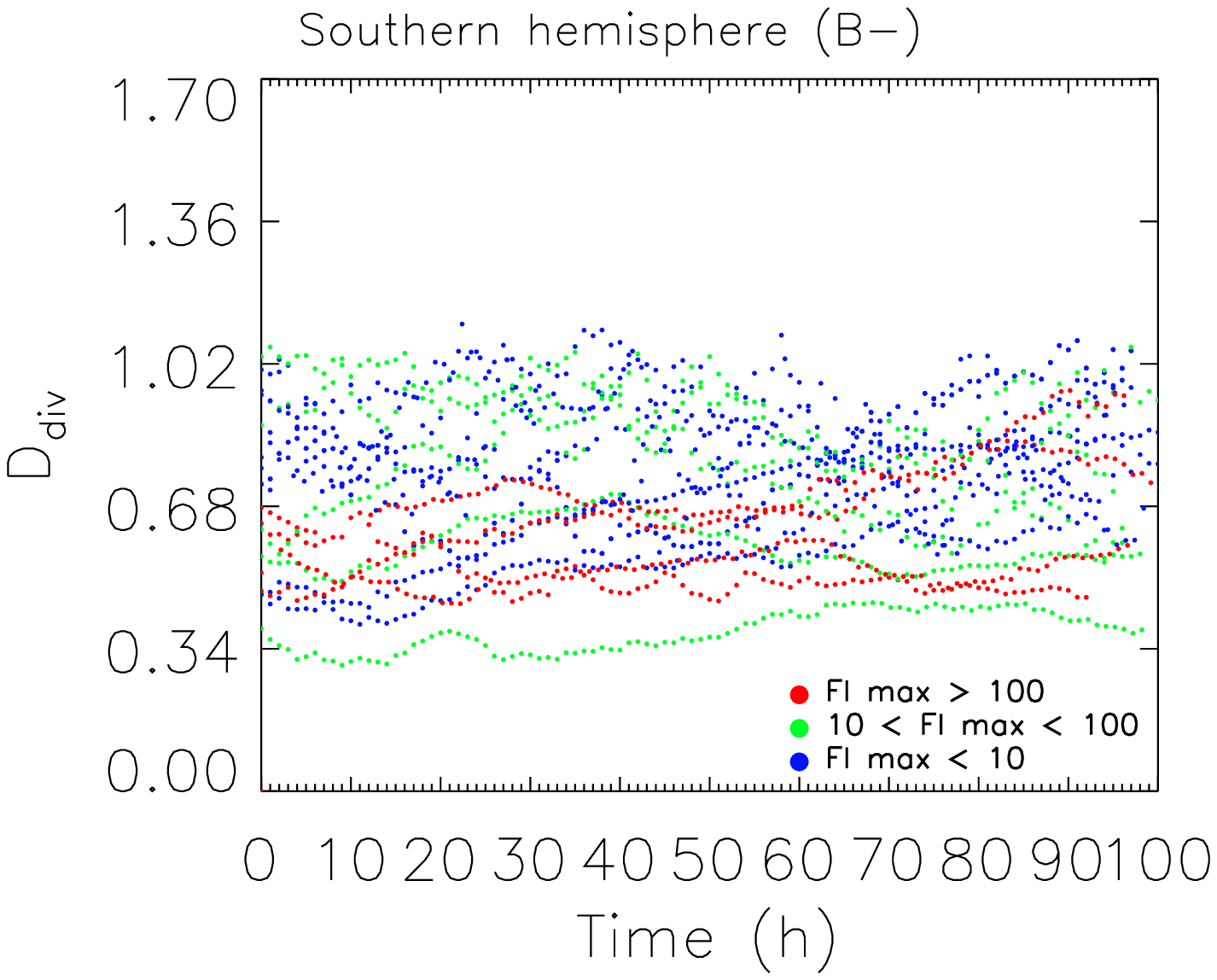}}
              \caption{{ Comparison among results of the fractal and multifractal parameters measured on the whole sample of analyzed ARs. Time series of fractal ($D_0$ and $D_8$) and multifractal ($C_{\mathrm{div}}$ and $D_{\mathrm{div}}$) parameters. Shown are  results from signed magnetic flux measurements of the trailing polarity in the analyzed ARs that  were divided according to their flaring level and the hemisphere hosting them,  northern (left panels) and southern (right panels). For clarity,  each symbol shows the average of results over 5 data points, which  correspond  to 1 h time interval.  Legend as in Figure \ref{f5}. }
 }
   \label{f5b}
   \end{figure*}

      \begin{figure*}
     \centerline{\includegraphics[width=6.5cm]{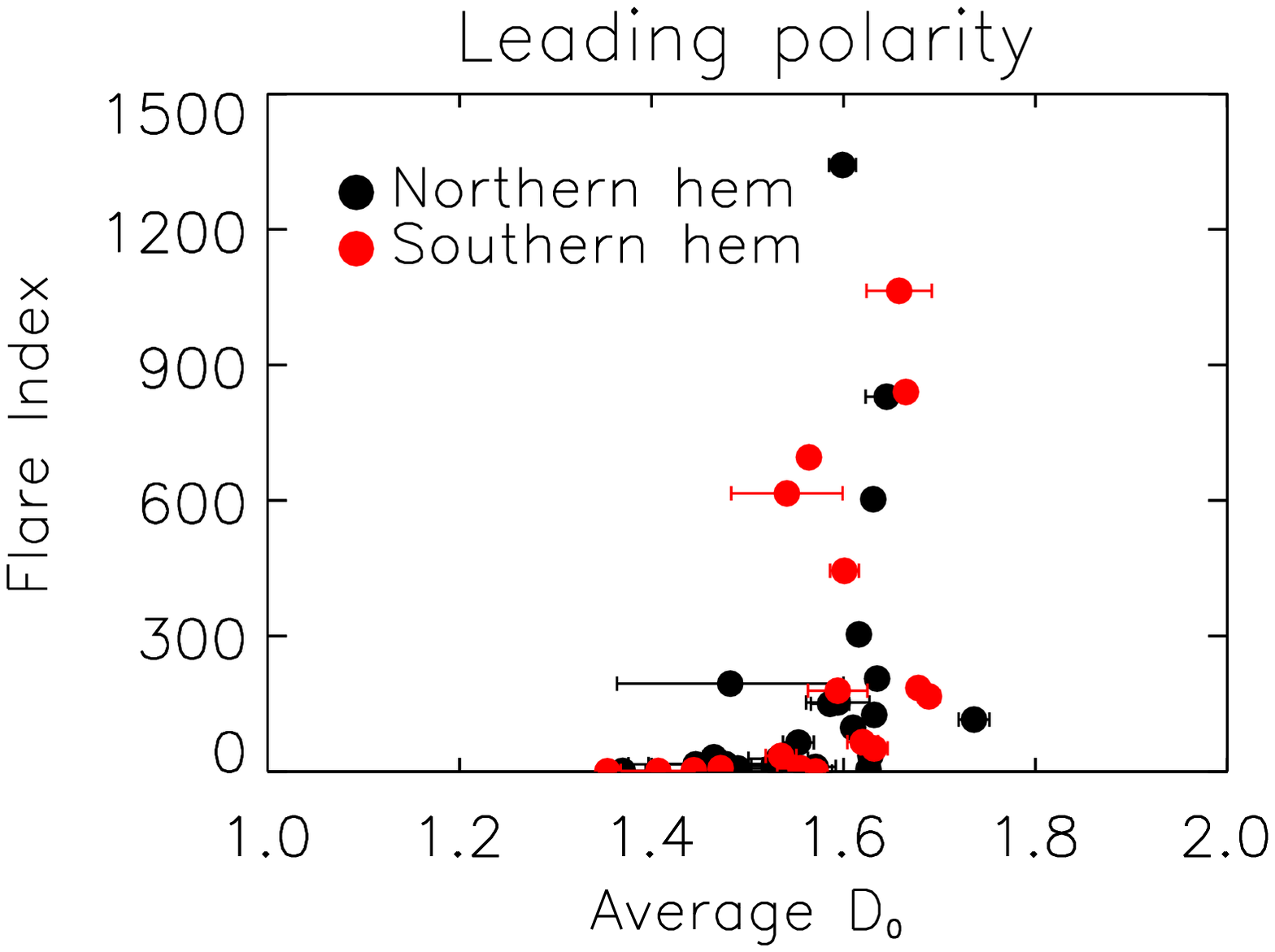}\includegraphics[width=6.5cm]{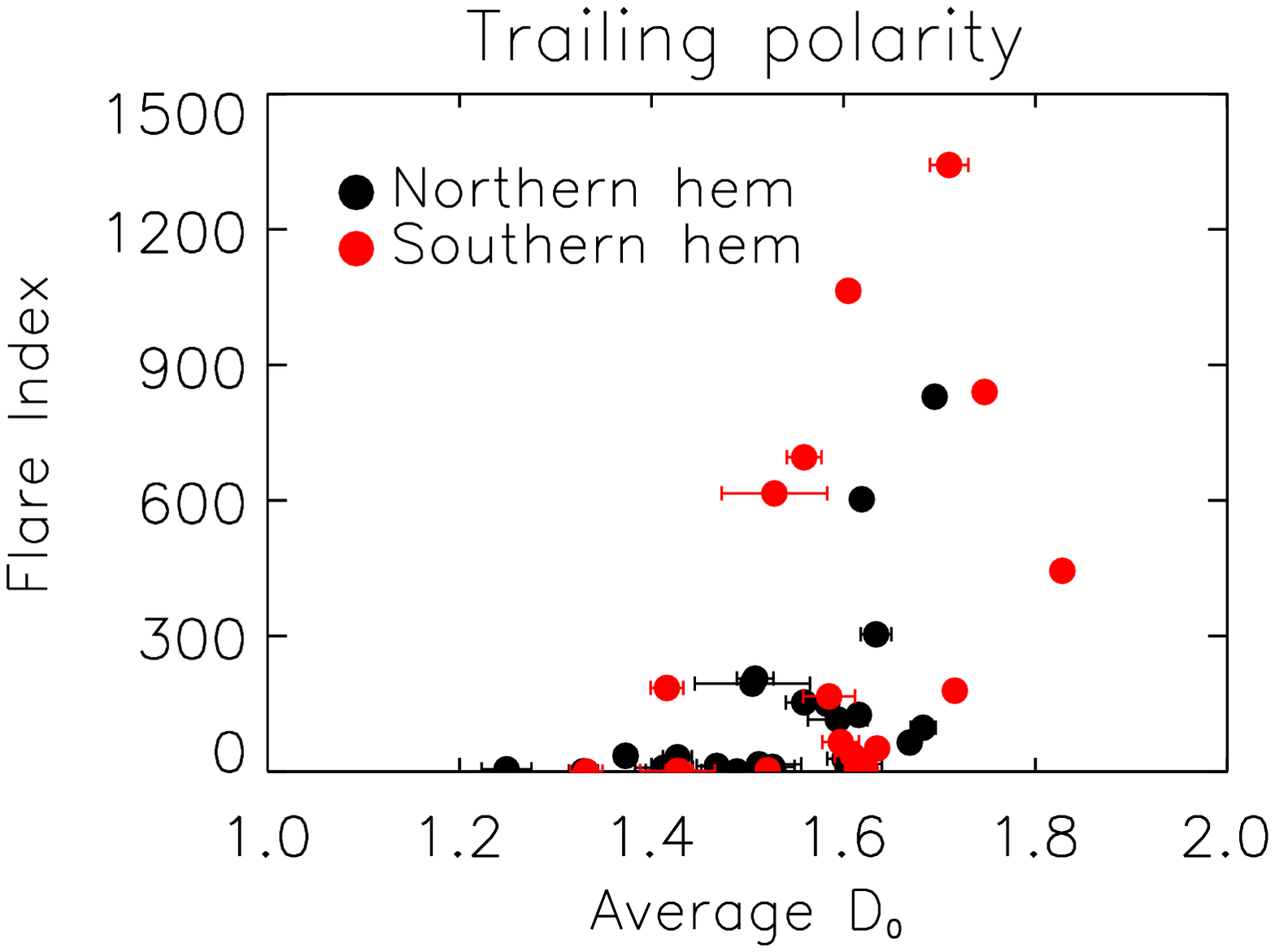}}
    \centerline{\includegraphics[width=6.5cm]{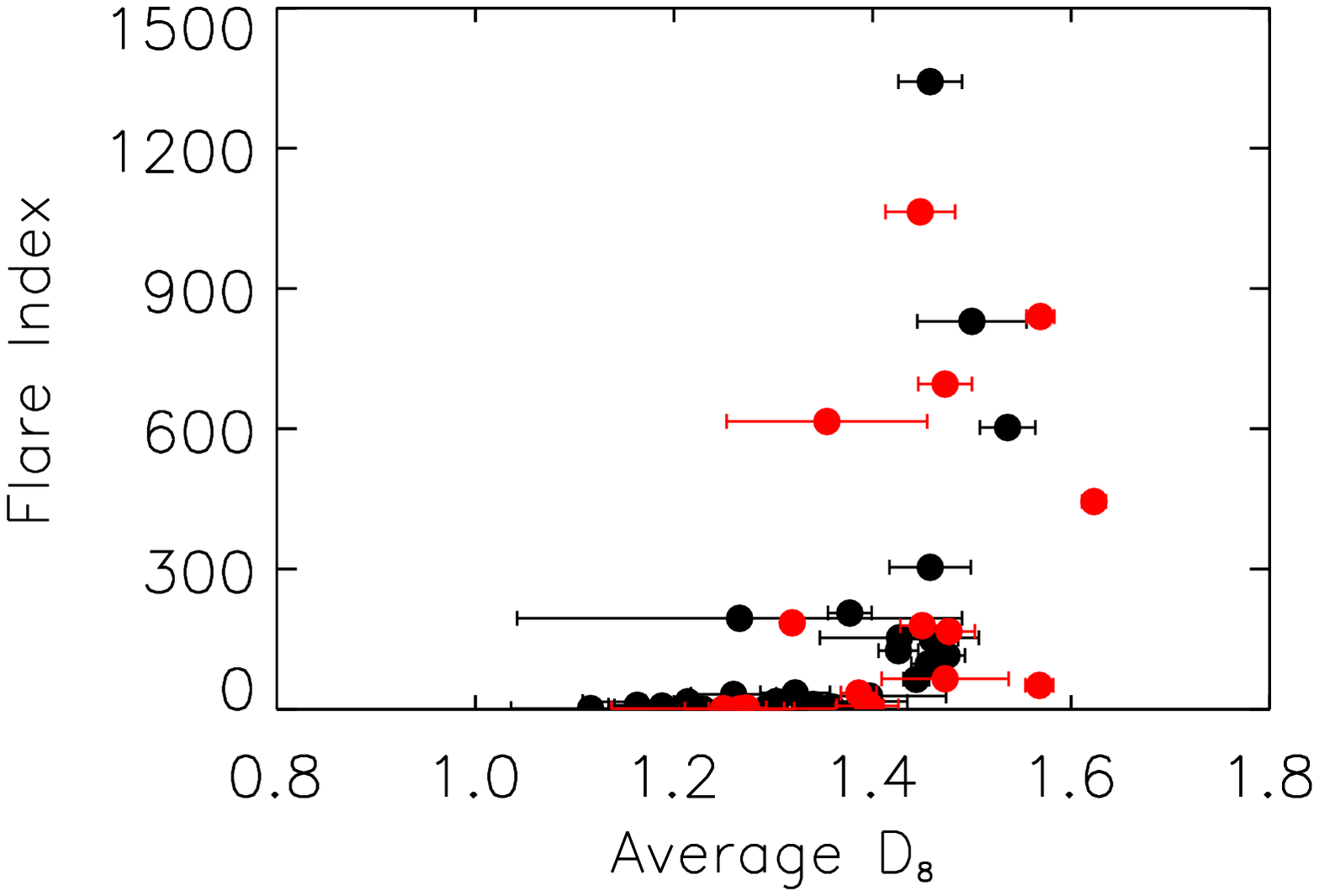}\includegraphics[width=6.5cm]{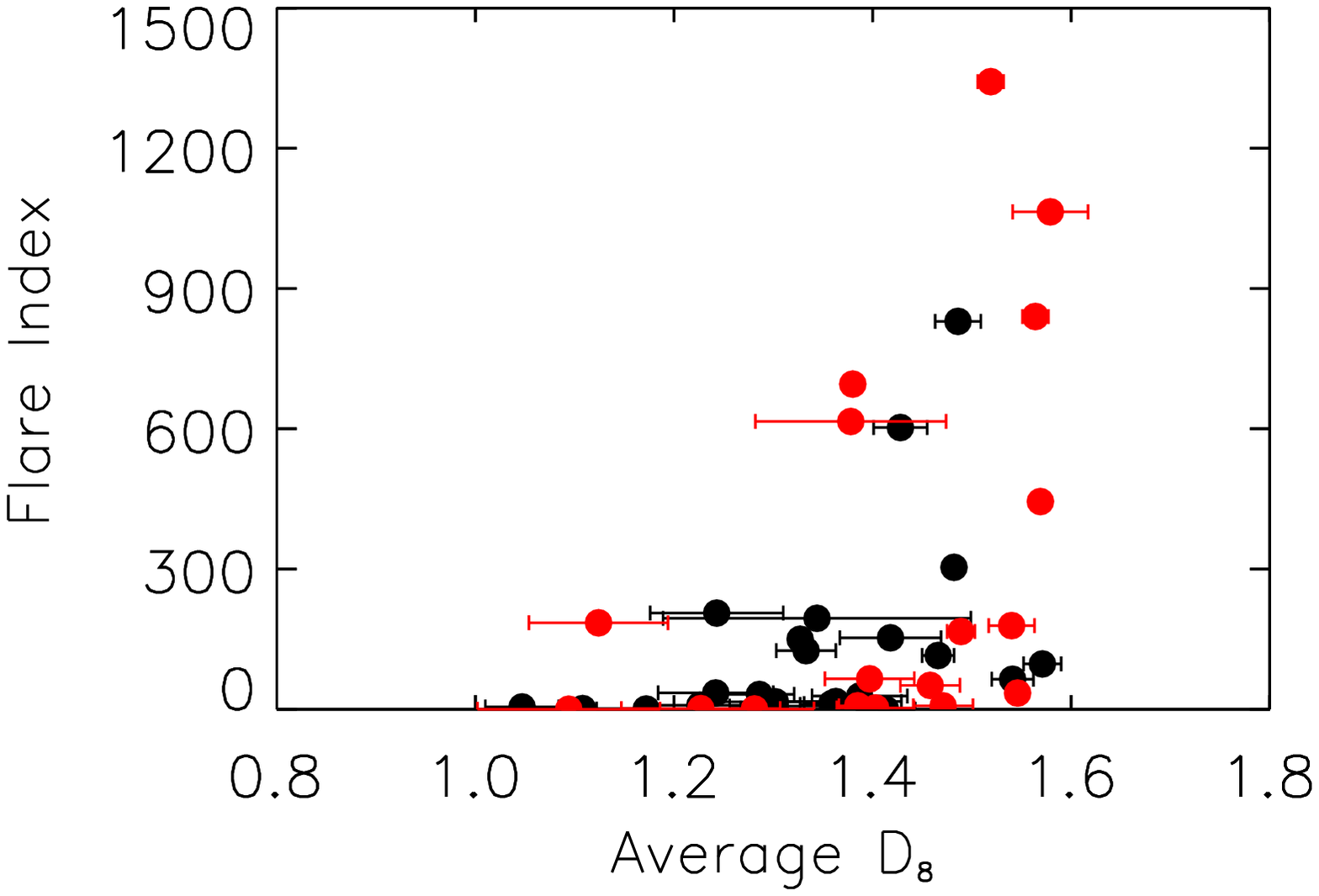}}
    \centerline{\includegraphics[width=6.5cm]{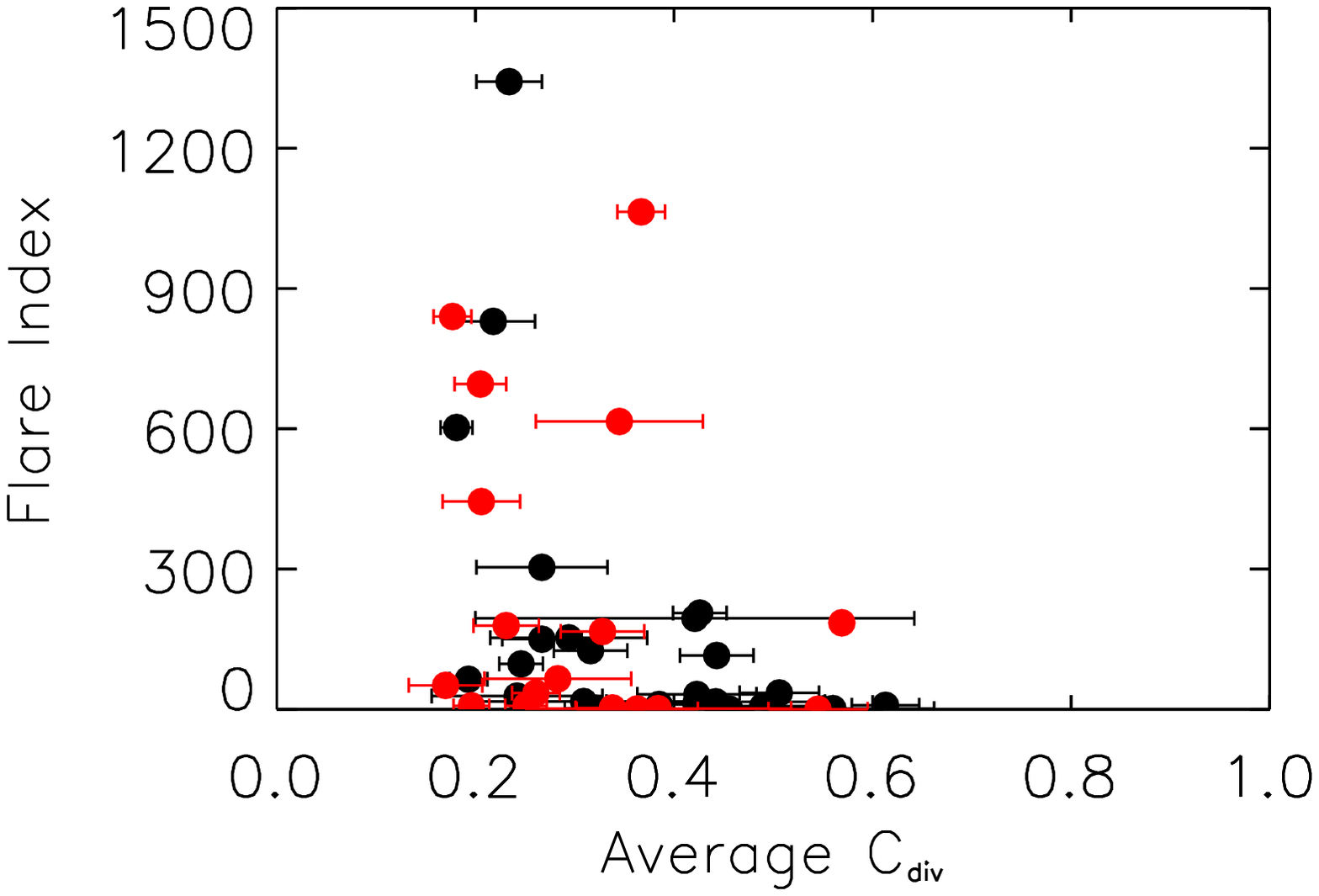}\includegraphics[width=6.5cm]{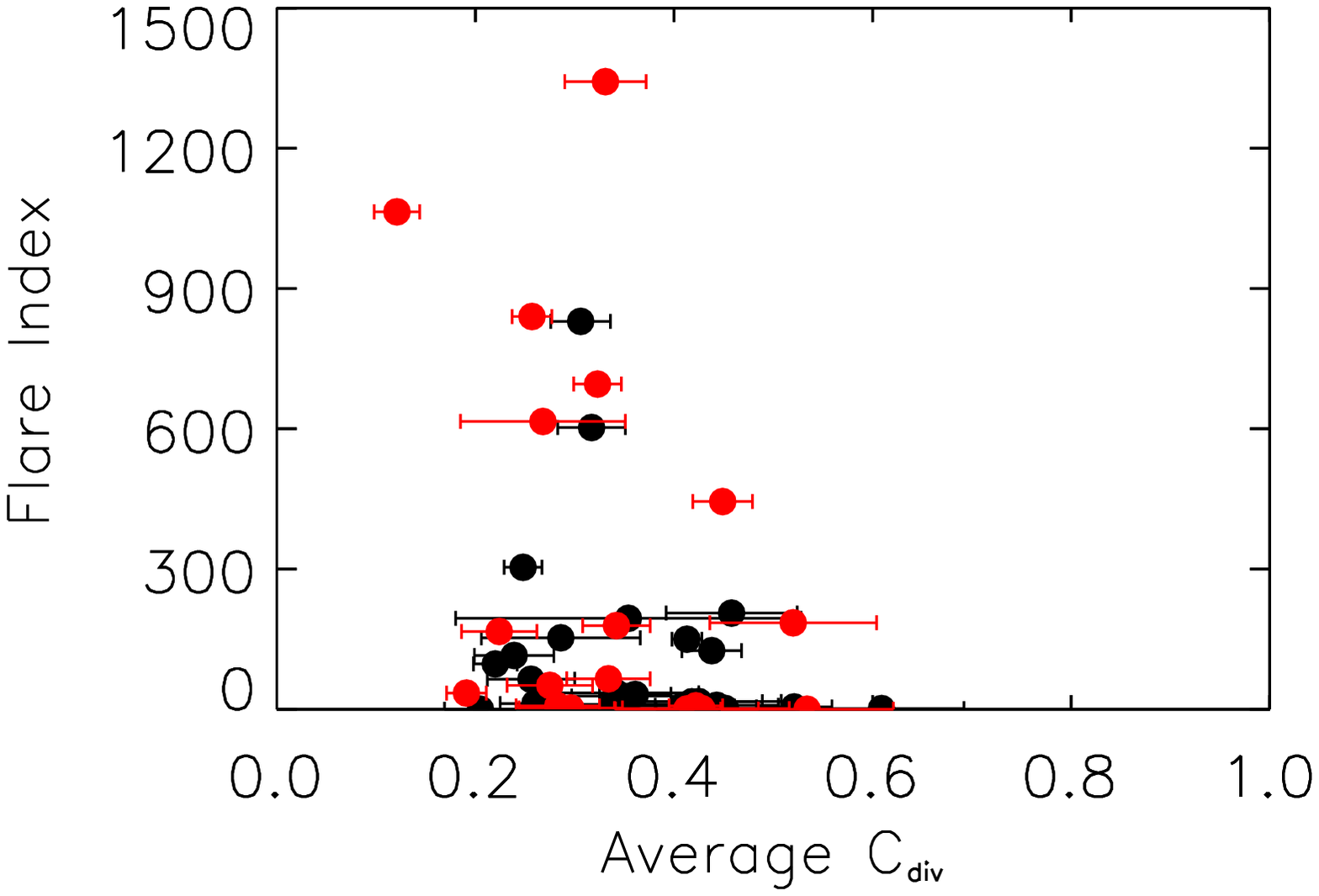}}
    \centerline{\includegraphics[width=6.5cm]{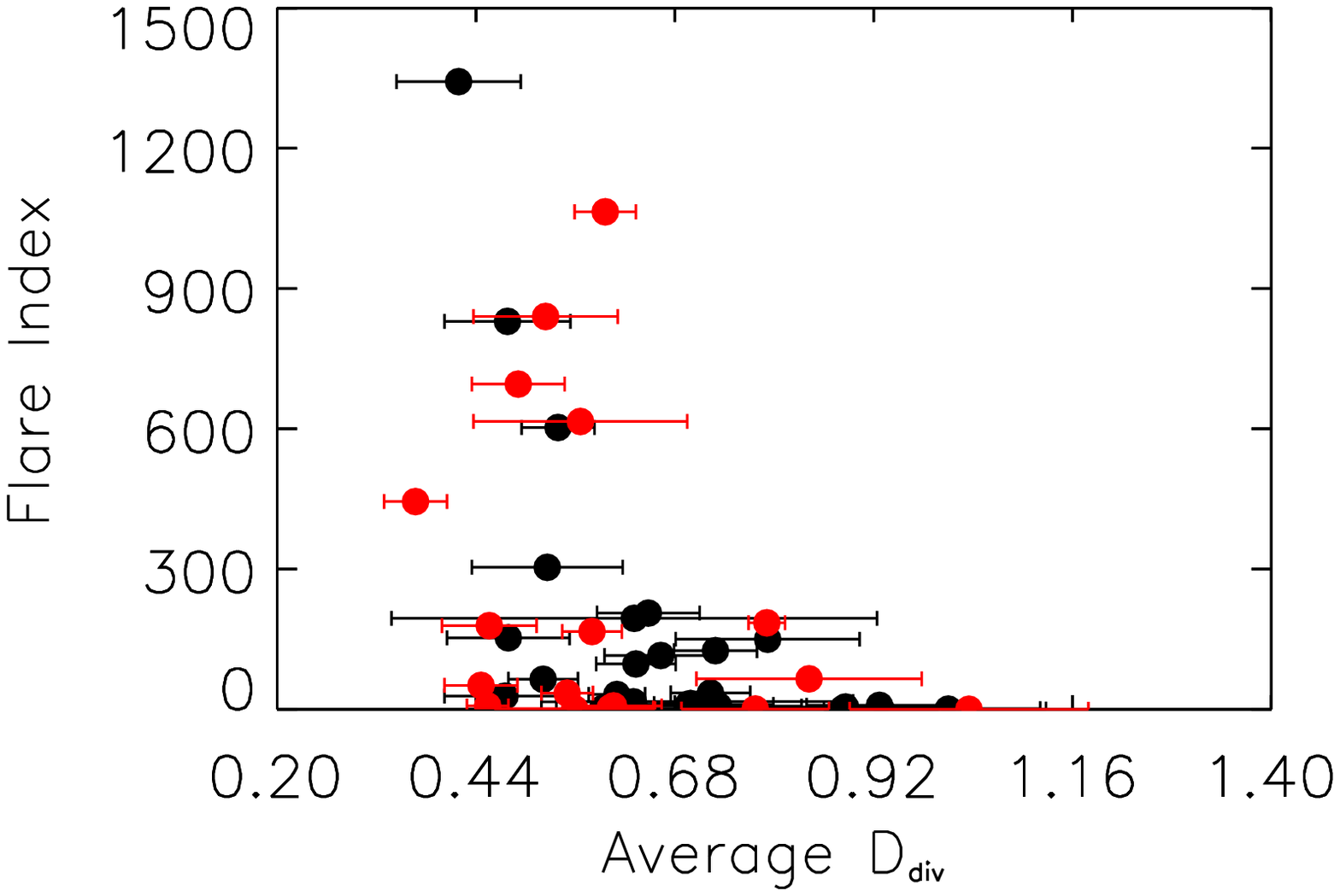}\includegraphics[width=6.5cm]{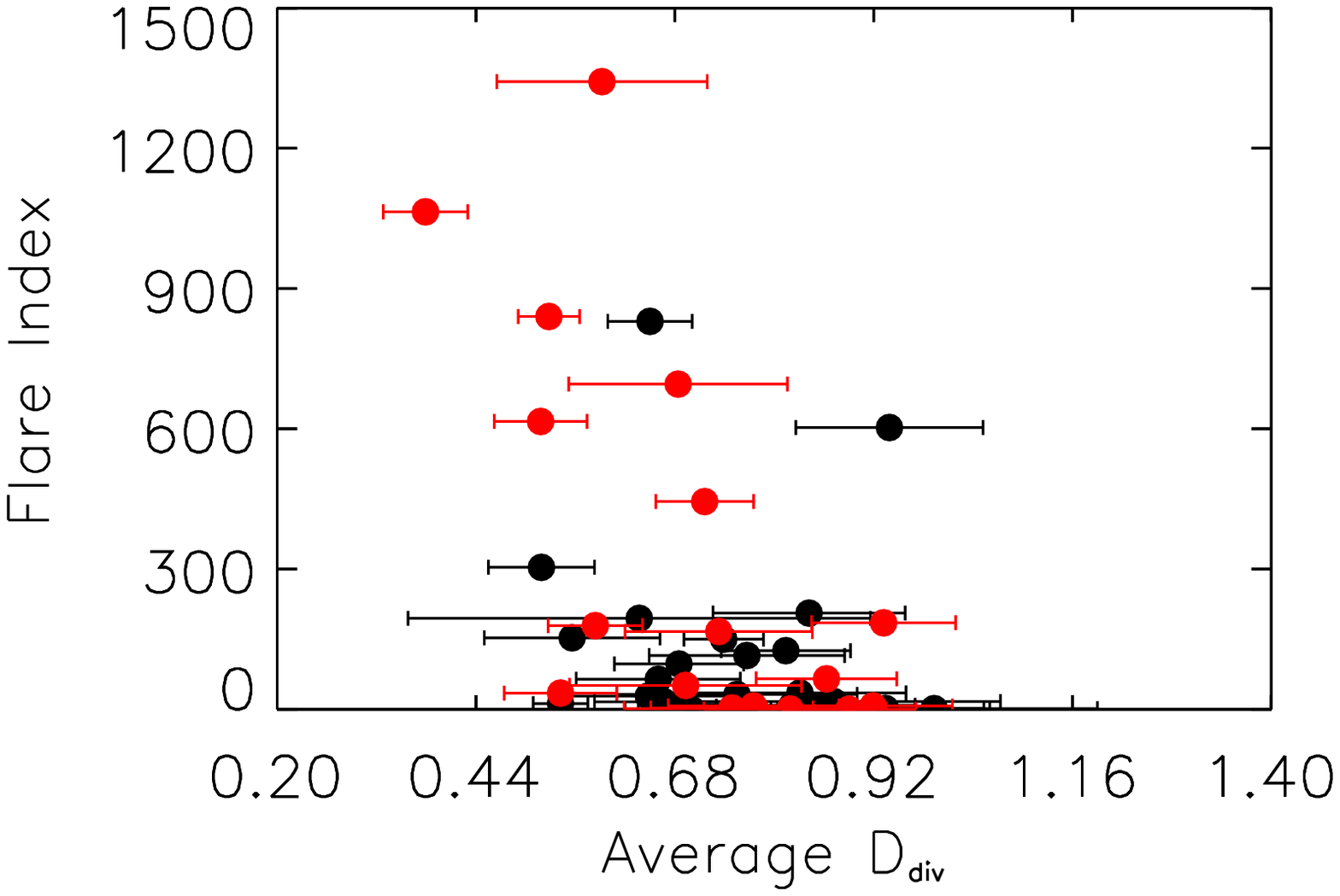}}
              \caption{{ Flare index of the analyzed ARs versus average values of the fractal 
  and multifractal  parameters measured on our sample of { 43}  ARs, which  includes  B- or C-,  M-, and X-class flaring regions. Shown are  results from signed magnetic flux measurements of the leading (left panels) and trailing (right panels) polarity in the analyzed ARs of the generalized fractal dimension  $D_0$ and $D_8$  and multifractal parameters $C_{\mathrm{div}}$ and $D_{\mathrm{div}}$. The ARs  were divided according to  hemisphere hosting them; black and red symbols show results for the ARs in the  northern and southern hemispheres, respectively. }
 }
   \label{f7b}
   \end{figure*}

{ 
Figures \ref{f5a} and  \ref{f5b} show the time series of fractal and multifractal parameters  derived from signed flux measurements of  the analyzed ARs. 
The regions were divided according to the three classes of flaring ARs mentioned above and depending on the hemisphere hosting the region, aiming at  investigating potential  signatures on the measurements of the leading and trailing flux polarities in  the studied ARs, as well as of the hemisphere hosting the regions. In this respect, it is worth noting that all analyzed ARs follow the 
main polarity orientation in the hemisphere, but AR NOAA 11429 that  showed  opposite  polarity orientation than the other regions in the northern hemisphere.

For each AR, we found that  the fractal $D_0$ values estimated on unsigned flux data are  higher than those obtained by taking into account signed flux measurements in the same region, whereas  the fractal $D_8$ and multifractal $C_{\mathrm{div}}$ and $D_{\mathrm{div}}$  values  derived from the region do not vary significantly with the flux data employed to estimate the parameter.
}

{ 
We ascribe the distinct results obtained for the fractal parameter $D_0$ by taking into account unsigned and signed flux data  to the different morphological configuration of magnetic flux concentration in an AR, by considering the region as a whole and its leading and following polarities separately. Indeed, observations  (see {\it e.g.} Section 1 of \opencite{Fan_2009})  show that the flux of the leading polarity in an AR tends to be concentrated in large well-formed sunspots, whereas the flux of the following polarity tends to be more dispersed and to have a fragmented appearance. At any  finite spatial resolution, the latter implies lower average flux values measured  in the analyzed region than actual ones and thus a higher sensitivity of morphological measurements based on flux data to threshold selection effects. 
We found  signatures  in our measurements of these observational evidences  in {\it e.g.} 
the lower average values of the parameters derived from signed flux data with respect to results from unsigned data, as well as of the values from the trailing polarity in the analyzed ARs,   with respect to the values obtained from the data of the leading polarity flux in  the ARs. 
The lower sensitivity of the fractal $D_8$ and multifractal $C_{\mathrm{div}}$ and $D_{\mathrm{div}}$  measurements on the flux data employed for the study than obtained from $D_0$ suggests us that measurements of the former parameters on signed flux data may benefit from analysis of higher resolution observations than employed in this study. In fact, the former parameters, by corresponding  to a higher order estimates of the fractal dimension than $D_0$, may require larger data set to allow similar statistical significance of results than obtained for  $D_0$.
}

{
Figure \ref{f7b} shows the relation  between the average values of the parameters estimated by taking into account the signed flux data of  each analyzed AR and its FI. The results in this Figure confirm the findings derived from the measurements of the parameters based on unsigned flux data of the analyzed ARs. In particular, they confirm both the weak relation between the compared quantities and lack of distinctive hemispheric signatures on the measurements.
}

{ Tables \ref{tbl:2} and \ref{tbl:3} quantify the  results discussed above. In particular, 
the average and standard deviation of  values measured by discriminating the flux data in the analyzed ARs reported in Table \ref{tbl:2}, as well as the Pearson correlation coefficients between the measured values and FI of the analyzed ARs  listed in Table \ref{tbl:3}, confirm previous conclusions by \inlinecite{Georgoulis_2005} on that  discriminating between the flux polarities in the ARs do not alter  the results derived from  analysis of total unsigned flux data of the regions. 
Indeed, in  this study, the lower values  obtained for the Pearson coefficients of  the parameters based on signed flux data in  the analyzed ARs show that discriminating between the flux polarities  has even reduced the weak correlation between the compared  quantities, with respect to results based on unsigned flux data of the ARs.
}



\subsection{ Temporal Evolution of the  Parameters}

  { 
 We then focussed our attention on the temporal evolution of the various parameters measured on the whole sample of analyzed ARs, aiming at identifying any clear  patterns on the parameter  series   that can be associated with the flare activity of the ARs. 
 Here in the following the term variation indicates the deviation of measured values with  respect to the average and  long-term trends of the parameters. These trends  were  defined as specified in the following. The variation of measured values has been  quantified by the variance of the distribution of measured values, as well as by the distinct patterns identified on the analyzed series. 
}

     \begin{figure*}
        \centerline{\includegraphics[width=10.5cm]{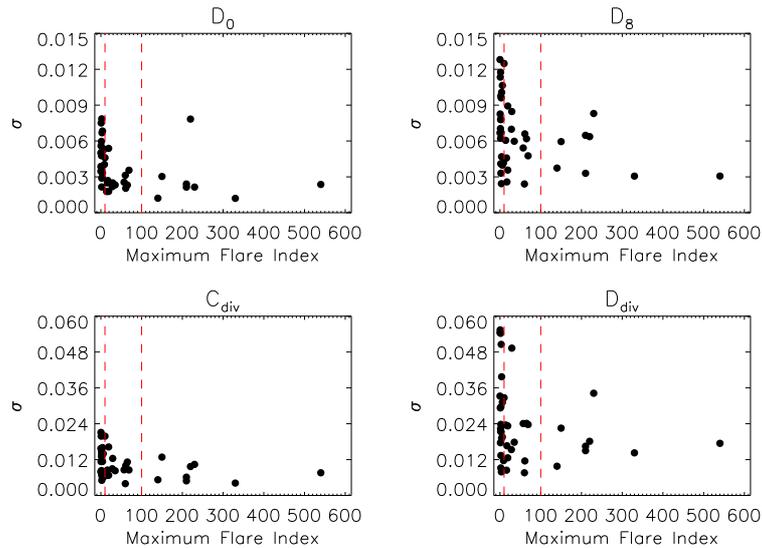}}
              \caption{{ 
              Variance of the distribution of the fractal and multifractal parameters measured on each region versus the maximum flare index of the studied AR. The parameter values refer to the deviation of measurement results with respect to the  average trend of the parameter that was obtained with a running average of the values measured over a 3 h time interval. Shown are  results from unsigned magnetic flux measurements in the ARs of the generalized fractal dimension  $D_0$ and $D_8$ (top panels), and multifractal parameters $C_{\mathrm{div}}$ and $D_{\mathrm{div}}$  (bottom panels). The ARs were divided into three classes (vertical dashed lines), depending on the value of their maximum flare index.}}
   \label{f10}
   \end{figure*}

{ 
First, we  studied the distribution of parameter values measured for each region during its  transit over the solar disc and the  dependence of the distribution variance $\sigma$ on the
flaring activity of the AR. In particular, we computed the distribution of parameter  values measured for each region with respect to the average and  longer-term trends of the parameters, which were obtained with a running average of values measured over 3 and 12 h time interval, respectively. We then estimated the width of the Gaussian function that best fits the measured distribution and assumed this width as an estimate of the distribution variance $\sigma$. 

Figure  \ref{f10} shows the $\sigma$ measured for each AR in the analyzed sample with respect to its  FI max. We found that  the distribution of values derived from the less flaring regions are characterized by a larger variance with respect to the ones obtained from the more flaring ARs. 
However,  the variance of the values derived from all analyzed ARs does not show any clear dependence on the flaring activity of the regions, by disclaiming the idea that more flaring ARs may leave the signature of their activity in more varying time series of the measured parameters. This result holds likewise for the variance of the values measured with respect to both the average and longer-term trends of measured values. 
}

{ We then analyzed the time series derived from each AR and their main features.  A sample of  the analyzed data has been already shown in Figures \ref{f8} and \ref{f9}. 
 We found that the time series of  the $D_0$ and $D_8$ fractal parameters measured on the whole sample of ARs show smaller and more regular variations during the analyzed time intervals than the corresponding trends of the $C_{\mathrm{div}}$ and $D_{\mathrm{div}}$ multifractal values. 
}
This holds for results derived from both unsigned and signed flux data  in the studied ARs. { We found that $D_8$ shows slightly larger changes than $D_0$, as $D_{\mathrm{div}}$ also does with respect to $C_{\mathrm{div}}$, and particularly when results obtained from signed flux data are taken into account. 
This result suggests us that measurements of these parameters on signed flux data may benefit from  analysis of higher resolution data than employed in this study.}

We also found that most of the studied  series  show different evolution  during the analyzed time intervals depending on the flux data employed to estimate the parameters. 
In particular, we found that the time series of the parameters derived from a given field polarity  are less consistent with those obtained from unsigned flux data on the same regions, which resulted to be  rather representative of the parameter trends deduced from the opposite flux polarity in the  ARs. Specifically, the most varying trends for all measured parameters were found to be the ones deduced from the trailing flux data of the analyzed ARs, {  i.e. the data of the flux concentration with higher fragmentation and thus also higher sensitivity of measurement results to the finite spatial resolution of the analyzed data. 
}
 The trends  of the parameter estimates based on  the prevailing leader  polarity in the  hemisphere hosting the AR were found to be  more consistent with  those obtained from unsigned flux data. { These features  of the time  series, specifically the similarity between the trends deduced from unsigned and leading polarity flux of the analyzed ARs and the deviation between  the same trends  and  those from the trailing polarity flux,  were found to  hold for $\approx$ 90\% of the studied regions. It is worth noting that the parameter time series of AR NOAA 11429, which is the only region in the analyzed data set  that showed opposite polarity orientation than the other regions, do no follow the results derived from most of the  data analyzed.}

From analysis of the parameter series we also found that several flares occur during a decreasing phase of the$D_{\mathrm{div}}$ and an increasing phase of the $D_8$ values estimated by considering  unsigned and signed flux data of the leading polarity of the AR hemisphere. However, those features of the parameter trends were not found to be a consistent pre-flare signature on the whole sample of ARs and events.  Indeed, this event signature was found on $\approx$ 50\% of the analyzed ARs and   on $\approx$ 50\%  of the $\ge$ M-class events considered. We also found that the$D_{\mathrm{div}}$ trends obtained by considering the trailing polarity of the ARs show  changes lasting few hours ($<$ 10) and occurring both prior ($\approx$ 15 h) and after major events, as  shown in Figure \ref{f9}  for {\it e.g.} AR  11158 and AR 11875, where  B- and B+ is the trailing polarity in the region, respectively.   These changes in $D_{\mathrm{div}}$, whose amplitude ranges 0.20\% to 0.25\% of the parameter value independently of the event class, are  compatible with a  simplification of the photospheric magnetic field  in flaring ARs through sudden changes of the trailing flux topology. However, they 
were found on $\approx$ 50\% of the analyzed ARs and events.


\section{ Conclusion}

{ The present study  shows that the  analysis of a large data set of higher spatial and temporal resolution, as well as higher flux sensitivity, observations than employed in previous studies does not give the fractal and multifractal parameter measurements  a higher efficiency to discriminate ARs depending on their flaring activity than reported  in the literature.   The 
measurements derived from our study hint to  a signature of the flare activity in the fractal and multifractal parameters of ARs and a potential  differentiation depending on their  flare activity, but this does not hold in general for all the  analyzed ARs. In particular, we found distinct average values of  the parameters measured on ARs that have hosted flares of different class, in agreement with the statistical trend reported by \inlinecite{Mcateer_etal2005}. At a first glance,  our findings contrast with results presented by {\it e.g.} \inlinecite{Georgoulis_2013}. However, we also show that the dispersion of values deduced from each class of flaring AR considered in our study is such that results from extreme classes of flaring ARs can also overlap, as shown in {\it e.g.} Figure 8 of   
 \inlinecite{Georgoulis_2013}. 
}

 The  results obtained by our study  also show that the parameter estimates based on signed flux measurements do not constitute 
 a better predictive tool of the flare activity than derived by using  unsigned flux data as in previous studies. 
 In particular, we  found that  the time series of the parameters obtained  from  unsigned flux data of the analyzed ARs  are rather representative of  those deduced from the signed flux measurements of the prevailing leader  polarity in the  hemisphere hosting the ARs. 
Both  time series 
  show  that the  flares occur during a decreasing phase of the$D_{\mathrm{div}}$ and an increasing phase of the $D_8$ parameters  on $\approx$ 50\% of the analyzed ARs and on  $\approx$ 50\% of the $\ge$ M-class events considered. The times series of parameters deduced from signed flux data of the trailing polarity flux in  the analyzed ARs are less consistent with those derived from unsigned flux data of the studied region. 
    The former time series show  changes of the  evolution of the $D_{\mathrm{div}}$ values  
 lasting few hours and occurring both prior and after major events. These  changes, which  are compatible with a  simplification of the magnetic field in the ARs,  were found   on $\approx$ 50\% of the analyzed regions and events.

The  results obtained in this study give reasons to  the conflicting conclusions reported in the literature on the efficiency of the fractal and multifractal parameter measurements based on unsigned flux data  of ARs to discriminate them depending on their flaring activity.  
They also  extend  the  validity of the previous findings  to results derived from signed flux measurements of the analyzed ARs. 
It remains to be shown whether  other parameters describing the magnetic field evolution and morphological complexity  based either on signed  or other flux data of  the analyzed regions  prove to be more ideal tools for discriminating flaring ARs than those analyzed in this study.

%
 \begin{acks}
The research leading to these results has received funding from the European Commission's Seventh Framework Programme (FP7/2007-2013) under the grant agreements eHEROES (project n 284461, www.eheroes.eu) and SOLARNET (n 312495, www.solarnet-east.eu). This work was also supported by the Istituto Nazionale di Astrofisica (PRIN-INAF-2010). The authors acknowledge
useful discussions from Giuseppe Consolini and Gherardo Valori.
 \end{acks}

%
%
 \bibliographystyle{spr-mp-sola}
 \bibliography{fract_eheroes_bib}  
%
%
%
%

\end{article} 
\end{document}